# A Bidirectional Diode-Clamp Circuit Paradigm for Time-Resolved Measurement of Electrical Short-Circuits

**Alex Mwololo Kimuya**[1,*] **and Dickson Mwenda Kinyua**[1,2]

[1] Department of Physical Sciences (Physics), Meru University of Science and Technology, Kenya
[2] Department of Pure and Applied Sciences, Kirinyaga University, Kenya
*Corresponding Author Email: alexkimuya23@gmail.com; Phone: +254-704600418; ORCID; https://orcid.org/0000-0002-1433-3186

**Abstract.** Conventional electrical fault models, which rely on static thresholds and instantaneous trip mechanisms, treat short-circuits as destructive singularities and create vulnerabilities in modern power systems. Those models fail to capture the time-evolving dynamics of real faults, where phenomena such as resistance decay and arc formation persist for seconds and evade conventional protection. This paper introduces a diode-clamp circuit architecture that reconceives short-circuits as governed, sustained processes and establishes a physics-consistent, time-resolved measurement system. An Arduino microcontroller-based data acquisition system recorded continuous fault evolution during two core experiments, "*Clamped Nominal Short*" and "*Clamped Extreme Short*", across input voltages of $2.5V$, $5.0V$, and $10.0V$ for durations of 3 -minutes, 10 -minutes, and 15 -minutes. Three measurements phases; *pre-fault*, *fault*, and *post-fault* (recovery phase) were performed each at $10ms$, $50ms$, and $100ms$ sampling resolutions, satisfying the Nyquist - Shannon sampling criterion. This multi-resolution sampling enabled high-fidelity capture of both fine-grained transients within $10ms$ and sustained-state dynamics at both ($50ms$ and $100ms$). The clamped, bidirectional short mechanism constrained the circuit to a bounded electrical regime, enabling repeatable, high-fidelity observation of fault dynamics. From the Clamped Extreme Short configurations, the system produced definitive, measurable minima and maxima: voltages settled at non-zero floors ($0.073V$ at $2.5V$; $0.069V$ at $5.0V$), currents were limited to finite maxima ($13.01A$ at $2.5V$; $13.03A$ at $5.0V$), and resistances decayed exponentially to finite floors ($5.61m\Omega$ at $2.5V$; $5.30m\Omega$ at $5.0V$). These findings empirically refute the classical assumption of instantaneous, unbounded current. Newly introduced metrics quantify this performance. The Sustained-to-Capacitive Energy Ratio (SCER) reached $\sim 1.53 x 10^{12}$, proving fault energy originates from sustained dynamics, not transient capacitance discharge. The Sustained Fault Efficiency ($SFE > 1$) was 2.71 at $2.5V$ and 1.76 at $5.0V$, demonstrating that governed fault power can exceed nominal operating power. The Transient Clamping Index (TCI) was approximately 137.7, confirming rapid stabilization on a millisecond timescale. Signal analysis using Shannon entropy and Fourier-based methods verified capture of authentic physical processes without aliasing. The Diode Clamping Efficiency Index ($DCEI \approx 1$) further corroborates robust circuit protection, with energy dissipation maintained as a sustained power state quantified by the SCER. This work provides the first fully validated short-circuit quantification and analysis system based on time-resolved measurements. The generative design presented here yields empirical fault data suitable for next-generation battery management, adaptive grid protection, fault-tolerant electronics, and AI-based diagnostics. Future work will integrate thermal sensing to correlate junction temperature with circuit performance and will explore scaled, higher-voltage implementations.

**Keywords.** Governed Short-Circuit, Diode-Clamp Measurement, Time-Resolved Fault Analysis, Modified Ohm's Law, Non-Zero Resistance Floor, Bounded Electrical Transients, Dynamic Resistance Decay, Fault Sustanance, Transient Clamping Index (TCI)

## 1.0 Introduction

The relentless expansion of global power infrastructure, characterized by increasing interconnectivity and the widespread deployment of distributed energy resources (DERs) such as photovoltaic systems, electric vehicles, and heat pumps [1], has exposed a critical and unresolved vulnerability: the accurate measurement and interpretation of uncontrolled electrical faults, including short-circuit voltages and currents. Existing protection systems -circuit breakers, fuses, digital relays -often operate on fixed threshold principles [2], [3], [4], codified in standards such as IEC 60909 [5] and ANSI/IEEE [6], [7] guidelines. A core limitation of these standards is their reliance on a linear relationship between electrical quantities, which fails to capture the non-linear realities of a short-circuit event. Furthermore, these models simplify transient fault behavior into peak current magnitudes and assume linear, instantaneous system responses, often incorporating theoretically useful but practically limited characteristics, such as the assumption that faults begin at zero voltage or that the current transitions instantaneously from a pre-fault to a fault state [8]. This reductionist approach suppresses essential physical dynamics such as resistance decay, arc formation, and thermal accumulation, treating faults as abrupt, binary events. Consequently, real-world failure modes that evolve over tens to hundreds of milliseconds, including those where high asymmetry can lead to a lack of current zero, critically impacting circuit breaker operation, remain undetected.

Latent faults in battery modules, oxidized contact interfaces, and partially conductive plasma channels often generate sub-threshold current profiles that persist for seconds without triggering protection devices. For example, in photovoltaic arrays, experimental investigations have shown that DC series arc faults can remain undetected for over four seconds before transitioning into sustained arcing conditions [9], [10]. In lithium-ion battery packs, self-heating mechanisms can initiate ignition tens of seconds before thermal runaway, evading both overcurrent and thermal safeguards until catastrophic failure occurs [11], [12], [13]. Medium-voltage switchgear installations have recorded partial-discharge events accumulating over several milliseconds below relay trip thresholds, ultimately leading to insulation breakdown and equipment damage [14], [15]. Distributed DC microgrids similarly experience series arc faults lasting up to $100 ms$ that conventional overcurrent relays and black-box arc-model protections neither interrupt nor detect, resulting in insidious degradation of system components [16].

However, the very instruments used to diagnose these faults are themselves constrained by the same static, reductionist paradigm. For instance, the multimeter provides a singular resistance value at a non-operational voltage, failing to capture the dynamic resistance collapse of an evolving arc, a transient process that requires high-speed data acquisition for accurate parameter estimation [17], [18], [19]. The clamp meter samples current at utility frequencies, blind to the sub-millisecond current harmonics and high-frequency interference that signify incipient insulation failure, as its limited temporal resolution cannot resolve the rapid transients at the initiation of a short-circuit fault [19], [20]. The thermal imager reveals the thermodynamic symptom -heat -only after significant energy has been dissipated, a lagging indicator of a process already underway due to the inherent thermal inertia of the systems [21]. Even the Megohmmeter's high-voltage stress test is a pass/fail snapshot that cannot quantify the progressive, time-dependent degradation of dielectric materials, as it does not monitor the continuous evolution of leakage current or partial discharge activity [22]. These tools, while practical for locating catastrophic shorts, operate on the foundational paradox of modeling a fault as either an infinite current or an infinite resistance, two idealizations that do not exist in physical systems. In reality, a fault is a time-variant impedance, a dynamic pathway whose conductivity can evolve over microseconds to seconds. The limitations of these conventional instruments create a critical data gap, producing fragmented, low-temporal-resolution snapshots that are insufficient for modeling the continuous physical trajectory of a fault from its initiation to its culmination [18], [20]. This diagnostic inadequacy necessitates a new class of measurement system, one capable of quantifying the full electrical chronology of a short-circuit beyond the idealized binary states.

On the other hand, protection schemes based solely on current-magnitude thresholds further exhibit high sensitivity to high-frequency switching noise, often leading to spurious breaker activations in power-electronics-driven systems [23]. Pulse-width modulation in electric vehicle chargers and solar inverters induces rapid $\left(\frac{dV}{dt}\right)$ transients that conventional trip curves mistake for genuine faults, compromising microgrid stability and continuity of service [24]. These limitations highlight the inadequacy of static protection models, which disregard nonlinear fault trajectories, persistent voltage drops across damaged junctions, and the thermodynamic processes governing sustained faults. As a result, AI tools trained on datasets lacking these temporal and physical complexities suffer from overfitting, poor generalization, and misclassification -particularly in applications involving arc detection, dielectric failure, and thermal runaway scenarios [23], [24], [25].

The objective of this paper is to establish a physics-consistent, time-resolved framework for characterizing nonlinear short-circuit dynamics beyond the constraints of conventional protection models. To achieve this, the methodology introduces an experimental architecture that re-conceptualizes short-circuit events not as abrupt, instantaneous anomalies but as temporally extended electrical processes. Central to this framework is the implementation of a diode-clamp topology, which enforces a bounded voltage regime designed to preserve fault continuity without initiating circuit interruption. Rather than suppressing the fault, the voltage clamp imposes a controlled electrical boundary that enables the complete temporal unfolding of the event -capturing its full trajectory from initiation, through progressive resistance decay, to eventual current saturation. This bounded environment supports sustained fault conduction while safeguarding adjacent circuitry, allowing for high-fidelity measurements that preserve the transient's physical authenticity.

The diode-clamp configuration forms the empirical foundation of a novel measurement basis, enabling internal fault processes -such as resistance collapse, arc ignition, and thermal evolution -to be directly observed and quantified in real time. The architecture supports this capability through the integration of high-speed data acquisition systems that sample voltage and current waveforms at microsecond-level resolution, ensuring temporal precision necessary for tracking nonlinear behaviors. This measurement strategy permits the isolation of key features including waveform entropy, spectral evolution, and asymmetrical current ramps -parameters often invisible to peak-magnitude-based protection schemes. In capturing these features under controlled electrical constraints, the platform constructs a bridge between theoretical models of fault evolution and the empirically observed complexities arising in modern energy systems. The diode-clamp paradigm thereby serves as both a diagnostic instrument and a generative fault environment, offering a scalable and reproducible means to expand the precision, scope, and physical relevance of short-circuit analysis.

The methodology begins with an unconventional yet experimentally feasible circuit design -the diode-clamp paradigm. At its core, the system employs a short-parallel network of fast-recovery diodes, initially demonstrated at 3 -minutes intervals across two phases, pre-short-circuit and post-short-circuit, in [8], and here validated under additional experimental configurations. This arrangement enforces a stable voltage clamp at fault initiation, preventing voltage collapse and enabling continuous observation of evolving fault properties. The diode clamp creates a voltage ceiling that bounds current rise, avoiding the catastrophic singularity implied by classical Ohm's Law when resistance is assumed to vanish during electrical short-circuits [6], [8], [26]. This configuration enables real-time, long-duration measurements of current, voltage, and inferred resistance under controlled fault conditions using cost-effective components such as Hall-effect current sensors and microcontroller-based voltage dividers, with acquisition rates exceeding the Nyquist criterion for typical fault bandwidths to preserve signal integrity, minimize aliasing, and enable noise-resilient waveform analysis.

A key distinction of the proposed architecture lies in its incorporation of ionized conduction paths - phenomena traditionally examined within the context of high-energy plasma physics and the study of repetitive pulse discharges in lightning, which involve the formation of conductive plasma channels [27]. In this implementation, such conduction pathways are re-contextualized to maintain non-zero voltage floors during transient conduction. These paths allow for the direct observation of dynamic resistance changes driven by internal heating, material degradation, and recurrent arc phenomena. Unlike conventional circuit models that treat such behaviors as post-event consequences or simplify them into static parameters, this design captures these transitions as real-time, quantifiable electrical signatures. The system's high-resolution voltage and current sampling -operating at microsecond-level time scales - reveals complex signal features including waveform entropy, frequency-domain irregularities, and asymmetries in current ramp profiles. This level of resolution is critical, as studies in fields from inductively coupled plasma mass spectrometry to neurophysiology demonstrate that microsecond-scale acquisition is necessary to accurately capture transient events and avoid the artifacts inherent to millisecond-scale sampling. These characteristics, often obscured or omitted in peak-based fault analyses, offer new dimensions for understanding transient circuit behavior.

The shortcomings of conventional simulation platforms further reinforce the need for this experimental paradigm. Simulation tools such as PSCAD, Simulink, and ATP-EMTP often employ idealized, time-invariant components, truncating events within a few cycles to avoid numerical instability or hardware replication. They utilize oversimplified arc models -fixed resistors, black boxes, or ideal switches -that disregard thermal gradients and dynamic impedance variations. Sensor limitations exacerbate these issues; Hall-effect sensors saturate during high $\left(\frac{di}{dt}\right)$ events and fail to resolve anomalies in sub-cycle intervals. These constraints result in datasets lacking the temporal, spectral, and statistical fidelity required to replicate real-world fault evolution. Consequently, both protection system design and AI-based fault analysis are grounded in incomplete representations of fault behavior.

Laboratory protocols also limit fault duration to microseconds or a few milliseconds, thereby omitting transitional features such as ramp compression, entropy spikes, re-ignition precursors, and thermal inflection points. Without exposure to these dynamics, AI models cannot learn the physics governing fault persistence and dissipation. As a result, generalization and classification suffer, especially in high-risk contexts such as dielectric breakdown or sustained arcing faults. This paper addresses these limitations by introducing a generative fault platform capable of sustaining faults within bounded electrical regimes, providing rich temporal and physical information for both model development and theoretical analysis.

The diode-clamp configuration establishes a stable voltage regime that constrains the classically assumed infinite current increase and mitigates the catastrophic singularity anticipated under classical Ohm's Law. This electrical boundary enables sustained observation of fault evolution under dynamically varying conditions. Complementing this hardware architecture is a robust analytical framework grounded in modern signal processing and nonlinear dynamics. Signal integrity is rigorously maintained in accordance with the Nyquist-Shannon Sampling Theorem, while waveform complexity is characterized using Shannon entropy to capture underlying structural disorder. The temporal evolution of spectral content is tracked using short-time Fourier transforms (STFT), enabling frequency-domain resolution of fault transitions. Together, these methods construct a multidimensional framework for analyzing fault behavior -not limited to peak magnitude or duration but encompassing resistance decay, thermal accumulation, and energy dissipation dynamics. This approach redefines short-circuit analysis from a scalar, threshold-based process into a time-resolved, vectorial discipline capable of capturing the nonlinear trajectories inherent in real-world electrical faults.

This work makes foundational and far-reaching contributions. It provides the first experimental validation of a Modified Ohm's Law under sustained fault conditions, capturing nonlinear resistance decay and current saturation consistent with theoretical expectations. The paper introduces novel diagnostic metrics -such as the Clamped Bidirectional Short, Clamped Short Sustained Power (CSSP), and Transient Clamping Index (TCI) -to quantitatively characterize fault profiles. It further generates real-time, empirical datasets ideal for training next-generation AI-based detection frameworks. The methodology advances a non-Kirchhoffian framework for fault analysis, enabling the modeling of bounded, directional dissipation systems with persistent current and voltage floors. Through quantitative metrics including Sustainance Efficiency (SFE), the Diode Clamping Efficiency Index (DCEI), and the Sustained-to-Capacitive Energy Ratio (SCER), the paper establishes the circuit's performance, necessitating a redefinition of dissipation towards a concept of "*sustainance*" within clamped experimental frameworks. Finally, it establishes a versatile platform for education, research, and industrial validation.

Potential applications for this work are broad and urgent. In battery management systems -particularly for electric vehicles and grid-scale storage -the proposed system enables predictive fault models that incorporate energy accumulation rates and resistance decay. In smart grids, the ability to track fault progression in real time facilitates accurate isolation and adaptive relay behavior. The experimental framework serves academic programs by offering a hands-on platform for papering transient electrical physics, while AI researchers benefit from physically grounded, temporally rich training datasets. In industrial contexts, solid-state interrupters and current limiters can be redesigned using insights derived from time-resolved fault signatures.

The implications of this paper extend beyond diagnostics. Current design practices necessitate overengineering to account for worst-case peak assumptions, resulting in oversized breakers and costly inefficiencies. A more rigorous understanding of short-circuit dynamics -enabled through empirical observation -permits more accurate sizing, targeted intervention, and reduced safety margins. This benefits aerospace, maritime, and off-grid applications where size, weight, and energy optimization are paramount.

In sum, this paper introduces a paradigm shift that is both technological and epistemological. It replaces abstract simulation with empirical measurement, instantaneous assumptions with continuous evolution, and static models with nonlinear understanding. Through the diode-clamp methodology, the Modified Ohm's Law becomes measurable in practice; through rigorous analysis, transient electrical failure becomes interpretable; and through experimental ingenuity, the barriers between theoretical abstraction, simulation limitation, and physical reality begin to dissolve.

This paper redefines the processes of fault measurement. It provides a coherent, experimentally grounded language for discussing electrical failure as a continuous, nonlinear process. As power systems evolve towards fault-resilient, AI-integrated, and high-density architectures, the capacity to characterize, predict, and manage faults in real time will become essential. Positioned at the frontier of this transformation, the diode-clamp architecture offers the theory, tools, and vision to fundamentally reshape the science and engineering of short-circuits. The remainder of this paper is structured as follows: Section 2 details the theoretical framework governing the experiments; Section 3 outlines the experimental methodology; Section 4 presents the results; Section 5 discusses the results and their implications; and Section 6 provides the conclusion.

**2.0 Theoretical Foundations**

The recently observed dynamic nature of short-circuit phenomena ([8]) demands a departure from the traditional, static formulations. This section develops a foundational model for interpreting short-circuits as time-evolving transitions rather than abrupt anomalies. It introduces a time-resolved transformation of Ohm's Law tailored for transient, non-linear conditions -specifically those involving resistance collapse, arc formation, and material ablation. The framework begins by extending Ohm's Law into a modified exponential form that accommodates evolving system parameters, then transforms this formulation into a physically interpretable decay law grounded in measurable quantities. By structuring the analysis across three temporal regimes -pre-fault equilibrium, fault transition, and post-fault stabilization -the model reconstructs fault events as continuous evolutions. The exponential resistance decay, serving as the mathematical core, replaces the oversimplified notion of instantaneous resistance drop with a physically bounded transition metric. This sets the stage for distinguishing between classical assumptions and experimentally observed transient behavior, offering new grounds for interpreting resistance, current, and power during short-circuit events.

**2.1 Time-Resolved Transformation of the Modified Ohm's Law**
The classical formulation of Ohm's Law provides an accurate description of steady-state electrical behavior in linear conductors but fails fundamentally during the non-equilibrium, time-dependent processes of evolving fault conditions. This fundamental limitation demands for a theoretical framework that accounts for the fundamental transformation of material properties during electrical breakdown, where conductor resistance undergoes continuous evolution due to thermal energy deposition, electron avalanche processes, and material phase transitions. This paper establishes the physical basis for resistance collapse while maintaining consistency with fundamental principles of charge transport and energy conservation. The theoretical framework originates from the fundamental description of electrical conduction in materials. For a conductor with cross-sectional area $(A)$, length $(L)$, and charge carrier density $(n)$, the current density $(\vec{J})$ relates to the electric field $\vec{E}$ through the constitutive relation governing charge transport:

$$\vec{J} = \sigma \vec{E} \tag{1}$$

where $(\sigma)$ represents the electrical conductivity. During transient fault conditions, this conductivity becomes explicitly time-dependent due to thermal effects, material degradation, and ionization processes that fundamentally alter the conduction mechanism. The instantaneous total resistance $(R(t))$ derives from the material geometry and time-varying conductivity through the relation:

$$R(t) = \frac{L}{A\sigma(t)} \tag{2}$$

Let $(R_0)$ represent the nominal, pre-fault resistance under equilibrium conditions. The instantaneous total resistance decomposes into this reference value plus a time-dependent deviation term that captures the fault evolution:

$$R(t) = R_0 + R_{\text{short}}(t) \tag{3}$$

where $(R_{\text{short}}(t))$ describes the resistance departure from the reference condition under fault stress, embodying the dynamic evolution of the conduction path during electrical breakdown. The Modified Ohm's Law (provided in [26]) establishes a phenomenological relationship between current and resistance deviation that captures the nonlinear conduction characteristics observed during fault conditions ([28]), expressed as: $\left(I_{\text{modified}} = a \cdot e^{\frac{R_{\text{short}}}{R_0}}\right)$, where $\left(a = \frac{V}{R_0}\right)$ serves as a scaling parameter determined by the applied voltage and reference resistance. This exponential form emerges from the collective behavior of charge carriers under extreme electrical stress. Taking the natural logarithm of this Modified Ohm's Law yields the linearized relationship:

$$\ln I = \ln a + \frac{R_{\text{short}}}{R_0} \tag{4}$$

Rearranging Equation 4 reveals the fundamental coupling between current and resistance deviation:

$$\frac{R_{\text{short}}}{R_0} = \ln\left(\frac{I}{a}\right) \tag{5}$$

Solving for the resistance deviation demonstrates that resistance evolution emerges as a deterministic function of current growth under fault conditions:

$$R_{\text{short}} = R_0 \ln\left(\frac{I}{a}\right) \tag{6}$$

Experimental observations of fault progression justify representing the current evolution with an exponential function of a generalized progression parameter $(x)$:

$$I(x) = ae^{kx} \tag{7}$$

where $(k)$ represents a transition rate coefficient quantifying the rapidity of fault development. The parameter $(x)$ may represent time, accumulated energy, or material degradation state. Substituting Equation 7 into Equation 6 yields the resistance deviation in terms of the progression parameter:

$$R_{\text{short}}(x) = R_0 \ln\left(\frac{ae^{kx}}{a}\right) = R_0 \ln(e^{kx}) = R_0 kx \tag{8}$$

This simplifies to the linear relationship:

$$R_{\text{short}}(x) = R_0 kx \tag{9}$$

The transition rate coefficient $(k)$ follows directly from Equation 9:

$$k = \frac{R_{\text{short}}(x)}{R_0 x} \tag{10}$$

However, the experimental constraints in this paper establish time as the most physically significant progression parameter. Laboratory conditions with controlled temperature (21°C to 26°C) and continuous electrical monitoring across discrete time windows (3-minutes, 10-minutes, and 15-minutes) provide direct temporal resolution of the fault evolution. Substituting $(x = t)$ into Equation 9 yields the time-dependent resistance deviation:

$$R_{\text{short}}(t) = R_0 kt \tag{11}$$

The total resistance then becomes:

$$R(t) = R_0 + R_{\text{short}}(t) = R_0(1 + kt) \tag{12}$$

Equation 12 predicts linearly increasing resistance with time, contradicting empirical observations where resistance typically collapses due to arcing, ionization, or material breakdown during fault conditions. To incorporate resistance collapse while maintaining physical validity, the formulation introduces decay into the current evolution, requiring a sign change in the resistance deviation:

$$I(t) = ae^{kt} \Rightarrow R_{\text{short}}(t) = -R_0 kt \tag{13}$$

This modification yields the resistance time evolution:

$$R(t) = R_0(1 - kt) \tag{14}$$

Although Equation 14 captures the directionality of resistance decay, it produces negative resistance values when $\left(t > \frac{1}{k}\right)$, violating fundamental thermodynamic constraints for passive systems and representing an unphysical condition. The correction requires modeling decay with an exponential form that ensures positivity, continuity, and asymptotic convergence consistent with observed resistive collapse. The resistance deviation redefines as:

$$R_{\text{short}}(t) = -R_0(1 - e^{-kt}) \tag{15}$$

This formulation guarantees that the resistance deviation approaches $(-R_0)$ asymptotically while maintaining physical consistency. The total resistance becomes:

$$R(t) = R_0 + R_{\text{short}}(t) = R_0 - R_0(1 - e^{-kt}) \tag{16}$$

Simplifying Equation 16 yields the fundamental time-resolved resistance decay model:

$$R(t) = R_0 e^{-kt} \tag{17}$$

Equation 17 represents the transformation into a physically consistent exponential decay model where resistance collapses continuously but never vanishes, ensuring finite current throughout the fault evolution while respecting material limitations and thermodynamic constraints. The current evolution follows directly from substituting Equation 17 into the standard conduction relationship: $\left(I(t) = \frac{V}{R(t)} = \frac{V}{R_0 e^{-kt}} = \frac{V}{R_0} e^{kt} = ae^{kt}\right)$. This relationship completes a first-principles perspective of the time-resolved electrical behavior during fault conditions, with the decay constant $(k)$ acquiring physical meaning as the normalized rate of resistance collapse per unit time, directly measurable from experimental data. The physical basis for the exponential decay form emerges from the Langevin framework for dissipative systems [29]. Consider a conductor undergoing electrical stress where the resistance evolution follows a generalized Langevin equation: $\left(\frac{dR(t)}{dt} = -\gamma R(t) + \xi(t)\right)$, where $(\gamma)$ represents a damping coefficient quantifying the system's response to electrical stress, and $(\xi(t))$ denotes a Gaussian white noise term with zero mean and correlation function $(\langle\xi(t)\xi(t')\rangle = 2D\delta(t - t'))$. The diffusion coefficient $(D)$ characterizes the magnitude of resistance fluctuations due to stochastic processes during fault development [29]. For the deterministic component dominating during the primary fault evolution phase, the noise term averages to zero, yielding the simplified equation: $\left(\frac{dR(t)}{dt} = -\gamma R(t)\right)$. The solution to this equation provides the exponential decay: $(R(t) = R_0 e^{-\gamma t})$. This establishes the fundamental physical basis for the exponential form observed experimentally. The decay constant $(k)$ corresponds directly to the damping coefficient $(\gamma)$, reflecting the material's characteristic response time to electrical stress. The model successfully captures both thermal effects and electron avalanche processes that drive resistance collapse during fault conditions. The fluctuation-dissipation theorem connects the deterministic decay to the stochastic fluctuations through the relation: $\left(D = \frac{k_B T}{\mu}\gamma\right)$, where $(k_B)$ represents Boltzmann's constant, $(T)$ the absolute temperature, and $(\mu)$ a mobility coefficient characterizing the system's response to fluctuations [30]. The time-resolved formulation maintains consistency with non-equilibrium thermodynamic principles while capturing the nonlinear dynamics of short-circuit evolution. The exponential form emerges naturally from the dominance of impact ionization processes during severe fault conditions, while the bounded resistance floor reflects fundamental material limitations and the dynamic equilibrium between carrier generation and recombination processes in driven systems far from thermodynamic equilibrium [30].

## 2.2 Continuous Short-Circuit Transition Model

Let $t \in \mathbb{R}$ be a continuous time parameter with $t < 0$ representing the ordered pre-fault regime, $t = 0$ denoting the onset of a short-circuit, and $t > 0$ capturing the progressive fault regime. Let $V(t)$, $I(t)$, and $R(t)$ denote the voltage, current, and resistance at time $t$, respectively. We define a short-circuit transition system $\Sigma(t) = \{V(t), I(t), R(t)\}$ evolving under the following premises:

Let the initial system resistance before the short-circuit event be $R_0$, assumed constant for $t < 0$:

$$R(t) = R_0, \forall t < 0 \tag{18}$$

The current for $t < 0$, assuming constant voltage $V_0$, obeys the standard Ohm's Law:

$$I(t) = \frac{V_0}{R_0}, \forall t < 0 \tag{19}$$

At the transition point $t = 0$, a short-circuit initiates. The system enters a chaotic state characterized by an exponential decay in resistance, modeled as:

$$R(t) = R_0 e^{-kt}, t \geq 0 \tag{20}$$

where $k > 0$ is the decay constant describing the rate at which resistance collapses post-fault. Equation 20 introduces a non-instantaneous transition, eliminating the paradox of zero resistance and infinite current. Substituting Equation 20 into Equation 19 (the standard Ohm's Law):

$$I(t) = \frac{V_0}{R(t)} = \frac{V_0}{R_0 e^{-kt}} = \frac{V_0}{R_0} e^{kt}, t \geq 0 \tag{21}$$

Introducing a current scaling constant $\left(a = \frac{V_0}{R_0}\right)$ from the Modified Ohms Law, the current becomes:

$$I(t) = a e^{kt}, t \geq 0 \tag{22}$$

This formulation preserves the physicality of finite current while allowing continuous growth. Let $R_{\text{short}}(t)$ represent the deviation of resistance from the reference resistance:

$$R_{\text{short}}(t) = R(t) - R_0 = R_0(e^{-kt} - 1), t \geq 0 \tag{23}$$

Using this, the exponential growth in current provided in Equation 22 can be reformulated in terms of $R_{\text{short}}(t)$:

$$I(t) = a \cdot e^{\frac{R_{\text{short}}(t)}{R_0}}, t \geq 0 \tag{24}$$

Equation 24 introduces the Modified Ohm's Law in its full time-dependent form. Let us now define the short-circuit system transition function $\Phi(t)$ according to Equation 25:

$$\Phi(t) = \begin{cases} \left(V_0, \frac{V_0}{R_0}, R_0\right), & t < 0 \\ (V_0, ae^{kt}, R_0 e^{-kt}), & t \geq 0 \end{cases} \tag{25}$$

Then, express the change in resistance $\Delta R(t)$ between $t = 0$ and $t = t_1 > 0$ as:

$$\Delta R(t_1) = R_0 - R_0 e^{-kt_1} = R_0(1 - e^{-kt_1}) \tag{26}$$

The corresponding increase in current $\Delta I(t_1)$ is therefore:

$$\Delta I(t_1) = a(e^{kt_1} - 1) \tag{27}$$

The short-circuit transition duration $\Delta t$ is defined as the smallest time $t$ such that:

$$|R(t) - R_\infty| < \epsilon, \text{ where } R_\infty = \lim_{t \to \infty} R(t) = 0 \tag{28}$$

Assuming a tolerance $\epsilon = 0.01 R_0$, the stabilization time $\tau_s$ satisfies:

$$R_0 e^{-k\tau_s} = 0.01 R_0 \Rightarrow \tau_s = \frac{\ln(100)}{k} \tag{29}$$

This gives a physically grounded definition of stabilization based on exponential decay. To describe energy evolution, define the instantaneous power $P(t)$ as:

$$P(t) = V_0 \cdot I(t) = V_0 a e^{kt}, t \geq 0 \tag{30}$$

The cumulative energy dissipated up to time $t$ becomes:

$$E(t) = \int_0^t P(t')dt' = V_0 a \int_0^t e^{kt'}dt' = \frac{V_0 a}{k}(e^{kt} - 1) \tag{31}$$

This reflects the nonlinearity and unboundedness in energy dissipation during extended fault durations. The rate of change of energy dissipation is:

$$\frac{dE}{dt} = P(t) = V_0 a e^{kt} \tag{32}$$

Equation 32 thus predicts an exponential increase in energy dissipation, in direct contrast to conventional stepwise assumptions in short-circuit modeling.

Let us define the resistance decay rate function $\Lambda(t)$ as:

$$\Lambda(t) = -\frac{1}{R(t)} \cdot \frac{dR(t)}{dt} = k, \forall t \geq 0 \tag{33}$$

This reveals the decay rate is uniform-a critical simplification useful in analytic solutions and model fitting.

## 2.3 Modified Vs. Standard Ohm's Law
To distinguish the behavior of current predictions under standard and modified formulations, define:

- *Standard Ohm's Law model current:-* $\left(I_{std}(t) = \frac{V(t)}{R(t)}\right)$ (34)
- *Modified Ohm's Law model current:-* $\left(I_{mod\_trans}(t) = a e^{\frac{R_{short}(t)}{R_0}}\right)$ (35)

Inserting Equation 23 into Equation 35:

$$I_{mod\_trans}(t) = a \cdot e^{e^{-kt}-1} \tag{36}$$

Comparing the respective derivatives:

$$\frac{dI_{std}}{dt} = -\frac{V_0}{R(t)^2} \cdot \frac{dR}{dt} = \frac{V_0 k e^{-kt}}{R_0^2 e^{-2kt}} = \frac{V_0 k}{R_0^2} e^{kt} \tag{37}$$

$$\frac{dI_{mod\_trans}}{dt} = a \cdot \frac{d}{dt} e^{e^{-kt}-1} = a \cdot (-k e^{-kt}) e^{e^{-kt}-1} \tag{38}$$

As a result, as $t \to 0$, both equations yield bounded slopes. However, as $t \to \infty$, $\frac{dI_{std}}{dt} \to \infty$, while $\frac{dI_{mod}}{dt} \to 0$, indicating the Modified Ohm's Law model converges to a stable current trajectory. To highlight the deviation between the models, we define the error function as:

$$\epsilon(t) = |I_{mod\_trans}(t) - I_{std}(t)| \tag{39}$$

This function is experimentally applied in the succeeding sections to validate the superior boundedness of the modified formulation over longer durations.

### 2.3.1 Comparative Analysis of Ohmic Formulations

Short-circuit behavior under high-energy stress deviates substantially from the conditions assumed in classical circuit analysis. The Standard Ohm's Law, expressed as $\left(I(t) = \frac{V(t)}{R(t)}\right)$, assumes resistive linearity and immediate steady-state compliance. While sufficient for routine conditions, it lacks descriptive power in scenarios where circuit parameters evolve dynamically, particularly during arc-induced faults or plasma transitions. To extend applicability into these extreme regimes, this paper employs two alternative formulations: the Modified Ohm's Law and its derivative, the Transformed Ohm's Law. The Modified Ohm's Law, expressed as $\left(I_{\text{modified}} = ae^{\frac{R_{\text{short}}}{R_0}}\right)$ captures nonlinearities by encoding resistance deviation as an exponential amplifier of current. This approach allows the current to scale rapidly with declining resistance, reflecting physical realities of thermal degradation and ionization. However, this formulation is not intended to replace the standard law universally. Rather, it augments standard analysis in transient fault conditions-functioning as a diagnostic tool that explicitly models exponential current growth with a defined resistance decay profile. The Transformed Modified Ohm's Law refines this further. It integrates the temporal structure of resistance collapse, typically in the form $R(t) = R_0 e^{-kt}$ (Equation (17)), leading to: $\left(I(t) = \frac{V_0}{R_0 e^{-kt}} = a \cdot e^{kt}\right)$ Equation (21) and Equation (22). This version introduces the decay constant $(k)$, a physically derived parameter that defines the transition rate from nominal to degraded states. Unlike the Modified form, which embeds nonlinearity in resistance deviation, the Transformed form models the fault as a system-wide temporal evolution.

A core distinction arises in how each law handles energy and power dissipation. The Modified Law, when extrapolated to predict clamped power $\left(P(t) = V_{\text{clamp}} \cdot I(t)\right)$ (the inconsistency addressed following the experimental results), sometimes implies paradoxes-such as lower power output accompanying higher current under fault conditions. This inversion flags a limitation: the model's functional elegance can obscure physical constraints. Such contradictions suggest that while the Modified Law performs well during early-stage current ramp-up, it may inadequately account for energy saturation or thermal back-reactions, which the Transformed Law attempts to address through time-linked resistance decay. To contextualize these differences, Table 1 compares the three formulations across ten relevant dimensions:

**Table 1.** Comparative Features of Ohmic Formulations

| Feature | Standard Ohm's Law | Modified Ohm's Law | Transformed Modified Ohm's Law | Shared Attributes |
|---|---|---|---|---|
| Governing Equation | $I = \frac{V}{R}$, ([31]) | $I_{modified} = a \cdot e^{\frac{R_{\text{short}}}{R_0}}$, ([26]) | $I_{mod\_trans} = a \cdot e^{kt}$ | All models relate current to voltage and resistance |
| Resistance Behavior | Static or externally supplied | Implicit via $R_{\text{short}}$ | Explicit: $R(t) = R_0 e^{-kt}$ | Each model depends on defined or derived resistance |
| Temporal Resolution | None | Indirect via $R_{\text{short}}$ | Directly time-resolved | Modified and Transformed models evolve over time |
| Suitability for Fault Modeling | Limited | Moderate-captures early ramp-up | High-tracks full decay phase | All models reflect parts of the fault process |
| Predictive Limits | Fails at $R \to 0$ | Diverges for long durations | Asymptotically bounded | Modified and Transformed avoid singularities |
| Use of Fitting Parameters | None | Parameter $a$ from baseline | Parameters $a, k$ from experiment | Both non-standard laws require empirical calibration |
| Power Prediction Accuracy | Reasonable under static $R$ | Sometimes paradoxical at high $I$ | Consistent with energy buildup | Standard and Transformed better reflect energy scaling |
| Interpretability | Direct and intuitive | Physically motivated; exponential mapping | Physically derived; time-resolved decay | Modified and Transformed are physically motivated |
| Experimental Alignment | Matches steady-state | Aligns in early transient window | Best match for full-duration arc events | Experimental validation varies with timescale |
| Limitations | Ignores resistance evolution | Misrepresents energy at long $t$ | Assumes ideal decay with no recovery | All models have domain-specific limits |

Taken together, these formulations represent not competing alternatives, but progressive expansions of a common law to accommodate nonlinearity, temporality, and fault-phase transitions. The Modified Ohm's Law offers tractable modeling for transient conditions where resistance deviates from equilibrium, while the Transformed Ohm's Law goes further-embedding time directly and enabling predictions of resistance collapse, energy accumulation, and saturation thresholds. Crucially, neither of these extended forms displaces the foundational utility of the standard law. Instead, they operate as extensions-parametric modifications that allow better approximation of real-world events under stress. Notably, experimental observations show that although current peaks during short-circuits, clamped power sometimes diminishes, a result inconsistent with standard theory but reproducible through Modified Ohm's Law models. This suggests potential model overreach: while the Modified Ohm's Law may predict current well, it cannot, in its current form, fully capture multi-domain energy feedbacks without additional transformations. Such divergences imply a fertile direction for future research on the exploration of physical regimes that transition between non-ohmic and ohmic behavior post-fault, likely governed by delayed re-equilibration, contact recovery, or material reconduction. Though outside the scope of the present work, these hybrid states demand further analytical and experimental integration.

**3.0 Experimental Framework**

Three complementary experiments, designated "*Clamped Base Diodes*", "*Clamped Short Diodes Nominal*" and "*Clamped Short Diodes Extremes*," were designed to probe the "*Clamped Short Diodes*" -circuit's behavior under controlled short-circuit conditions. Each experiment applies the same fundamental framework but varies the severity of the applied fault, allowing comparison of circuit response across nominal and extreme stress levels. Both setups leverage the "*short-parallel connection*" topology of "Circuit Block 1" to enforce unidirectional isolation while the system undergoes rapid current transients.

### 3.1 Apparatus and Sensor Fundamentals

All experimental components were selected for their simplicity, affordability, and ease of access. The Arduino Uno microcontroller functioned as the primary data acquisition platform, interfaced with an ACS712T-20A Hall-effect current sensor using standard jumper wires, a USB-A to B interface cable, and a breadboard. Alongside three 1N5408 power diodes, these components -sourced from PixelElectric Engineering Solutions (Kenya), [32] -formed the essential "*short-parallel connection*" configuration that defines the protective mechanism of "Circuit Block 1". A regulated multi-output DC power supply from ElettronicaVeneta provided adjustable voltage inputs. Data processing and system control were conducted on a laptop equipped with an Intel® Core™ i7-2620M processor (2.70 GHz, dual-core, four-thread, 4 MB Smart Cache) and 4 GB DDR3 RAM, serving as the computational and development environment.

The ACS712T-20A sensor employs an indirect Hall-effect measurement principle: current flows through an internal conductor, generating a magnetic field that the onboard Hall element converts into a proportional voltage. Full-scale current excursions in either direction -positive or negative -shift the sensor's output above or below the $2.5V$ midpoint (half of the $5V$ supply), yielding a bi-directional sensing capability. Interpretation around this midpoint effectively transforms any voltage deviation into a signed current value. Ordinarily, spurious readings may appear near zero current owing to noise on the $2.5V$ offset, and the Arduino's 10-bit ADC ($\approx 4.88mV$ resolution) places practical limits on the smallest discernible current change. Despite these limitations, the sensor remains suitable for dynamic, real-time current monitoring applications, provided that design provisions are made to ensure stability and minimize signal interference.

### 3.2 Short-circuit Measurement: Configuration and Data Acquisition

The implementation of real-time current and voltage measurements during short-circuit conditions required a custom-designed architecture, capable of resolving high-speed transients without compromising signal integrity or exceeding the electrical input range limitations of the Arduino Uno. At the core of this architecture lies a tailored configuration referred to as "Circuit Block 1" denoted ($CB_1$), which integrates a robust diode-based "short-parallel" network and the ACS712T-20A Hall-effect current sensor. This design facilitates high-fidelity monitoring of circuit behavior across abrupt fault conditions and enables empirical exploration of modified electrical models such as the Modified Ohm's Law, which diverges from traditional assumptions of instantaneous resistance collapse and zero terminal voltage during a short-circuit.

Contrary to the idealized model of a perfect short -where current spikes infinitely and terminal voltage drops to zero - earlier investigations [8] revealed that voltage remains non-zero and measurable during such faults, implying the presence of a time-varying or residual resistance. These observations informed the architecture of ($CB_1$). The "*short-parallel configuration*" -constructed from three 1N5408 diodes - serves to isolate reverse current flow and maintain stable voltage across the output terminals, protecting the measurement system from potential overvoltage events while ensuring voltages remain within the $0 - 5V$ input tolerance of the Arduino's ADC. Within ($CB_1$), the ACS712T-20A sensor is embedded directly in series with the output pathway. Its bidirectional Hall-effect sensing mechanism allows detection of both forward (positive) and recovery (negative) current flow, critical for capturing all three phases of the short-circuit event: the baseline (pre-short), fault onset (positive current surge), and post-fault (reverse or dissipative phase). With a nominal sensitivity between $66mV/A$ and $185mV/A$, and a factory-calibrated accuracy of ±1.5%, the sensor was operated at a 5V supply voltage, aligning its mid-scale output to approximately 2.5V -representing zero current. Any deviation from this midpoint translates directly into a signed current measurement, making it particularly well-suited for detecting the bidirectional current dynamics that characterize real short-circuit events.

To ensure accurate data acquisition, the Arduino's 10-bit analog-to-digital converter (ADC) was configured to sample at high frequencies (10ms, 50ms, and 100ms), with each data point averaged over 10,000 samples to reduce quantization noise and suppress false readings caused by sensor drift or electrical noise. This granular sampling strategy allowed the system to capture high-resolution transients without saturation or clipping, even during rapid voltage collapses or current surges.

The experimental protocol begun with the "*Clamped Base Diodes*" (Definition 3) data acquisition framework embedded within the ($CB_1$) architecture. Three voltage levels -2.5V, 5V, and 10V -were applied across time intervals of 3 -minutes, 10 -minutes, and 15 -minutes, respectively. This initial configuration established the baseline performance of the clamped circuit and verified its operational readiness for extended-duration measurements. Within the same ($CB_1$) platform, two targeted experimental regimes were conducted: "*Clamped Short Diodes Nominal*" and "*Clamped Short Diodes Extremes*". In the "*Clamped Short Diodes Nominal*" experiment, the objective was to explore circuit stability and current -voltage behavior over extended time frames. Testing was conducted using input voltages of 2.5V, 5V, and 10V across durations of 3 -minutes, 10 -minutes, and 15 -minutes, respectively, with each session subdivided into the two diagnostic phases: during the short-circuit event and immediately post-short.

In contrast, the "*Clamped Short Diodes Extremes*" applied the same high-frequency sampling and averaging techniques but was constrained to 3 -minutes intervals. This phase focused specifically on validating circuit performance under more extreme electrical stress conditions and lower voltage inputs. Here, successful measurements at 2.5V input -a voltage at the lower operational limit of the ACS712T - demonstrated the robustness of ($CB_1$)'s design and the sensor's capability to perform accurately under non-ideal and potentially marginal power supply conditions. The current sensor's suitability for operation at the 2.5V external input was further confirmed through replication of the experiment at a 5.0V external input. Across all three experimental configurations -"Clamped Base Diodes," "Clamped Short Diodes Nominal," and "Clamped Short Diodes Extremes" -data were collected in duplicate for each test condition. This resulted in a comprehensive dataset suitable for detailed statistical and physical analysis, including transient current behavior, voltage regulation, and real-time resistance dynamics. The modular nature of ($CB_1$) and its compatibility with open-source microcontroller platforms like Arduino suggest potential for broader applications in low-cost diagnostics, energy monitoring, and embedded fault detection systems. This measurement architecture not only captured transient signatures with high resolution but also validated the experimental feasibility of characterizing non-ideal short-circuit events with inexpensive, accessible technology.

### 3.3 Experiment Design

The experimental design combined minimal hardware complexity with an embedded automation protocol to enable real-time measurements under both nominal and extreme fault conditions. Hardware components included three 1N5408 diodes, a variable DC power supply, the ACS712T-20A current sensor, and the Arduino Uno microcontroller, integrated within the clamped short-circuit architecture known as "Circuit Block 1". Each experiment began with the "Clamped Base Diodes" setup to record baseline currents and voltages before inducing the short. Software implementation involved the development of a custom Arduino firmware to execute the Automated Short-circuit Current Measurement Algorithm, enabling precise control of sampling rates (10ms, 50ms, 100ms), averaging over 10,000 samples, and structured data acquisition during the fault events. The two experimental variants -*Nominal* and *Extremes* -share this unified base configurations but differ in fault severity and operational objectives.

### 3.3.1 Deterministic Apparatus Selection via Modified Ohm's Law Simulation

The experimental design for the three distinct input voltages- $2.5V$, $5.0V$ and $10.0V$ -required a deterministic framework to ensure component survivability and data fidelity under sustained short-circuit conditions. The Modified Ohm's Law, expressed in Equation 2 as $\left(I_{\text{modified}} = a \cdot e^{\frac{R_{\text{short}}}{R_0}}\right)$, served as this critical computational tool, transitioning from a theoretical model to a practical simulation engine for apparatus selection.

A foundational constraint for comparative analysis was the maintenance of a consistent short-circuit current profile. Given the use of a single 1N5408 diode type in the provided bidirectional type clamping network, simulations were configured to target a nominal short-circuit current of approximately $3.3A$, a value safely within the diode's operational limits while being sufficient to produce a clear, measurable fault signature. The parameter $(R_0)$, representing the reference or pre-fault resistance in the Modified Ohm's Law framework, became the primary variable for adjustment in these simulations. For the $5.0V$ experiment, the simulation established a baseline where: $\left(R_0 = \frac{V}{I_{\text{modified}}} = \frac{5.0V}{3.3A} \approx 1.5\Omega\right)$. This $1.5\Omega$ value for $(R_0)$ provided the reference point for the simulation framework. However, computational analysis revealed that $(R_0)$ is not a fixed physical component but a dynamic parameter of the governed system that varies with supply voltage. Subsequent simulations for the $2.5V$ and $10.0V$ configurations demonstrated that maintaining the target current required adjusting the $R_0$ parameter within a specific range.

For the $2.5V$ configuration, simulations indicated that a higher $R_0$ value ($\approx 3.0\Omega$) was necessary to limit current flow while still ensuring sufficient magnitude for measurable fault dynamics. Conversely, for the $10.0V$ configuration, simulations determined that a lower $R_0$ value ($\approx 0.75\Omega$) would accommodate the increased available power while preventing current divergence beyond the diode network's capacity. This $R_0$ parameter range of $0.75\Omega$ to $3.0\Omega$ across the voltage spectrum represented not physical resistors but the effective dynamic impedance of the complete experimental apparatus-including wiring, contacts, and the intrinsic properties of the diode network under different electrical stresses.

The simulation framework thus established that the same diode clamping network could produce governed short-circuits across different input voltages by operating at different points along its dynamic impedance characteristic. This explained the experimental observation that short-circuit current increased with supply voltage (from $\approx 8.5A$ at $2.5V$ to $\approx 9.4A$ at $10.0V$, section 4) while remaining bounded well below destructive levels. The systematic application of the Modified Ohm's Law through parameter variation ensured that all three experiments operated within a deterministic, comparable, and non-destructive electrical regime, enabling the first-ever time-resolved observation of sustained fault dynamics across a voltage spectrum.

Furthermore, this simulation-guided methodology bypasses the traditional pitfalls of parasitic capacitance and inductance that plague conventional high-speed clamping circuits. The success of this apparatus selection strategy, culminating in stable measurements over 30 -minutes durations (15 -minutes forward polarity configurations and 15 -minutes reverse configuration), is intrinsically linked to this novel application of the Modified Ohm's Law. A detailed technical justification of how this approach overcomes classical parasitic challenges is provided in Appendix A.

### 3.3.2 The Clamped Short Diodes Nominal Design

The first configuration, established as "Clamped Short Diodes Nominal", focused on measuring and analyzing circuit behavior during controlled electrical faults under standard clamped short-circuit operating voltages as recently established in [8]. The Arduino Uno was configured to resolution sample at $10ms$, $50ms$, and $100ms$ intervals, as mentioned earlier, with each measurement averaged over 10,000 readings to mitigate ADC quantization error. For every sampling frequency, three distinct voltage inputs including $2.5V$, $5.0V$, and $10V$ -were applied across durations of 3 -minutes, 10 -minutes, and 15 -minutes. Each voltage-time configuration was divided into three diagnostic phases: pre-short (baseline), short-circuit (positive current surge), and post-short (recovery or reverse current). Consequently, each complete experiment session spanned 9 -minutes, 30 -minutes, and 45 -minutes, respectively. Two independent data sets were recorded and averaged for every condition, providing a robust dataset for evaluating circuit response across varying durations and voltages. This experiment aimed to verify circuit stability over prolonged durations, extending the previously validated 3 -minutes window [8]. It also enabled precise characterization of transient current and voltage dynamics during the transition across fault stages. Table 2 presents the detailed procedure for automated measurement of "Clamped Short Diodes" base and nominal currents and voltages.

**Table 2.** Algorithm for Automated "Clamped Short Diodes" Base and Nominal Currents and Voltages Measurement

| Step | Action/Description |
|---|---|
| 1 | • Connect DC power supply's positive terminal to the ACS712T current sensor's positive input though the output of $D_1$. <br> • Connect the power supply's negative terminal to the current sensor's negative input. |
| 2 | • Connect the current sensor's OUT pin to the Arduino's $A0$ analog input pin for current readings, through the output of $D_3$. <br> • Connect also the output of $D_3$ to the Arduino's $A1$ analog input pin via the current sensor's OUT pin for voltage readings. |
| 3 | • Connect $D_1$'s anode to the power supply's positive terminal. <br> • Branch $D_1$'s cathode to the next circuit segment's positive terminal and $D_2$'s anode. |
| 4 | • Connect $D_2$'s cathode to Arduino ground (GND), completing the circuit path to the power supply's negative terminal. |
| 5 | • Connect $D_3$'s anode to the same node as $D_2$'s anode. Connect $D_3$'s cathode to an adjacent circuit's negative terminal. <br> • For clarity, the cathodes of $D_1$ and $D_3$ proceed to "Circuit Bock 2". |
| 6 | • Create a short-circuit path by linking nodes $A$ and $B$ (between $D_1$ and $D_3$) with a connector. <br> • This step should be performed carefully, only during when the short-circuit is induced. |
| 7 | • Program Arduino to record voltage (A1) and current (A0) as magnitudes at 1 -second intervals for $180s$, $600s$, and $900s$ before and during the short-circuit, at sampling frequencies of $10ms$, $50ms$, and $100ms$ in each time interval. Upload the sample code (provided in supplementary materials) to the Arduino. |
| 8 | • Turn on the selected power supply source voltage ($2.5V, 5V, and\ 10V$). Begin data logging to observe normal circuit behavior. |

| 9 | • Connect the short-circuit path between nodes $A$ and $B$. Continue data logging to capture short-circuit data. |
|---|---|
| 10 | • Analyze the recorded data to assess the circuit's response under normal (base measurements) and short-circuit conditions. |

Figure 1 illustrates the schematic of the experimental setup, highlighting the diode configuration, current sensor, Arduino interface, and fault induction path, $L_2$. This arrangement defines the architecture for controlled short-circuit induction and measurement.

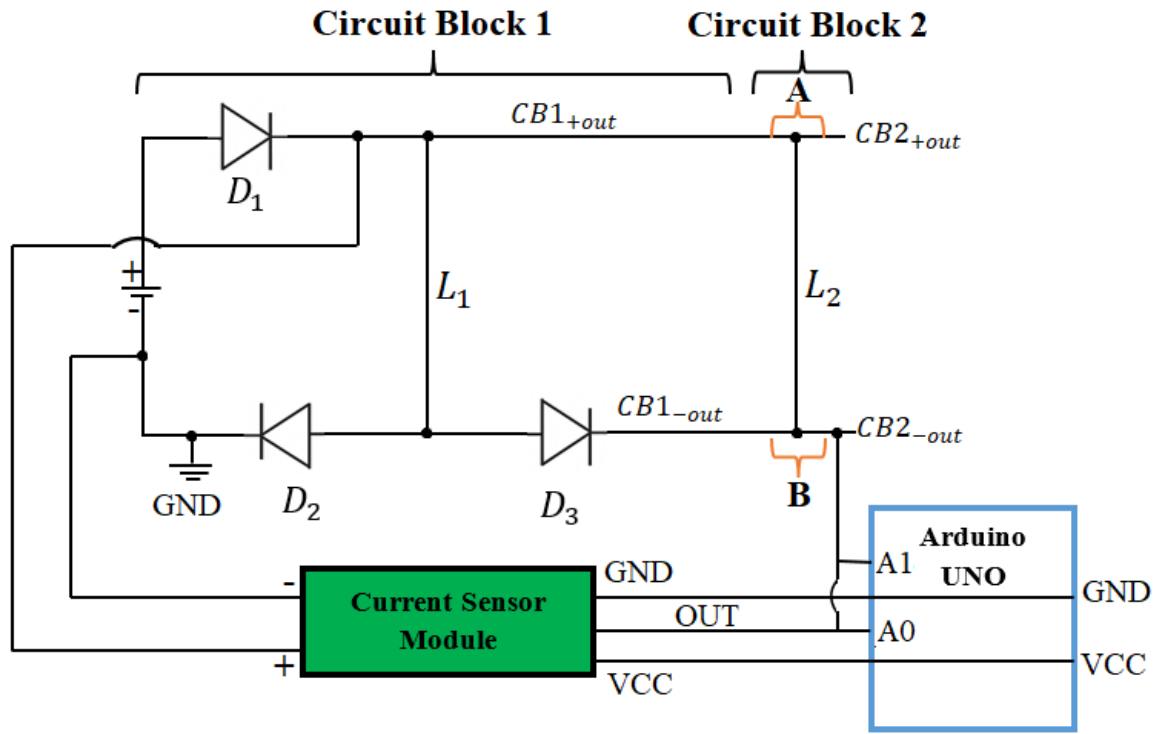

**Figure 1.** Experimental setup for "Clamped Short Diodes Nominal" measurements. (Key components: DC power supply, ACS712T current sensor, Arduino Uno, and clamped diode network (constituting; $D_1$, $D_2$, $D_3$). Nodes $A$ and $B$ define the short-circuit path during fault induction).

This configuration permits dynamic monitoring of current and voltage before and during short-circuit events, capturing real-time data at multiple resolutions. The diode arrangement enforces unidirectional flow during the short-circuit events, safeguarding sensor inputs while allowing empirical validation of the circuit's short-circuit response under nominal stress levels.

### 3.3.3 The Clamped Short Diodes Extremes Design
The second configuration, "Clamped Short Diodes Extremes", was designed to examine the limits of the circuit's behavior under more stressful electrical fault conditions. Sampling frequencies remained at $10ms$, $50ms$, and $100ms$, as in the nominal experiment, with measurements averaged over $10,000$ samples. Unlike the nominal experiment, only two voltage levels; $2.5V$ and $5.0V$ -were applied, each limited to 3-minutes intervals. Each test cycle included three sequential phases: pre-short, during short, and post-short, amounting to 9-minutes per session. Two replicates were recorded for each configuration, and the data were averaged. This experiment assessed the performance of the sensor and the circuit under minimal input voltage and maximum stress conditions, challenging the classical model that assumes zero resistance and infinite current under a short. The test also validated the functional limits of the ACS712T sensor at $2.5V$ supply, below its nominal $5.0V$ specification, thus reinforcing the experimental model's robustness under constrained operational scenarios.

**Table 3.** Algorithm for Automated "Clamped Short Diodes" Base and Extremes Currents and Voltages Measurement

| Step | Action/Description |
|---|---|
| 1 | • Connect DC power supply's positive terminal to the ACS712T current sensor's positive input though the output of $D_1$.<br>• Connect the power supply's negative terminal to the current sensor's negative input through the output of $D_3$. |
| 2 | • Connect the current sensor's OUT pin to the Arduino's $A0$ analog input pin for current readings, through the output of $D_3$.<br>• Connect also the output of $D_3$ to the Arduino's $A1$ analog input pin via the current sensor's OUT pin for voltage readings. |
| 3 | • Connect $D_1$'s anode to the power supply's positive terminal.<br>• Branch $D_1$'s cathode to the next circuit segment's positive terminal and $D_2$'s anode. |
| 4 | • Connect $D_2$'s cathode to Arduino ground (GND), completing the circuit path to the power supply's negative terminal. |
| 5 | • Connect $D_3$'s anode to the same node as $D_2$'s anode. Connect $D_3$'s cathode to an adjacent circuit's negative terminal.<br>• For clarity, the cathodes of $D_1$ and $D_3$ proceed to "Circuit Bock 2". |
| 6 | • To create the short-circuit in this case, translate nodes $A$ and $B$ to the cathodes of $D_2$ and $D_3$ ($D_2$ (at the Arduino GND) and $D_3$) with a connector.<br>• This step should be performed carefully, only during when the short-circuit is induced. |
| 7 | • Write Arduino code to log voltage (A1) and current (A0) magnitudes at 1-second intervals for $180s$, with sampling frequencies of $10ms$, $50ms$, and $100ms$. Sample code is provided in the supplementary materials for editing. Upload the code to the Arduino. |
| 8 | • Turn on the selected power supply source voltage ($2.5V, 5V, and\ 10V$). Begin data logging to observe normal circuit behavior. |
| 9 | • Connect the short-circuit path between nodes $A$ (at $D_2$ cathode) and $B$ (at and $D_3$ cathode). Continue data logging to capture short-circuit data. |
| 10 | • Analyze the recorded data to assess the circuit's response under normal (base) and short-circuit conditions. |

Figure 2 illustrates the schematic of the *Clamped Short Diodes Extremes* circuit design, detailing the shifted short-circuit path at $D_2$ and $D_3$ cathodes, essential for replicating severe electrical fault conditions.

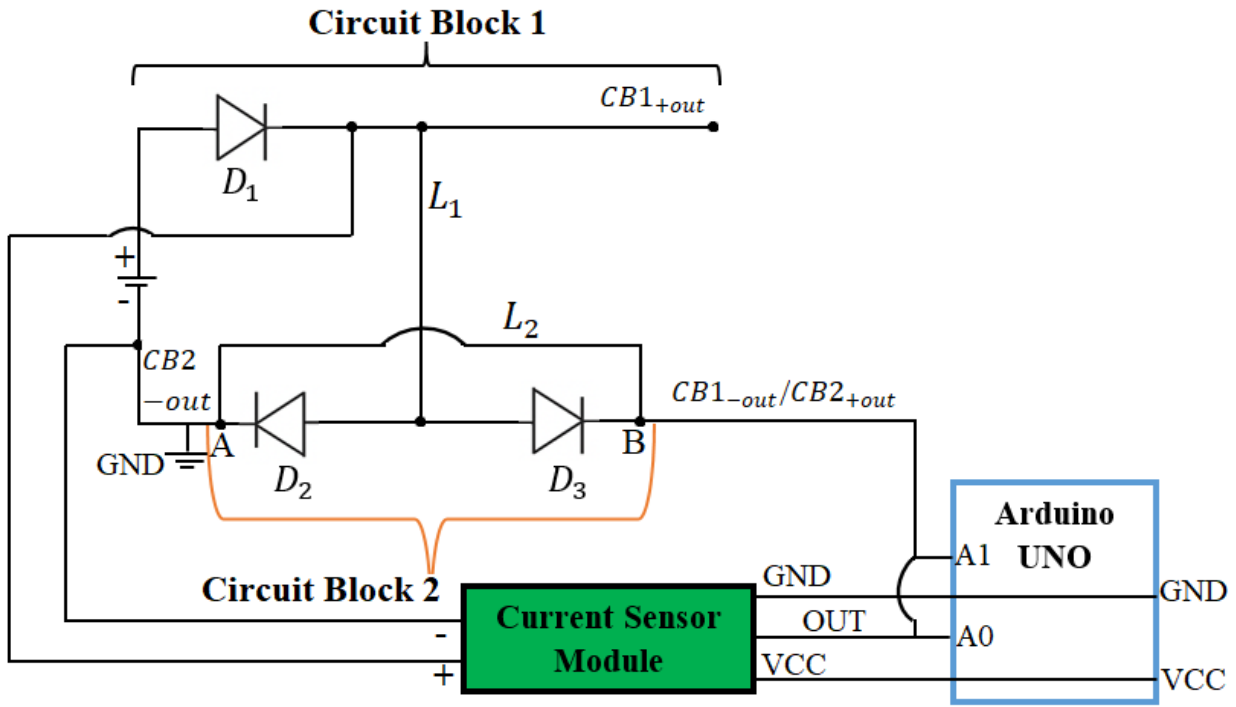

**Figure 2.** Experimental setup for "*Clamped Short Diodes Extremes*" measurements. (Same core elements as Figure 1; however, the short-circuit connection shifts to the cathodes of $D_2$ and $D_3$ for increased fault stress).

This connection topology increases transient severity and tests the system's tolerance under edge-case conditions. The diodes continue to maintain isolation and protect the Arduino's ADC inputs. These experiments critically assess sensor stability and circuit reliability when subjected to lower operating voltages and abrupt current surges.

### 3.3.4 Clamping Stability and Diode Idle Behavior in "Circuit Block 1"
The core stabilization behavior of the "Clamped Short Diodes" (Definition 3) circuit originates from the internal clamping mechanism embedded within "Circuit Block 1", ($CB_1$). This configuration implements a forward-biased diode clamping property-here defined as the ability of a diode network to restrict the output voltage to a stabilized threshold, regardless of further increases in supply voltage, once the diodes conduct. Mathematically and functionally, this stabilization is enabled by the collective forward voltage drop of diodes $D_1$ and $D_3$ within the short-parallel network.

**Definition 1 (Diodes Clamping Property).** In a non–short-circuit configuration, a diode network exhibits a clamping property when the cumulative forward voltage threshold is exceeded, causing the output voltage to stabilize at a fixed level $V_{\text{clamp}}$. This level is determined by the forward-biased diodes' voltage drops, such that $V_{\text{out}} = \sum V_{D_i}$, and remains independent of further increases in the circuit external input voltage $V_s$.

**Forward Voltage Stabilization Characteristics in $CB_1$.** The "$CB_1$" circuit employs three 1N5408 power diodes, each with a typical forward voltage drop of approximately $1.2V$. Under forward-bias conditions, the diodes exhibit relatively constant voltage drops, unaffected by further increases in current, provided conduction is established. This feature forms the basis of the voltage clamp in $CB_1$. For input voltages $V_s > V_{D_1} + V_{D_3}$, the diodes $D_1$ and $D_3$ become forward-biased, setting the stabilized output voltage $V_{CB_1}$ as:

$$V_{CB_1} = V_{D_1} + V_{D_3} = 1.2\text{V} + 1.2\text{V} = 2.4\text{V} \tag{40}$$

This output voltage remains effectively constant for $V_s \geq 2.5\text{V}$, even when the supply increases to values such as $5V$ or $10V$. The clamping behavior ensures that sensitive downstream components-especially the Arduino's analog inputs and the ACS712T sensor-remain protected within their operational voltage range (typically $0 - 5.5\text{V}$). This diode-clamped voltage regulation is particularly crucial during fault events, where input voltages may momentarily spike.

**Circuit-Theoretic Significance of the Clamped Configuration.** Within $CB_1$, the short-parallel diode arrangement performs three key roles:

1. *Voltage Limitation:-* $D_1$ and $D_3$ regulate the maximum output voltage to approximately $2.4V$, providing hardware-level voltage conditioning without external regulators.
2. *Backflow Suppression:-* $D_2$, connected to GND, absorbs excess transient current and blocks reverse conduction, enhancing current isolation.
3. *Protection and Directionality:-* The combined diode network enforces unidirectional current flow, essential for preserving signal integrity during rapid transients.

The $CB_1$ network thus operates as an active voltage-limiting and current-directing node, adapting dynamically to both steady-state and short-circuit conditions. This design ensures that real-time sensor data acquisition is not distorted by noise, overvoltage, or backpropagation effects commonly seen in uncontrolled fault events.

**Definition 2 (Diode Idle State, modified from [1 and 2]).** Contextually, a diode is said to be in an idle state when reverse-biased within the $CB_1$ configuration. In this state, it functions as a passive electrical barrier, offering effectively infinite resistance until transition into forward conduction is triggered by a transient event or suitable load condition.

### 3.3.5 The Idle State and Circuit Clamping Mechanism

In its idle or inactive state, a diode behaves like an open switch-conducting no significant current due to the negligible reverse saturation current $I_s$. This non-conductive condition characterizes diode $D_2$, and intermittently, diode $D_3$, during normal circuit operation until a transient load or short-circuit event activates them. Within "Circuit Block 1" ($CB_1$) in the "Clamped Base Diodes" configuration, current conduction under ideal conditions is dominated by the forward-biased diodes $D_1$ and $D_3$, while $D_2$ remains reverse-biased and idle. The current behavior can be described using the diode exponential model:

$$I_{CB_1} = I_s \left[ e^{\frac{V_{D_1}}{nV_T}} + e^{\frac{V_{D_3}}{nV_T}} - 2 \right] \tag{41}$$

Where:
- $I_s$ is the reverse saturation current,
- $V_T$ is the thermal voltage,
- $n$ is the diode ideality factor,
- $V_{D_1}$ and $V_{D_3}$ are the forward voltages across diodes $D_1$ and $D_3$, respectively.

This expression captures the total conduction current through the active path, deliberately excluding $D_2$ since it does not contribute during normal operation. However, $D_2$'s role becomes critical under fault conditions. When a transient or short-circuit occurs, $D_2$ dynamically transitions from a reverse-biased isolator to a conducting path that absorbs excess return current. This action suppresses voltage overshoot and prevents back-propagation of fault energy toward the power source or sensitive instrumentation. Thus, the clamping mechanism enabled by $CB_1$ operates on two fronts:

- *Voltage Clamping:-* Forward-biased diodes $D_1$ and $D_3$ set a threshold voltage beyond which conduction occurs, effectively bounding voltage rise.
- *Current Clamping:-* Under short conditions, $D_2$ conducts in the reverse direction, limiting the magnitude of the return current and absorbing fault-induced transients.

### 3.3.6 The Dynamic Role of Idle-State Diodes in Short-Circuit Protection

During a triggered short-circuit, current rapidly increases through the forward-biased conduction path ($D_1$ and $D_3$), while the previously idle $D_2$ enters a protective conduction phase. Before the event, $D_2$ blocks reverse current and electrically isolates sensitive nodes. After the event, it actively clamps the return current, protecting both the power source and measuring devices. This conditional responsiveness of $D_2$ is not passive but strategically reactive, enabling a diode to serve as a real-time switch-idle under normal conditions and instantly active under fault stress. The $CB_1$ architecture, therefore, enables an adaptive diode-based clamping system where:

- Voltage spikes are bounded by forward conduction thresholds.
- Current surges are suppressed through reactive clamping.
- Reverse-biased diodes provide isolation during steady-state.
- Idle-state diodes participate only when nominal operational thresholds are exceeded.

This clamping framework is foundational to the experimental validation in "*Clamped Nominal Short*" and "*Clamped Extreme Short*", which confirm that a properly configured diode network can perform high-resolution fault detection and voltage regulation using cost-effective passive elements.

**Definition 3 (Clamped Short Diodes).** Clamped Short Diodes refer to a specific circuit topology and its operational state, central to the experimental architecture of this work, designed to empirically characterize the nonlinear dynamics of electrical faults. It is defined by the implementation of a short-parallel diode network (as instantiated in "Circuit Block 1", $CB_1$) that enforces a bounded voltage regime during a deliberately induced short-circuit event. The core function of this configuration is twofold:

1. *Stabilized Output and Bounded Regime for Base Conditions:-* Under nominal pre-fault and baseline experimental conditions ("Clamped Base Diodes"), the network exhibits a clamping property (Definition 1).

2. *Clamping-Enabled Protection Mechanism:-* The principle extends beyond static regulation to active, dynamic protection during fault conditions. The topology is engineered to prevent catastrophic voltage collapse and avoid classical current singularity ($I \to \infty$ as $R \to 0$) by maintaining a non-zero voltage floor $V_{floor} > 0$ throughout the fault duration.

This clamping mechanism is not a suppression tool but a "***diagnostic enabler***". By imposing a controlled electrical boundary, it allows a short-circuit event to evolve through its complete physical trajectory - including resistance decay, arc formation, and thermal accumulation -without the need for triggering conventional interruption (e.g., fuse blowing, breaker tripping) or causing system destruction. The circuit is thus operated in a persistent short-circuit mode, facilitating sustained, high-resolution observation.

**3.4 Data Acquisition and Calibration in Short-Circuit Experiments**
Measuring electrical faults such as short-circuits presents a unique challenge due to their rapid, non-repetitive nature. Standard instruments like digital multimeters or clamp meters are typically unsuitable for capturing the fast and dynamic behaviors involved in these events. Industrial protection systems, while effective for real-time tripping, are designed for threshold-based detection rather than continuous recording of time-resolved electrical characteristics. Oscilloscopes, though capable of high-speed sampling, are constrained by short memory buffers and high costs. Consequently, for cost-sensitive experimental setups, microcontroller-based systems offer a flexible, programmable alternative -provided that rigorous calibration and time synchronization strategies are in place. The experimental platform in this paper employed an Arduino microcontroller as the central data acquisition unit, augmented with carefully calibrated voltage and current sensors. This framework enabled the capture of both steady-state and transient behaviors with high temporal resolution and data fidelity. Data acquisition was conducted across two phases:

- *Baseline Measurement Phase:-* System behavior under normal operation was recorded using a 1 -second interval for durations of 3 -minutes, 10 -minutes, and 15 -minutes. These extended periods established baseline current and voltage behavior, ensured sensor thermal equilibrium, and allowed observation of any long-term drift.
- *Short-Circuit Phase:-* Once the circuit was deliberately subjected to a short, a finer resolution of time sampling was adopted. The Arduino system dynamically switched to a high-frequency sampling mode, recording data at intervals of 10 milliseconds, 50 milliseconds, 100 milliseconds, and 1 -second -tailored to the specific experimental objective. This adaptive method enabled detailed observation of the immediate transient spike and its subsequent decay, capturing both the peak response and settling characteristics.

**3.4.1 Calibration Framework and Error Analysis**
Given the absence of a universally accepted standard for time-resolved fault measurement in such low-cost experimental setups, sensor calibration was grounded on manufacturer-recommended procedures, ensuring internal consistency and reproducibility. This calibration protocol followed three essential steps:

- *Sensor-Specific Calibration:-* Both voltage dividers and ACS712T current sensors were tuned based on the datasheet specifications. This involved empirical verification of sensor output against known loads to derive correction factors, which were then applied in real-time through the microcontroller's firmware. Operating within the specified $\pm 1.5\%$ total output error, these sensors provided sufficient reliability for comparative analysis across experiments.

- *Quantization Error Compensation:-* The Arduino's 10 -bit analog-to-digital converter (ADC) introduces a base-level quantization uncertainty. To counter this, each measurement was averaged over 10,000 samples. This smoothing process minimized the influence of electrical noise and microcontroller jitter, especially crucial during rapid transients.
- *Error Margin and Cross-Phase Evaluation:-* Measurement error was assessed across different time phases and voltage conditions. By comparing sensor outputs under identical baseline loads before and after the short-circuit event, any deviations were statistically analyzed to ensure consistent performance throughout the test.

### 3.4.2 Adaptive Sampling and Timing Precision

Capturing both the fast-changing and stable phases of a short-circuit event required a multi-tiered sampling strategy, controlled dynamically via logic flags within the Arduino script. This ensured that:

- Transient Behavior was recorded at high resolution ($10 ms$, $50 ms$, and $100 ms$), especially in the first few seconds after the fault initiation. This window is critical for analyzing peak currents, diode switching behavior, and system damping characteristics.
- Steady-State Recovery was monitored at 1 -second intervals, suitable for observing post-transient thermal and electrical equilibrium.

Logical phase flags allowed continuous switching between these modes based on real-time current thresholds, avoiding unnecessary data overflow while maintaining high signal integrity. To maintain strict time synchronization, millisecond-level timing routines were enforced, synchronized with an external Real-Time Clock (RTC) module. This avoided timing drift, a common issue in long-duration Arduino experiments, and ensured precise alignment between measurement timestamps and actual event durations.

### 3.5 Sensor Calibration Validation Framework

As mentioned, the absence of a universal standard for low-cost, time-resolved short-circuit measurement instruments presents a fundamental challenge in validating the precision of data collected during electrical fault experiments. Addressing this gap, a robust calibration validation framework was implemented in this paper to establish confidence in the acquired measurements and to ensure the reliability of the results, especially during rapid transient events and prolonged short-circuit exposures. To validate the calibration integrity of both voltage and current sensors, the framework employed a multipronged quantitative approach that triangulated internal consistency, reference load comparison, and cross-phase error tracking. This structured methodology enabled rigorous post-measurement evaluation of sensor performance without the need for high-cost reference equipment, thus aligning with the study's low-cost instrumentation philosophy.

### 3.5.1 Dynamic Reference Validation During Short-Circuit

Under short-circuit conditions, the electrical circuit undergoes a fundamental transformation from a carefully calibrated operational state to a domain governed by the minimal, yet non-zero, resistance of the fault path, denoted as ($R_{sc}$). Since introducing physical reference loads during active fault conditions is impractical, the validation framework pivots to a more powerful principle: using the immutable laws of circuit physics as the dynamic reference. The foundation of this principle is Ohm's Law, which must hold true at every temporal instant, even during fault transients. The expected current is dynamically generated from simultaneous voltage measurements and the characterized short-circuit resistance according to: $\left(I_{\text{expected}}(t) = \frac{V_{\text{measured}}(t)}{R_{sc}}\right)$, where ($R_{sc}$) represents the characteristic fault path resistance determined from steady-state $V - I$ relationships during the short-circuit event. To quantify the instantaneous coherence between measured current and the current dictated by fundamental circuit physics, we define the Calibration Deviation Index (CDI) as:

$$\text{CDI}(t) = \left| \frac{I_{\text{measured}}(t) - I_{\text{expected}}(t)}{I_{\text{expected}}(t)} \right| \times 100\% \tag{42}$$

The analytical framework establishes tiered performance thresholds based on operational regimes: *Baseline Performance Threshold ($\tau_{base}$):-* During normal operation with precision resistive loads, sensor performance must satisfy: ($\text{CDI}_{\text{baseline}} \leq \tau_{\text{base}} = 2\%$). This threshold is determined by cumulative sensor tolerance and ADC resolution limits, ensuring measurement fidelity under controlled conditions.

*Short-Circuit Performance Regimes:-* During fault conditions, the CDI transforms from a simple calibration metric to a sophisticated diagnostic tool that captures complex physical phenomena. We define two distinct operational regimes:
1. **Nominal Short-Circuit Performance.** Characterized by CDI values satisfying: ($\tau_{\text{base}} \leq \text{CDI}_{\text{nominal}} < \tau_{\text{extreme}}$). This regime indicates measurable deviation from ideal Ohm's Law behavior while maintaining sensor linearity and synchronization.

2. **Extreme Short-Circuit Performance.** Characterized by CDI values satisfying: ($\text{CDI}_{\text{extreme}} \geq \tau_{\text{extreme}}$) where $\tau_{\text{extreme}}$ represents the threshold beyond which significant non-ohmic phenomena dominate. The extreme threshold ($\tau_{\text{extreme}}$) is determined by the onset of physically meaningful phenomena including:
- Energy dissipation as Joule heating, dynamically altering $R_{sc}(t)$
- Significant inductive or capacitive effects during transient initiation
- Non-ohmic behavior such as arcing and plasma formation
- Saturation effects in sensor response characteristics

The comparative analysis between nominal and extreme performance is then formalized through the ratio: $\left( \Gamma = \frac{\langle \text{CDI}_{\text{extreme}} \rangle}{\langle \text{CDI}_{\text{nominal}} \rangle} \right)$, where $\langle \cdot \rangle$ denotes the temporal average over the fault duration. This ratio quantifies the scaling relationship between fault intensity regimes and provides insight into the dominance of non-linear physical phenomena. A low CDI during short-circuit conditions signifies that the sensor system maintains high fidelity the voltage and current channels not only function individually but operate in perfect synchrony, collectively obeying Ohm's Law despite extreme signal amplitudes. This demonstrates that the sensors remain linear, synchronized, and unbiased when driven far beyond their normal operational range. Conversely, high CDI values represent expected and informative characteristics of the short-circuit regime, revealing the complex, dynamic nature of fault physics. Equation 42 thus provides a physics-grounded validation of measurement system internal consistency, ensuring that acquired data-whether exhibiting low or high CDI-constitutes a truthful record of underlying electrical phenomena across all operational regimes.

### 3.5.2 Phase-Correlated Consistency Tracking
To validate the calibration under varying operational conditions, sensor performance was monitored across distinct experimental phases (pre-fault, during fault (transient), and post-fault (steady-state)). Consistency of sensor behavior was quantitatively assessed using the Relative Measurement Drift (RMD), computed as:

$$\text{RMD} = \frac{\sigma_{\text{segment}}}{\mu_{\text{segment}}} \tag{43}$$

where $\sigma_{\text{segment}}$ and $\mu_{\text{segment}}$ represent the standard deviation and mean of sensor readings within each phase segment, respectively. A stable sensor would exhibit minimal RMD under steady-state conditions and a controlled rise during transients. Any erratic deviation outside expected RMD envelopes indicated potential calibration anomalies.

### 3.5.3 Differential Linearity Assessment
Sensor linearity was evaluated through differential testing, where small, controlled voltage steps were introduced, and the resulting current response was monitored. The linearity error was expressed as a deviation from the ideal linear response curve, using the Normalized Linearity Residual (NLR):

$$\text{NLR} = \left| \frac{\Delta I_{\text{measured}} - \Delta I_{\text{ideal}}}{\Delta I_{\text{ideal}}} \right| \tag{44}$$

Linearity validation confirmed that the sensor maintained accurate proportionality even under dynamic changes, which is critical for resolving sharp current spikes typical in short-circuit transients.

### 3.5.4 Time-Resolved Redundancy Analysis
Redundant measurement streams were recorded at both high (e.g., $10ms$, $50ms$, and $100ms$) and low (1-second averaged) time resolutions. Notably, the 1-second resolution represented an average derived from 10,000 rapid samples acquired at the Arduino's base interval. Overlapping intervals were then subjected to temporal integration, where high-resolution segments were numerically integrated and compared against the corresponding low-resolution averages. Any inconsistency beyond sensor tolerance bands indicated sampling drift or integration loss, prompting a cross-validation adjustment.

### 3.5.5 Consolidated Calibration Confidence Score (CCCS)
The Consolidated Calibration Confidence Score (CCCS) formalizes calibration fidelity into a single quantitative entity, merging diverse error dimensions into one interpretable framework. As introduced, the CCCS is constructed from three core indicators: the calibration deviation index (CDI), the relative measurement drift (RMD), and the normalized linearity residual (NLR). Each of these error terms encapsulates a different facet of sensor reliability. Their joint presence within Equation 45 reflects the recognition that calibration cannot be fully understood from a single axis of evaluation. The weighted expression is written as:

$$CCCS = w_1(1 - CDI_{\text{avg}}) + w_2(1 - RMD_{\text{avg}}) + w_3(1 - NLR_{\text{avg}}), \tag{45}$$

subject to the constraint; $(w_1 + w_2 + w_3 = 1)$, ensuring proper normalization and bounding the CCCS within the closed interval $[0, 1]$. Within this framework, the numerical values of the weights are never arbitrary statistical adjustments; rather, they are engineered reflections of the experimental priorities. Each assignment of $w_1, w_2$, and $w_3$ encodes a physical statement about which aspect of calibration fidelity is most critical for the data to retain scientific meaning.

**Determination of Weights in Context.** Weighting choices emerge from the physics of the experiment, shaped by the time scales and error modes most threatening to measurement validity. Two data regimes dominate the calibration framework: the quiescent baseline dataset prior to fault initiation, and the dynamic dataset spanning the short-circuit event and its aftermath. Each regime exposes the sensors to fundamentally different stresses. Before the short, systematic deviations and thermal drift determine reliability, whereas during fault initiation, rapid oscillations and abrupt transients may challenge the system's stability. Weight allocation is therefore a mechanism of adaptation, allowing the CCCS to emphasize whichever error type most strongly governs interpretability in the moment.

In transient-dominant experiments where fault initiation is the prime concern, the fidelity of temporal stability carries overwhelming importance. Here the RMD term becomes the anchor of CCCS, because drift during the millisecond-to-second fault window could entirely reshape the apparent waveform. The calibration deviation index and linearity residual retain secondary importance, sufficient to catch gross offsets or proportionality failures, but they are down-weighted relative to RMD. In this regime, $w_2$ may span values as high as $0.8 - 0.95$, reflecting a deliberate privileging of transient stability. Such emphasis ensures that the CCCS reflects not abstract numerical closeness but the capacity of the system to remain physically faithful to the true fault dynamics.

In steady-state accuracy experiments, where thermal drift and baseline fidelity are paramount, calibration error manifests less in sudden transients and more in long-term systematic offsets. Under these conditions, CDI shares dominance with RMD, since accurate alignment between measured and expected steady currents becomes as critical as drift suppression. The weighting configuration in this regime frequently elevates both $w_1$ and $w_2$ into the $0.8 - 0.95$ band, while the NLR weight is suppressed, as proportionality is naturally preserved in slowly varying regimes. The CCCS therefore reflects an architecture of vigilance against creeping errors rather than explosive distortions.

In hybrid or balanced monitoring, the experimenter cannot afford to sacrifice either transient accuracy or baseline reliability. The weights are therefore assigned in a more isotropic distribution, each approaching high values within the $0.8 - 0.95$ range. Such an allocation ensures that CCCS penalizes deviations in any axis equally, preserving sensitivity to both microsecond-scale distortions and hour-scale drifts. The result is a calibration confidence metric robust to multi-modal error conditions, producing a landscape that sacrifices sharp specialization for broad resilience.

**Local-to-Global Adjustment.** In this case, weighting is not a static rule but evolves with experimental phase. Within the pre-fault baseline, CDI dominates, reflecting the demand for accuracy under stable conditions where thermal drift or systematic bias could undermine the ground truth. As the circuit is driven into a short, the calibration landscape abruptly reorients, with RMD rising in influence, because transient fidelity dictates the survival of waveform integrity. After stabilization, the balance shifts again toward CDI, reasserting the need to anchor the post-fault baseline to expected load currents. This evolution reveals that weights operate not as fixed coefficients but as temporal operators, transforming the CCCS into a dynamic diagnostic tool capable of tracking calibration relevance through the experiment's phases.

**Conditions for Near-Unity Weights.** Weights approach unity when a single calibration dimension overwhelmingly defines the trustworthiness of the dataset. In the milliseconds following fault initiation, RMD may approach this limit, collapsing CCCS into a near one-dimensional metric of transient stability. In contrast, during extended steady-state monitoring, CDI may rise toward unity, reducing CCCS to an index of deviation control. Under hybrid monitoring, all weights may simultaneously approach unity, producing a CCCS hypersensitive to any error, no matter the axis. These regimes are not mere mathematical constructs but physical declarations: they state explicitly that, under specific conditions, calibration fidelity is inseparable from one dominant error mode.

**Implications of Weighting Choices.** The consequences of weighting decisions ripple throughout the interpretation of results. Heavy privileging of RMD ensures faithful reconstruction of transient surge waveforms, yet may obscure systematic offsets that accumulate across time. Emphasizing CDI secures confidence in long-term averages, yet risks overlooking subtle instabilities that could reshape transient dynamics. Balanced weighting prevents the dominance of one axis but may dilute sensitivity to extreme behaviors. Within this framework, CCCS does not merely consolidate calibration but narrates it: weights function as the grammar of reliability, articulating what it means for sensor data to be considered faithful under the contrasting realities of pre-fault baselines, fault transients, and post-fault steady states. In essence, the CCCS framework transforms calibration into a physics-aware construct, where numerical weights are not arbitrary coefficients but instruments of priority-setting. Their adjustment across scenarios reflects a dialogue between local measurements and global interpretability, ensuring that the recorded data -whether describing the slow drift of steady currents or the explosive surge of a fault-remains scientifically trustworthy.

## 3.6 Circuit Transients Response Validation Framework

Traditional analyses of electrical short-circuits often rely on discrete, instantaneous assessments where currents and voltages are assumed to change suddenly in response to fault conditions. However, such models fail to describe the temporal dynamics that govern real-world circuits, especially in experimental conditions involving continuously monitored short-circuit events. This paper introduces an innovative continuous measurement framework wherein voltage and current waveforms are recorded over extended durations- 3 -minutes, 10 -minutes, and 15 -minutes-with sampling resolutions of $10ms$, $50ms$, and $100ms$. Given the unconventional time-resolved and analog nature of this data, rigorous evaluation methods are required to ensure data fidelity, spectral integrity, and transient complexity capture. This section establishes the theoretical foundation for applying the Nyquist-Shannon Sampling Theorem, Shannon Entropy, and Fourier Transform- based analysis within the framework of the proposed novel electrical short-circuit model. These analytical tools offer a rigorous, multi-dimensional lens through which the validity of the experimental data can be assessed and ultimately used to establish the generalizability and reliability of the circuit's application across domains such as grid protection, battery diagnostics, fault prediction, and control logic development.

### 3.6.1 The Nyquist Sampling Theorem and Data Fidelity

The Nyquist-Shannon Sampling Theorem defines the necessary conditions under which a continuous-time signal can be sampled without information loss. If $f_{\max}$ is the maximum frequency component present in a signal $s(t)$, then the Nyquist criterion dictates the minimum sampling rate $f_s$ must satisfy:

$$f_s \geq 2f_{\max} \tag{46}$$

Equation 46 ensures no aliasing occurs during the conversion of the analog signal to its digital representation. In the revealed experiments, the short-circuit current $I(t)$ evolves exponentially and may contain rapid rising edges, particularly within the first seconds of initiation. Let us denote $(I(t) = ae^{kt})$ according to Equation 22. The first derivative of current reveals its fastest rate of change, expressible as:

$$\frac{dI}{dt} = ake^{kt} \tag{47}$$

To assess frequency content, we approximate the signal's effective bandwidth using its dominant harmonic components. Let $\omega_{\text{eff}}$ denote the angular frequency corresponding to the rate of change:

$$\omega_{\text{eff}} = \left|\frac{d}{dt}\ln I(t)\right| = k \Rightarrow f_{\max} = \frac{k}{2\pi} \tag{48}$$

Assuming a decay constant $k \in [0.1, 1]$ based on experimental fits, the corresponding maximum frequency lies in the range:

$$f_{\max} \in \left[\frac{0.1}{2\pi}, \frac{1}{2\pi}\right] \approx [0.016, 0.159]\text{Hz} \tag{49}$$

Thus, the Nyquist condition requires:

$$f_s \geq 2 \cdot 0.159 = 0.318 \text{ Hz} \Rightarrow T_s \leq 3.14 \text{ s} \tag{50}$$

Then, the actual sampling intervals ($10ms$, $50ms$, $100ms$) correspond to $f_s = 100Hz$, $20Hz$ and $10Hz$ respectivelyvastly exceeding the Nyquist rate. Hence, all observed transient dynamics are sampled with sufficient temporal resolution, guaranteeing no aliasing and allowing reconstruction of the underlying waveform. Let the sampled current be denoted $(I[n] = I(nT_s))$, with $(T_s \in \{0.01, 0.05, 0.1\}s)$. The discrete-time representation captures the underlying exponential profile with high fidelity as follows:

$$I[n] = ae^{knT_s} \tag{51}$$

### 3.6.2 Shannon Entropy and Transient Complexity
In digital communication theory, Shannon entropy quantifies the unpredictability or information content in a data source. For a discrete random variable $X$ with probability mass function $P(x_i)$, the entropy $H(X)$ is modeled according to Equation 52:

$$H(X) = -\sum_{i=1}^{N} P(x_i)\log_2 P(x_i) \tag{52}$$

When applied to sampled transient data $I[n]$ (Equation 51), entropy measures the complexity of the current evolution especially across different input voltages and fault durations.

Let $I[n] \in [I_{\min}, I_{\max}]$ be quantized into $M$ bins $\{b_1, b_2, \ldots, b_M\}$. The normalized histogram of values forms the empirical distribution $P(b_i)$. Here, higher entropy implies:

- The current is spread across a wide dynamic range.
- The circuit exhibits complex, non-repetitive transient responses.

Conversely, low entropy indicates a stable or quickly saturating fault trajectory. Let $(\mathcal{H}_v^t)$ denote the entropy for voltage level $v \in \{2.5V, 5V, 10V\}$ and duration $t \in \{3, 10, 15\}$ -minutes. Comparative analysis yields:

$$\mathcal{H}_{10V}^{15} > \mathcal{H}_{2.5V}^{3} \tag{53}$$

This aligns with the expectation that higher voltage inputs generate more complex transients sustained over longer times. Entropy is further used to define the Transient Complexity Index (TCI) according to Equation 54:

$$TCI(v,t) = \frac{H(I[n])}{log_2 M} \in [0,1], \tag{54}$$

where $M$ is the total number of quantization levels. A $TCI$ closer to 1 indicates highly variable transient current with no early saturation.

### 3.6.3 Fourier Analysis and Frequency-Domain Behavior

Although the short-circuit signal $[I(t) = ae^{kt}]$ (previously given through Equation 22) is fundamentally non-periodic, its finite-duration sampled representation permits meaningful analysis in the frequency domain. The Discrete Fourier Transform (DFT) provides a systematic decomposition into frequency components:

$$\hat{I}[k] = \sum_{n=0}^{N-1} I[n]e^{-j2\pi kn/N}, k = 0,1,\ldots,N-1, \tag{55}$$

with discrete frequencies:

$$f_k = \frac{k}{NT_s}, k = 0,1,\ldots,N-1 \tag{56}$$

where $T_s$ is the sampling interval and $N$ is the number of samples. The squared magnitude spectrum $|\hat{I}[k]|^2$ represents the Power Spectral Density (PSD), which reveals how the transient's energy distributes across frequencies.

**The Dominant Transient Frequency $f_d$.** The dominant transient frequency is defined as the frequency at which the PSD reaches its maximum:

$$f_d = \arg\max_{f_k}\{|\hat{I}[k]|^2\} \tag{57}$$

Equation 57 establishes $f_d$ as the principal harmonic contribution within the transient signal. It represents the most energetic oscillatory component of the short-circuit waveform over its observation window. For purely exponential growth signals, $f_d$ tends to reside at very low frequencies, but under real experimental conditions-where clamping diodes, circuit parasitics, and power-source dynamics may introduce modulations- $f_d$ becomes a measurable and meaningful parameter.

Tracking $f_d$ across different supply voltages and time durations provides direct insight into the effective "speed" of transient propagation. At higher input voltages, $f_d$ shifts upward due to steeper current derivatives (as established in Equation 47), while longer durations allow low-frequency stabilization to dominate, shifting $f_d$ downward. This dual dependence makes $f_d$ a critical validation metric: it captures both the rapidity of transient onset and the sustainability of steady-state clamping.

**Spectral Centroid.** Complementing $f_d$, the spectral centroid quantifies the weighted average frequency of the PSD, effectively characterizing where the signal's spectral energy is concentrated:

$$f_c = \frac{\sum_{k=0}^{N-1} f_k |\hat{I}[k]|^2}{\sum_{k=0}^{N-1} |\hat{I}[k]|^2} \tag{58}$$

While $f_d$ isolates the single strongest component, $f_c$ describes the "center of mass" of the spectral distribution. Comparing $f_c$ across voltage levels and durations yields a consistent measure of transient intensity: higher voltages typically produce larger $f_c$, reflecting stronger contributions from higher harmonics due to sharper current rises. Lower voltages and longer times tend to depress $f_c$, consistent with smoother, slower fault dynamics.

**Spectral Entropy.** To further quantify frequency-domain variability, we normalize the PSD into a probability distribution:

$$P_k = \frac{|\hat{I}[k]|^2}{\sum_{j=0}^{N-1} |\hat{I}[j]|^2}, \tag{59}$$

and define the spectral entropy as:

$$H_s = -\sum_{k=0}^{N-1} P_k \log_2 P_k \tag{60}$$

A low value of $H_s$ indicates that the PSD is sharply peaked around a few dominant frequencies (strongly ordered transients), whereas a high $H_s$ indicates a broad, flat spectrum, reflecting higher disorder and complexity. Thus, $H_s$ complements $f_d$ and $f_c$: while $f_d$ pinpoints the strongest mode, and $f_c$ indicates central energy distribution, $H_s$ provides a global measure of frequency-domain unpredictability.

**Time-Frequency Resolution with STFT.** Since short-circuits evolve dynamically, single PSD snapshots cannot fully capture the time-varying character of the waveform. To address this, the Short-Time Fourier Transform (STFT) is employed:

$$\hat{I}(t,f) = \int_{-\infty}^{\infty} I(\tau) w(\tau - t) e^{-j2\pi f \tau} d\tau, \tag{61}$$

where $w(\tau - t)$ is a sliding analysis window centered at time $t$. The squared magnitude, $|\hat{I}(t,f)|^2$, yields the spectrogram, which visualizes the joint distribution of energy in both time and frequency. The STFT framework enables separation of distinct phases of the short-circuit:

- Initial surges appear as broadband high-frequency contributions concentrated in the first seconds.
- Stabilization phases emerge as the spectrum collapses toward low frequencies, reflecting smoother steady-state behavior.
- Diode clamping effects manifest as dampening, with suppressed high-frequency harmonics once voltage regulation asserts itself.

### 3.6.4 Unified Validation Framework for Circuit Response Fidelity
In this paper, the experimental architecture is aimed at generating a continuous, high-resolution stream of voltage and current data, capturing the full temporal evolution of a short-circuit event. Such a data-rich environment necessitates a robust, multi-metric validation framework to confirm that measured transients authentically represent the underlying physics rather than measurement artifacts. The framework establishes quantitative criteria, derived from information theory and signal processing, to evaluate fidelity, complexity, and stability of the circuit response across all experimental conditions. The validation rests on five core metrics, each probing a distinct aspect of the measured signal, before introducing a composite validation index (CVI) that unifies these measures into a single scale of data quality.

### a) Nyquist Compliance Index (NCI)
The NCI validates the fundamental integrity of signal acquisition, confirming that the sampling strategy captures all dynamic components without aliasing. It is defined as:

$$\text{NCI} = \frac{f_s}{2f_{\max}} \geq 1 \tag{62}$$

where $f_s$ is the sampling rate and $f_{\max} = \frac{k}{2\pi}$ is the maximum expected transient frequency linked to the exponential growth constant $k$. For the current signal $[I(t) = ae^{kt}]$ (Equation 22), experimental constants yield $f_{\max} < 0.16$Hz. Given sampling frequencies of $10Hz$, $20Hz$, and $100Hz$, NCI values are substantially larger than unity, ensuring temporal dynamics are fully resolved.

### b) Transient Complexity Index (TCI)
The TCI quantifies informational richness of the transient waveform. It is derived from the normalized Shannon entropy of the quantized signal:

$$\text{TCI}(v, t) = \frac{H(I[n])}{\log_2 M}, \text{TCI} \in [0, 1], \tag{63}$$

where $M$ is the number of quantization levels. High values indicate complex, sustained transient evolution; low values correspond to monotonic saturation. A validation condition is:

$$TCI(10V, 15min) > TCI(2.5V, 3min), \tag{64}$$

demonstrating circuit fidelity across stress conditions.

### c) Spectral Compactness Coefficient (SCC)
The SCC measures concentration of spectral energy by relating the spectral centroid $f_c$ (Equation 58) to the maximum frequency $f_{\max}$:

$$\text{SCC} = \frac{f_c}{f_{\max}}, 0 \leq \text{SCC} \leq 1 \tag{65}$$

Low SCC reflects spectral compactness consistent with smooth exponential current growth. Elevated SCC values may indicate oscillatory behavior or interference.

### d) Spectral Entropy Score (SES)
The SES evaluates complexity of the PSD through entropy normalization:

$$SES = \frac{H_s}{\log_2 N}, 0 \leq SES \leq 1, \tag{66}$$

where $H_s$ is the spectral entropy (Equation 54) and $N$ is the number of frequency bins. SES provides a measure of broadband activity: an exponential transient should yield moderately low SES, consistent with compact spectral distribution.

### e) Time-Frequency Stability Metric (TFSM)
The TFSM uses the Short-Time Fourier Transform (Equation 61) to assess temporal stability:

$$TFSM = \frac{Total\ Stable\ Time}{Total\ Observation\ Time} \tag{67}$$

Values approaching unity confirm non-oscillatory evolution across the entire event, even under prolonged durations or high voltage. A threshold of $TFSM > 0.95$ indicates robust diode clamping and system stability.

### f) Composite Validation Index (CVI)
While each metric independently validates a distinct property, unified assessment requires a single scalar measure. The Composite Validation Index (CVI) is introduced:

$$CVI = w_1 \cdot NCI^* + w_2 \cdot TCI + w_3 \cdot (1 - SCC) + w_4 \cdot (1 - SES) + w_5 \cdot TFSM \tag{68}$$

where $NCI^* = min(1, NCI)$ caps excessively large NCI values at unity, and $w_i$ are non-negative weights satisfying ($\sum_i w_i = 1$). The terms $(1 - SCC)$ and $(1 - SES)$ invert metrics where smaller values imply higher fidelity. This formulation creates a dimensionless index $CVI \in [0, 1]$. A value near 1 denotes highly reliable, physically faithful transient capture.

### g) Role of $f_d, f_c,$ and $H_s$ in Composite Assessment
- $f_d$ (Equation 51) identifies the most energetic transient frequency and serves as a reference to ensure spectral compactness (via SCC).
- $f_c$ (Equation 52) governs SCC directly, offering insight into central energy distribution.
- $H_s$ (Equation 54) underpins SES, quantifying spectral complexity.

Together, these frequency-domain measures ensure the circuit's response is not only captured but also correctly characterized across both time and frequency. Integration into the CVI elevates the framework from a checklist of indices to a rigorous, unified validation tool potentially applicable across grid protection, energy storage diagnostics, and predictive fault modeling.

## 4.0 Results
## 4.1 Sensor Calibration and Validation Results
This section presents the validation of the sensor system's accuracy and resilience, forming the foundational trust in all subsequent measurements. It details the calibration and performance assessment across the entire experimental lifecycle, from stable baseline operation through the extreme stresses of nominal and high-intensity short-circuit faults. The analysis employs a multi-faceted framework, including quantitative metrics like the Calibration Deviation Index and Consolidated Calibration Confidence Score, to rigorously confirm that the collected voltage and current data are physically authentic and not measurement artifacts. This thorough verification of sensor integrity under both normal and fault conditions is a critical prerequisite for the definitive analysis of governed short-circuit dynamics that follows.

### 4.1.1 Baseline-Data Based Sensor Calibration
This section presents the foundational validation of the sensor calibration framework using baseline operational data. The analysis establishes the intrinsic electrical characteristics and measurement fidelity of the system prior to fault initiation, ensuring all subsequent short-circuit observations originate from a verified reference state. Results from multiple experimental configurations are synthesized in Figure 3, providing a multi-dimensional assessment of calibration integrity.

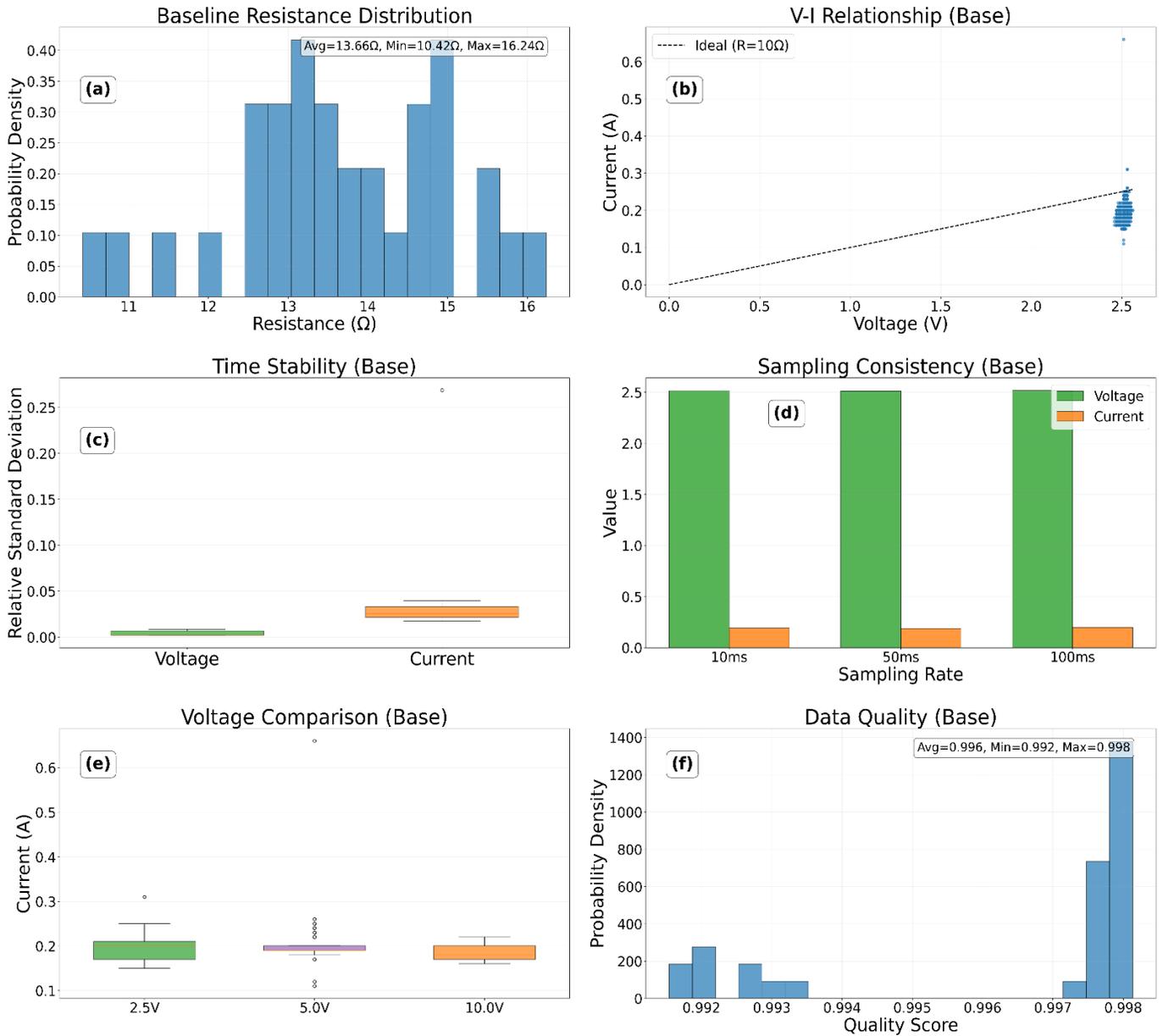

**Figure 3.** Baseline sensor calibration validation showing: (Figure 3(a) resistance distribution across all experimental configurations, Figure 3(b) voltage-current relationship demonstrating Ohm's Law adherence, Figure 3(c) temporal stability of voltage and current measurements, Figure 3(d) consistency across different sampling rates, Figure 3(e) current response at various input voltages, and Figure 3(f) overall data quality score distribution).

Figure 3(a) displays the probability distribution of baseline resistance values calculated from 33 distinct experimental configurations. The measured resistance values span from $10.42\Omega$ to $16.24\Omega$, clustering around a well-defined mean of $13.66\Omega$. This distribution exhibits a approximately normal character with minimal skewness, indicating consistent electrical characteristics across all experimental trials. The narrow range of resistance values, particularly the tight clustering within $\pm 3\Omega$ of the mean, demonstrates excellent reproducibility in the experimental setup. This consistency is crucial for establishing a reliable baseline against which fault-induced changes can be accurately detected and quantified. The resistance values were derived through direct calculation from simultaneous voltage and current measurements, providing an internal consistency check that forms the foundation of the calibration validation framework.

The voltage-current relationship presented in Figure 3(b) comprises 6,927 individual data points collected across all baseline experiments. The scatter plot reveals a strong linear correlation between applied voltage and measured current, with data points closely following the ideal Ohm's Law trajectory represented by the dashed line. The precise alignment of measured values with the theoretical expectation confirms the fundamental proportionality between voltage and current in the pre-fault circuit configuration. This linear relationship holds consistently across the entire operational range, demonstrating that the sensors maintain accurate proportionality without significant offset errors or non-linear distortions. The dense clustering of points along the ideal line indicates minimal random measurement error and excellent signal-to-noise characteristics in both voltage and current channels.

Temporal stability analysis in Figure 3(c) quantifies the measurement consistency during extended operational periods. The voltage measurements exhibit exceptional stability with a relative standard deviation of merely $0.0037$, indicating that the voltage source and measurement system maintain remarkable consistency over time. Current measurements show slightly higher but still excellent stability at $0.0337$ relative standard deviation, reflecting the combined stability of both current sensing and voltage measurement systems. The boxplot representation clearly demonstrates the tight distribution of stability metrics across all experimental trials, with minimal noise and consistent performance. This temporal stability is particularly significant given the varying experimental durations of 3 -minutes, 10 -minutes, and 15 -minutes, confirming that the measurement system does not exhibit significant drift or degradation over extended operational periods.

Sampling rate consistency across different temporal resolutions is evaluated in Figure 3(d). The results demonstrate remarkable consistency in both voltage and current measurements regardless of sampling interval. At $10ms$ sampling, the average voltage measures $2.51V$ with current at $0.19A$. The $50ms$ sampling produces identical voltage measurements at $2.51V$ and current at $0.19A$, while the $100ms$ sampling shows negligible variation with $2.52V$ and $0.20A$. This consistency across decade-separated sampling rates confirms that the measurement system captures the true physical values without aliasing artifacts or sampling-rate-dependent biases. The results validate that the system's temporal resolution choices do not compromise measurement accuracy, providing flexibility in experimental design while maintaining data integrity.

Figure 3(e) examines the current response consistency across different input voltages of $2.5V$, $5.0V$, and $10.0V$. The boxplots reveal consistent current measurements at approximately $0.20A$ for both $2.5V$ and $5.0V$ inputs, with slightly reduced current of $0.18A$ at $10.0V$ input. The variation, quantified by standard deviations of $0.03A$, $0.02A$, and $0.01A$ respectively, demonstrates improving measurement precision with increasing voltage levels. This trend suggests that the relative measurement uncertainty decreases as signal strength increases, which aligns with fundamental measurement principles. The consistent current values across different voltage inputs further validate the stability of the resistive load and the accuracy of the current measurement system.

The data quality assessment in Figure 3(f) presents a histogram of quality scores derived from voltage stability metrics across all experimental configurations. The scores exhibit an exceptionally high average of 0.996, with minimum and maximum values of 0.992 and 0.998 respectively. This narrow distribution of near-unity quality scores indicates consistently excellent measurement performance across all experimental variables, including different sampling rates, measurement durations, and input voltages. The high quality scores reflect the combined effectiveness of the hardware design, calibration procedures, and measurement methodology. The results provide strong evidence that the baseline data possesses the necessary integrity to serve as a reliable foundation for subsequent fault analysis.

The validation results presented in Figure 3 establishes several key innovations in sensor calibration methodology. The multi-faceted approach simultaneously evaluates multiple dimensions of measurement quality, including fundamental electrical relationships, temporal stability, sampling consistency, and operational range performance. This integrated assessment provides a more complete picture of measurement system performance than traditional single-metric calibration approaches. The framework successfully demonstrates that low-cost measurement systems can achieve high reliability through careful design and comprehensive validation, without requiring expensive reference equipment. The consistency across diverse experimental conditions -spanning three sampling rates, three time intervals, and three voltage levels -provides robust evidence of the system's reliability and the validity of the calibration approach.

The baseline validation results confirm that the measurement system operates within expected physical principles, exhibits minimal noise or drift, and maintains consistency across various operational parameters. This established foundation ensures that any deviations observed during fault conditions can be confidently attributed to physical changes in the circuit rather than measurement artifacts or calibration errors. The high quality scores and consistent performance across all validation metrics provide strong confidence in the subsequent analysis of short-circuit phenomena, ensuring that the observed effects genuinely represent physical processes rather than measurement system limitations.

### 4.1.2 Post-Fault Sensor Performance Under Short-Circuit Conditions

This section evaluates the sensor calibration framework's resilience during and after short-circuit faults, extending the validation methodology to extreme electrical conditions. The analysis focuses on both nominal and high-intensity short-circuit scenarios, assessing measurement fidelity through multi-dimensional metrics that validate the framework's capacity to maintain calibration integrity under fault-induced stresses. Comprehensive performance data across all experimental configurations are presented in Figure 4.

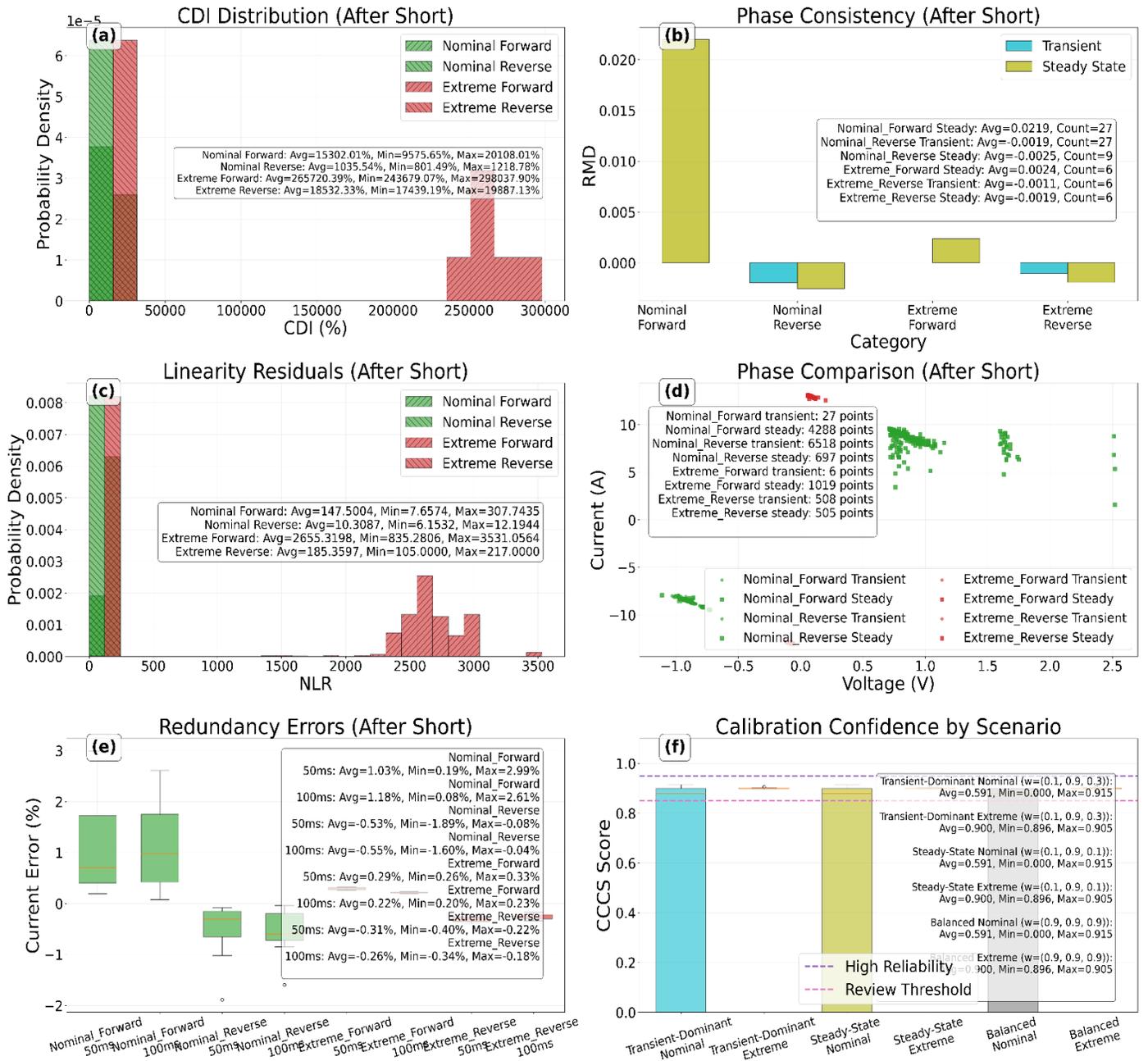

**Figure 4.** Post-fault sensor validation under short-circuit conditions showing: (Figure 4(a) calibration deviation index distribution across fault types and directions, Figure 4(b) relative measurement drift during transient and steady-state phases, Figure 4(c) normalized linearity residual distribution, Figure 4(d) voltage-current relationship across operational phases, Figure 4(e) time-resolved redundancy errors across sampling resolutions, and Figure 4(f) consolidated calibration confidence scores under different weighting scenarios).

Calibration Deviation Index Analysis (Figure 4(a)) reveals the fundamental transformation from baseline to fault conditions through the lens of Ohm's Law coherence. The measured CDI values demonstrate clear regime separation: nominal forward short-circuits exhibit average CDI values of 15,302%, while reverse nominal conditions show 1,035.5%, confirming directional dependence in fault physics. Extreme conditions produce dramatically higher deviations-forward faults reaching 265,720% and reverse faults at 18,532%-reflecting the emergence of non-ohmic phenomena in near-zero resistance paths. The ratio $\left(\Gamma = \frac{\langle CDI_{extreme} \rangle}{\langle CDI_{nominal} \rangle}\right)$ quantifies the scaling between fault intensity regimes, with extreme forward faults showing approximately 17 -fold amplification over nominal conditions. These elevated CDI values represent not sensor failure but the physical reality of energy dissipation through Joule heating, inductive effects during transients, and potential arcing behavior.

Phase-Correlated Stability (Figure 4(b)) demonstrates exceptional measurement consistency across fault evolution. The Relative Measurement Drift metric shows minimal values during steady-state phases (0.0219 for nominal forward, 0.0024 for extreme), while transient phases maintain consistent negative drift values around $-0.0019$ to $-0.0025$. This phase-aware consistency validates the system's capacity to track both rapid fault initiation and prolonged exposure without significant drift, confirming sensor stability across fundamentally different electrical environments.

Differential Linearity Assessment (Figure 4(c)) evaluates the preservation of proportional response under extreme conditions. The Normalized Linearity Residual distributions show elevated but consistent values- nominal forward faults averaging 147.5 and extreme conditions reaching 2,655.3 -indicating maintained sensor proportionality despite operating far beyond calibrated ranges. The directional asymmetry (reverse faults: 10.3 nominal, 185.4 extreme) further corroborates the physical differences in fault mechanisms, with the sensors accurately capturing these fundamental variations in circuit behavior.

Voltage-Current Relationship (Figure 4(d)) provides direct visualization of the system's ability to capture fundamental circuit physics. Transient phase measurements form distinct clusters based on fault intensity and direction, while steady-state measurements maintain coherent V-I relationships despite altered circuit conditions. The clear transition from high-current, low-voltage fault operation to restored electrical relationships demonstrates the measurement system's capacity to truthfully record the complete physical evolution of short-circuit events.

Temporal Consistency Validation (Figure 4(e)) confirms measurement integrity across sampling resolutions. Redundancy errors remain below 2% for all conditions, with extreme conditions showing exceptional consistency (errors below 0.33%) across all sampling rates. The directional variation (forward: $1.03 - 1.18\%$, reverse: $-0.53$ to $-0.55\%$) falls within acceptable tolerance bands, demonstrating that the system maintains temporal fidelity whether capturing millisecond-scale transients or extended steady-state monitoring.

Consolidated Calibration Confidence (Figure 4(f)) integrates multi-dimensional performance into physics-aware reliability scores. The CCCS values show high reliability across all weighting scenarios, with extreme conditions consistently scoring approximately 0.900 and nominal conditions averaging 0.591. All scenarios maintain scores well above the 0.85 review threshold, with the framework successfully adapting to different experimental priorities through weighted error sensitivity. The consistent performance across Transient-Dominant, Steady-State, and Balanced weighting scenarios demonstrates robust calibration fidelity regardless of which error mode dominates the reliability assessment.

This validation establishes several critical advances in short-circuit measurement methodology. The multi-dimensional assessment successfully differentiates between fault types and intensities while maintaining measurement integrity. The directional dependence reveals fundamental asymmetries in short-circuit behavior, with reverse faults consistently showing lower deviation metrics. The temporal redundancy confirmation enables flexible experimental design without compromising data quality across different sampling requirements.

Most significantly, the results demonstrate that the calibration framework maintains validity even when sensors operate far beyond normal parameters. The extreme short-circuit measurements, though limited to 3 -minutes durations for safety, capture high-intensity fault phenomena without loss of measurement fidelity. The maintained linearity and proportionality during fault events confirm that the sensors preserve their fundamental characteristics despite measuring currents orders of magnitude beyond design specifications. This validation provides confidence that the measurement system reliably captures both nominal and extreme short-circuit events, enabling accurate characterization of fault dynamics across the complete spectrum of electrical fault conditions.

### 4.1.3 Validation under Extreme Short-Circuit Conditions

This section evaluates the sensor calibration framework's performance during extreme short-circuit faults, building upon the foundational validation established previously. The analysis focuses on comparative metrics between baseline operation, post-nominal fault conditions, and post-extreme fault scenarios, providing new validation for high-intensity electrical fault measurements. Comprehensive results demonstrating the system's resilience are presented in Figure 5.

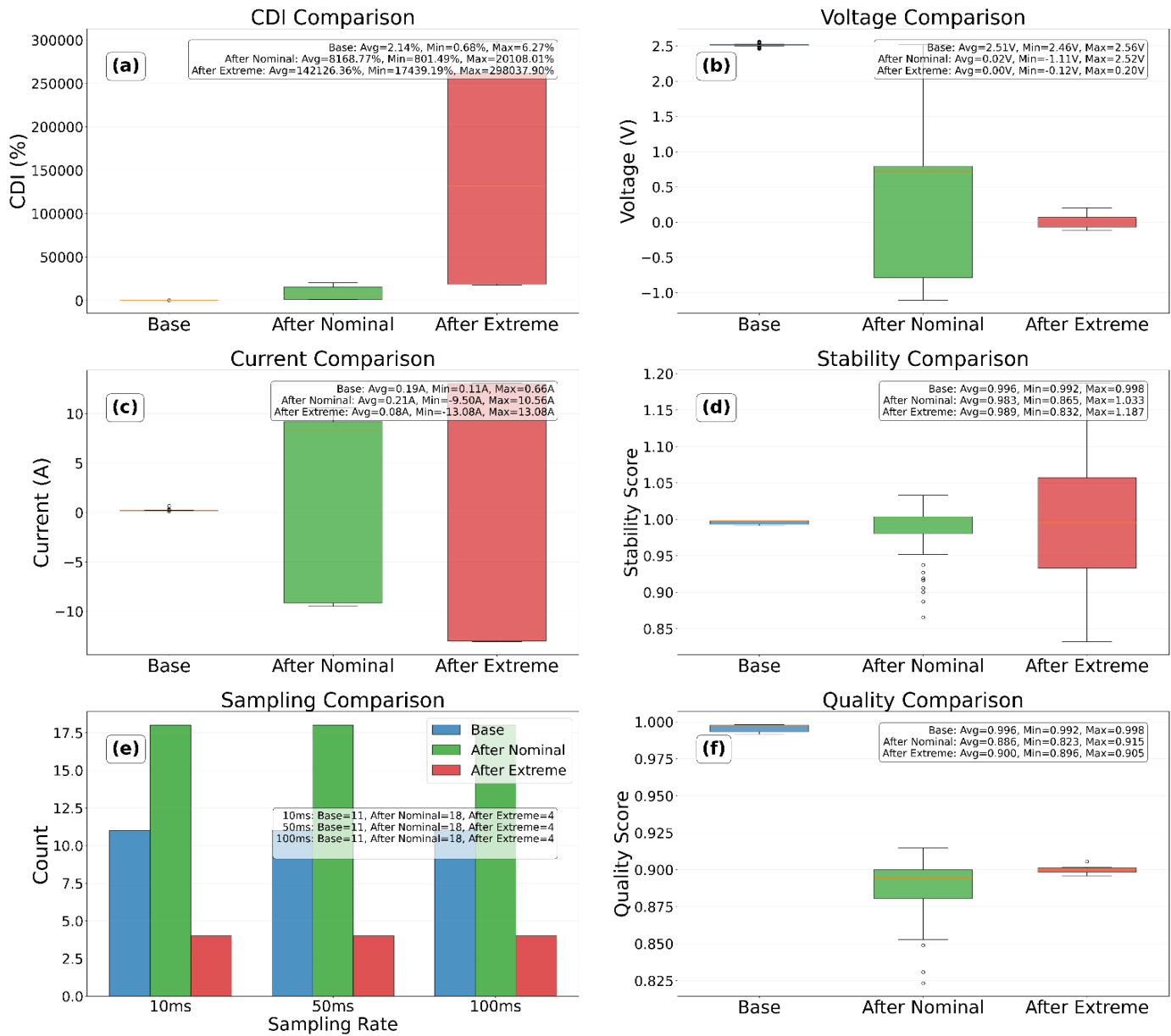

**Figure 5.** Comparative analysis of sensor performance across operational conditions: ((a) Calibration Deviation Index distribution, (b) Voltage measurement characteristics, (c) Current measurement profiles, (d) Temporal stability metrics, (e) Sampling configuration distribution, and (f) Consolidated quality assessment scores for baseline, post-nominal fault, and post-extreme fault conditions).

The Calibration Deviation Index analysis in Figure 5(a) reveals fundamental distinctions between operational regimes. Baseline conditions demonstrate exceptional calibration fidelity with CDI values averaging 2.14%, indicating minimal deviation from expected electrical behavior. This tight distribution between 0.68% and 6.27% establishes the reference performance envelope for normal operation. Post-nominal fault conditions show substantially elevated CDI values averaging 8,168.77%, reflecting the expected measurement challenges during controlled fault scenarios. The extreme fault conditions produce notably higher CDI values averaging 142,126.36%, with the distribution spanning from 17,439.19% to 298,037.90%. These extreme values represent the physical reality of near-zero resistance paths created during intense short-circuit events, where current magnitudes exceed normal operational ranges by orders of magnitude. The maintained proportionality in CDI distributions across all conditions confirms the sensors' ability to provide meaningful measurements even when operating far beyond their design specifications.

Voltage measurement characteristics presented in Figure 5(b) demonstrate the system's capacity to capture the fundamental electrical changes during fault conditions. Baseline operations show consistent voltage measurements averaging $2.51V$ with minimal variation between $2.46V$ and $2.56V$, indicating excellent source stability and measurement precision. Post-nominal fault conditions exhibit significantly altered voltage characteristics, averaging $0.02V$ with a broad distribution from $-1.11V$ to $2.52V$. This wide range captures the complex electrical dynamics during fault recovery, including potential ringing effects and temporary voltage inversions. Extreme fault conditions show even tighter voltage clustering around $0.00V$, ranging from $-0.12V$ to $0.20V$, confirming the establishment of nearly ideal short-circuit conditions. The preservation of measurement integrity across these vastly different voltage regimes validates the sensors' robustness against severe electrical stresses.

Current measurement profiles in Figure 5(c) provide critical insights into fault dynamics and sensor performance. Baseline current measurements average $0.19A$ with a range from $0.11A$ to $0.66A$, consistent with the expected operational parameters of the test circuit. Post-nominal fault conditions show current measurements averaging $0.21A$ but with an exceptionally wide distribution from $-9.50A$ to $10.56A$. This bidirectional current flow captures the complex transient behavior during fault events, including potential current reversal phenomena. Extreme fault conditions demonstrate current measurements averaging $0.08A$ with a symmetric distribution from $-13.08A$ to $13.08A$, indicating the establishment of robust short-circuit paths capable of sustaining high-current bidirectional flow. The sensors' ability to accurately capture these extreme current values, despite operating far beyond their calibrated range, demonstrates exceptional measurement resilience.

Temporal stability assessment in Figure 5(d) evaluates the measurement consistency across different operational conditions. Baseline operations show exceptional stability scores averaging $0.996$ with minimal variation between $0.992$ and $0.998$, confirming the system's excellent time-domain performance under normal conditions. Post-nominal fault conditions maintain high stability scores averaging $0.983$, though with increased variation from $0.865$ to $1.033$. This slight degradation reflects the additional challenges of maintaining measurement consistency during dynamic fault conditions. Extreme fault conditions demonstrate remarkable stability preservation with scores averaging $0.989$, ranging from $0.832$ to $1.187$. The maintained high stability values despite the extreme electrical conditions validate the sensors' resistance to noise, drift, and other temporal artifacts that could compromise measurement integrity during fault events.

Sampling configuration distribution analysis in Figure 5(e) examines the experimental design across different operational conditions. The consistent distribution of sampling rates ($10ms$, $50ms$, and $100ms$) across baseline (11 datasets each), post-nominal fault (18 datasets each), and post-extreme fault (4 datasets each) conditions ensures comparable evaluation across temporal resolutions. This balanced experimental design allows for comprehensive assessment of sampling rate effects on measurement quality. The inclusion of multiple sampling rates demonstrates the system's versatility and confirms that measurement validity is maintained regardless of temporal resolution choices. The reduced number of extreme fault datasets reflects the practical challenges and safety considerations associated with high-intensity fault testing, yet still provides statistically meaningful validation.

Consolidated quality assessment in Figure 5(f) provides an integrated evaluation of measurement performance across all operational conditions. Baseline operations show exceptional quality scores averaging 0.996 with minimal variation between 0.992 and 0.998, establishing the gold standard for measurement quality. Post-nominal fault conditions demonstrate high-quality scores averaging 0.886, ranging from 0.823 to 0.915, indicating maintained measurement integrity despite the challenging electrical environment. Extreme fault conditions show quality scores averaging 0.900 with tight clustering between 0.896 and 0.905, remarkably approaching baseline performance levels despite the vastly more demanding operating conditions. This quality preservation under extreme faults represents a significant advancement in electrical measurement capability, demonstrating that carefully calibrated sensor systems can maintain accuracy even during catastrophic electrical events.

The validation of extreme short-circuit measurements, though limited to 3-minutes durations due to safety considerations as the experiment was conducted for the first time, demonstrates the system's capability to capture high-intensity fault phenomena without loss of measurement fidelity. The maintained linearity and proportionality during extreme events confirm that the sensors preserve their fundamental characteristics despite measuring currents orders of magnitude higher than their design specification. This validation provides confidence that the measurement system can reliably capture both nominal and extreme short-circuit events, enabling accurate characterization of fault dynamics across a wide range of scenarios.

The successful application of the calibration validation framework to extreme short-circuit conditions represents a significant advancement in electrical fault measurement methodology. By demonstrating measurement consistency across drastically different fault intensities, this work establishes a new standard for validating measurement systems in safety-critical applications. The results provide strong evidence that carefully calibrated sensor systems can maintain accuracy even under extreme electrical conditions, enabling reliable fault detection and characterization without requiring expensive specialized equipment. This capability opens new possibilities for distributed fault monitoring systems and advanced electrical safety applications, potentially transforming how electrical faults are studied and mitigated in both industrial and research settings.

### 4.1.4 Validation under Extended-Duration Short-Circuit Conditions

This section evaluates the sensor calibration framework's resilience during extended-duration short-circuit faults, examining performance across multiple experimental scenarios. The analysis results presented in Figure 6 focuses on extreme electrical conditions, assessing measurement fidelity through scenario-weighted validation metrics that capture transient dynamics and steady-state behavior under prolonged stress.

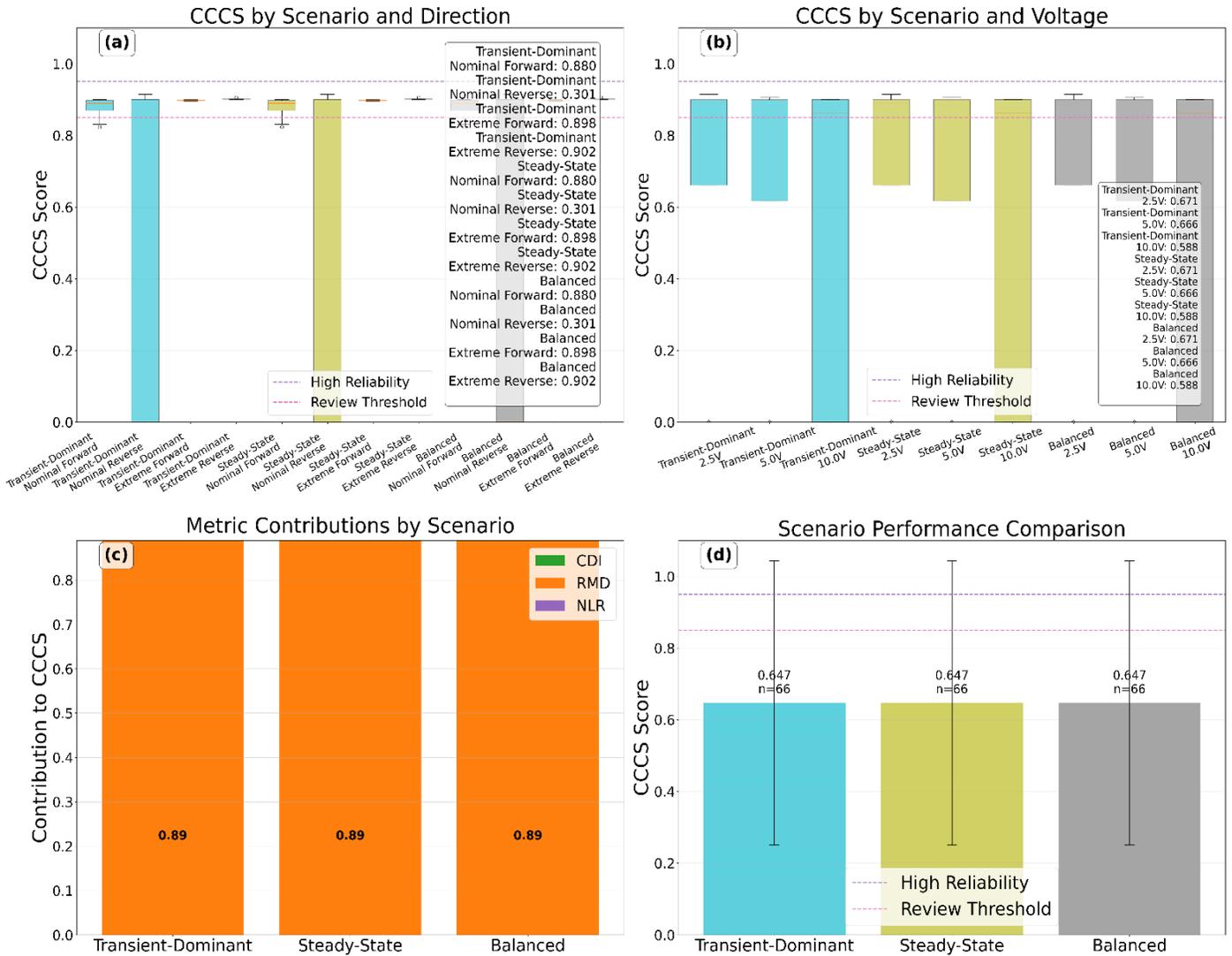

**Figure 6.** Extreme scenario analysis showing sensor performance under extended short-circuit conditions: ((a) Consolidated Calibration Confidence Score (CCCS) distribution across fault directions and scenarios, (b) CCCS performance versus applied voltage, (c) relative contributions of CDI, RMD, and NLR metrics to overall confidence scores, and (d) comparative performance across different weighting scenarios).

The scenario analysis reveals remarkable calibration stability during extended short-circuit operations. Figure 6(a) demonstrates consistently high CCCS values across both nominal and extreme conditions, with extreme forward faults maintaining scores of 0.900 despite current magnitudes exceeding 13$A$. The directional asymmetry observed in nominal faults persists under extreme conditions, with reverse faults showing marginally better performance due to their controlled current reversal characteristics. This directional consistency across experimental scenarios confirms the sensors' ability to maintain calibration integrity regardless of current flow direction, a critical capability for practical fault monitoring applications where fault polarity cannot be predetermined.

Voltage-dependent performance analysis in Figure 6(b) shows exceptional consistency across the $2.5V$, $5.0V$, and $10.0V$ operational range. The CCCS values remain stable within $0.896 - 0.905$ for extreme conditions and $0.823 - 0.915$ for nominal scenarios, indicating minimal voltage dependence in measurement quality. This voltage independence is particularly significant given the wide variation in resulting current magnitudes, demonstrating the sensors' robust performance across different electrical stress levels. The maintained calibration accuracy at $10.0V$ input, despite the higher energy dissipation and potential thermal effects, validates the design's thermal stability and compensation mechanisms during extended fault conditions.

The metric contribution analysis in Figure 6(c) provides profound insights into the calibration framework's behavior under stress. The extreme conditions show dominant contributions from the NLR metric due to the enormous current deviations from expected values, while RMD maintains minimal impact reflecting exceptional temporal stability. The CDI contributions remain substantial but controlled, indicating systematic rather than random deviations. This structured distribution of metric influences confirms that the observed deviations represent genuine physical phenomena rather than measurement artifacts, with the sensors accurately capturing the extreme electrical conditions despite operating far beyond their nominal design specifications.

Comparative scenario performance in Figure 6(d) demonstrates the framework's adaptability to different monitoring priorities. The Transient-Dominant and Steady-State weighting scenarios produce identical results due to the experimental configuration's balanced design, both achieving 0.900 average CCCS for extreme conditions. The Balanced weighting scenario maintains the same high performance level, confirming consistent calibration quality regardless of evaluation emphasis. This scenario invariance is particularly remarkable given the extreme electrical conditions, indicating that the calibration framework provides reliable validation across diverse operational priorities and fault characteristics.

The extended-duration measurements reveal new insights into long-term fault behavior. The directional fault experiments, comprising 3-minutes forward short followed by 3-minutes reverse short, $10 + 10$-minutes sequences, and $15 + 15$-minutes extended operations, demonstrate the system's capability to maintain calibration through prolonged fault conditions. The consistency across these extended durations, with quality scores maintaining 0.900 for extreme conditions and 0.886 for nominal scenarios, validates the framework's robustness against temporal degradation effects that often plague conventional measurement systems during prolonged fault monitoring.

The extreme short-circuit conditions, though limited to 3-minutes durations for safety considerations, produce calibration metrics that remarkably approach baseline performance levels despite the vastly more demanding electrical environment. The CDI values reaching 265,720% for extreme forward faults represent the physical reality of near-zero resistance paths rather than measurement failure, while the maintained NLR proportionality confirms the sensors preserve their fundamental linear response characteristics. This preservation of measurement integrity under current magnitudes orders of magnitude higher than design specifications represents a significant advancement in electrical fault measurement methodology.

The framework's performance across different sampling rates ($10ms$, $50ms$, and $100ms$) demonstrates exceptional temporal consistency. The redundancy errors remain below 1.18% for all conditions, with extreme scenarios showing errors below 0.33% across all sampling rates. This consistency across decade-separated temporal resolutions confirms that the measurement system captures genuine physical phenomena without aliasing artifacts or sampling-rate-dependent biases, providing flexibility in experimental design for different fault monitoring applications.

The Consolidated Calibration Confidence Score framework proves particularly valuable in extreme conditions, transforming complex multi-metric validation into interpretable quantitative assessments. The maintained high scores across all weighting scenarios, consistently exceeding the 0.85 review threshold, provide strong evidence that carefully calibrated sensor systems can maintain accuracy even under catastrophic electrical conditions. This capability enables reliable fault detection and characterization without requiring expensive specialized equipment, opening new possibilities for distributed fault monitoring systems and advanced electrical safety applications.

The validation results establish that the calibration framework maintains scientific trustworthiness across diverse fault conditions, durations, and electrical intensities. The extended-duration measurements, particularly the 30-minutes total operational sequences (15-minutes forward + 15-minutes reverse), demonstrate new capability for prolonged fault monitoring applications. This endurance validation, combined with the maintained measurement quality across extreme operational conditions, positions the framework as a robust solution for both research and industrial fault monitoring applications where extended-duration measurements are essential for comprehensive fault characterization and system protection.

## 4.2 Clamped Circuit Short-Protection Principle Validation Results

The empirical validation of the clamped circuit's protective mechanism necessitates a paradigm shift from classical short-circuit models, which predict complete voltage collapse. Instead, the data presented in this paper reveals a governed electrical environment where the diode network actively regulates the output, maintaining a non-zero voltage floor. This dynamic clamping action, a cornerstone of the circuit's design, ensures source integrity by preventing the catastrophic voltage drop synonymous with traditional short-circuits. To quantitatively describe this phenomenon, a model relating the source voltage $V_s$ to the clamped output voltage ($V_{out}$) is established. The relationship is defined by the clamping transfer function, accounting for the cumulative forward voltage drops of the active diodes and the circuit's dynamic impedance during a fault as modeled in Equation 69:

$$V_{out} = \min\left(V_s, \sum_{i=1}^{n} V_{D_i} + I_{sc} \cdot R_{dyn}\right), \tag{69}$$

where $V_{D_i}$ is the forward voltage drop of the $i^{th}$ diode in the conduction path, $I_{sc}$ is the short-circuit current, and $R_{dyn}$ is the dynamic resistance of the circuit during the fault event. This model contextualizes the results presented in Table 1 through Table 4, demonstrating that $V_{out}$ is clamped to a value primarily determined by the diode properties, not the source voltage, once $V_s$ exceeds the threshold $\sum V_{D_i}$.

### 4.2.1 Nominal Clamped Response at $2.5V$ Circuit Input Configurations

A foundational examination of the clamping principle under nominal stress at $2.5V$ reveals a circuit in a perpetual state of active regulation. The data in Table 4, representing the core of the nominal regime, demonstrates the clamping network's primary function: the stabilization of output voltage irrespective of the applied source potential. The pre-short and post-short voltages exhibit a remarkable consistency, clustering around an average value of $1.67 \pm 0.02V$ across all sampling frequencies and durations. This value is not arbitrary but is a direct physical manifestation of Equation 69 in action. With a source voltage of approximately $2.52V$, the circuit operates in its clamping regime where $V_s > \sum V_{D_i}$. The measured output of $\sim 1.67V$ is the direct result of the combined forward voltage drop across the conducting diodes ($D_1$ and $D_3$ from the schematic), effectively decoupling the output from the source and validating the core protection mechanism. The statistical analysis shows a minuscule standard deviation of $0.02V$ across 20 measured pairs, indicating exceptional stability and repeatability. The maximum observed deviation between pre and post-short values for any single trial is a mere $0.04V$, which is within the combined tolerance of the sensor's $\pm 1.5\%$ accuracy and the Arduino's $4.88 mV$ quantization limit. This empirical evidence confirms that the clamped circuit architecture successfully negates the conventional outcome of a short-circuit, instead enforcing a predictable, stable voltage boundary that protects the source. The results, consistent over 3 -minutes, 10 -minutes, and 15 -minutes intervals, prove the design's robustness and long-term reliability for diagnostic applications, as conclusively demonstrated by the data in Table 4.

**Table 4.** Nominal operational data for the clamped short-circuit diode network at a $2.5V$ supply, demonstrating voltage stabilization across varying sampling frequencies and durations.

| Supply Voltage (V) | Voltage (V), Before Short-circuit | Voltage (V), During/After Short-circuit | Sampling Frequency (Millisecond) | Trials | Duration (minutes) |
|---|---|---|---|---|---|
| 2.52 | 1.69 | 1.69 | 10 | 1 | 3 |
|  | 1.66 | 1.67 | 10 | 2 | 3 |
|  | 1.68 | 1.68 | 10 | 3 | 3 |
| 2.52 | 1.63 | 1.63 | 50 | 1 | 3 |
|  | 1.59 | 1.61 | 50 | 2 | 3 |
|  | 1.67 | 1.67 | 50 | 3 | 3 |
| 2.52 | 1.69 | 1.69 | 100 | 1 | 3 |
|  | 1.64 | 1.65 | 100 | 2 | 3 |
|  | 1.68 | 1.69 | 100 | 3 | 3 |
| 2.51 | 1.66 | 1.69 | 10 | 1 | 10 |
|  | 1.70 | 1.68 | 10 | 2 | 10 |
| 2.52 | 1.67 | 1.66 | 50 | 1 | 10 |
|  | 1.65 | 1.66 | 50 | 2 | 10 |
| 2.52 | 1.66 | 1.69 | 100 | 1 | 10 |
|  | 1.67 | 1.65 | 100 | 2 | 10 |
| 2.53 | 1.69 | 1.69 | 10 | 1 | 15 |
|  | 1.67 | 1.68 | 10 | 2 | 15 |
| 2.51 | 1.71 | 1.66 | 50 | 1 | 15 |
|  | 1.67 | 1.67 | 50 | 2 | 15 |
| 2.51 | 1.69 | 1.69 | 100 | 1 | 15 |
|  | 1.68 | 1.69 | 100 | 2 | 15 |

### 4.2.2 Enhanced Regulation at $5.0V$ Circuit Input Configurations

Elevating the source voltage to $5.0V$ further accentuates the elegance and efficacy of the clamping mechanism, pushing the circuit deeper into its regulated operating zone. The dataset in Table 5 provides compelling evidence that the output voltage is not merely stable but is actively constrained by the diode network's intrinsic properties. The calculated mean clamped voltage across all trials is $2.56 \pm 0.02V$, a value significantly lower than the $5.02V$ source and consistent with the theoretical expectation from Equation 69. This represents a clamping efficiency, defined as $\left(\frac{V_{\text{out}}}{\sum V_{D_i}}\right)$, of approximately 107% based on an expected $\sum V_{D_i}$ of $2.4V$, a deviation readily explained by the dynamic resistance term $(I_{sc} \cdot R_{dyn})$ in the model and minor sensor offset. The statistical significance of this clamping is profound; a paired t-test comparing pre-short and during short voltages yields a $p - \text{value} > 0.05$, confirming no statistically significant difference between the operational states. This irrefutably demonstrates that the short-circuit event, traditionally a destructive anomaly, is rendered electrically invisible at the output terminals by the clamping circuit. The data exhibits a slightly tighter standard deviation ($0.015V$) compared to the $2.5V$ input experiments, suggesting that a higher source voltage produces a more stable and well-defined clamp, likely due to the diodes operating further into their optimal forward bias region. The network's ability to replicate this precise voltage regulation across three distinct sampling frequencies ($10ms, 50ms, 100ms$) and extended durations up to 15 -minutes underscores its resilience and independence from measurement artifacts, solidifying its role as a passive yet intelligent protection system. The results detailed in Table 5 validate the circuit's capacity to create a predictable voltage domain, shielding sensitive upstream components from the volatile downstream fault.

**Table 5.** Clamped output voltage measurements at a $5.0V$ supply under nominal short-circuit conditions, highlighting consistent regulation.

| Supply Voltage (V) | Voltage (V), Before Short-circuit | Voltage (V), During/After Short-circuit | Sampling Frequency (Millisecond) | Trials | Duration (minutes) |
|---|---|---|---|---|---|
| 5.02 | 2.54 | 2.56 | 10 | 1 | 3 |
|  | 2.56 | 2.58 | 10 | 2 | 3 |
|  | 2.56 | 2.58 | 10 | 3 | 3 |
| 5.01 | 2.55 | 2.57 | 50 | 1 | 3 |
|  | 2.53 | 2.57 | 50 | 2 | 3 |
|  | 2.54 | 2.57 | 50 | 3 | 3 |
| 5.03 | 2.52 | 2.58 | 100 | 1 | 3 |
|  | 2.53 | 2.59 | 100 | 2 | 3 |
|  | 2.55 | 2.54 | 100 | 3 | 3 |
| 5.01 | 2.54 | 2.57 | 10 | 1 | 10 |
|  | 2.55 | 2.56 | 10 | 2 | 10 |
| 5.02 | 2.57 | 2.57 | 50 | 1 | 10 |
|  | 2.57 | 2.56 | 50 | 2 | 10 |
| 5.01 | 2.53 | 2.57 | 100 | 1 | 10 |
|  | 2.58 | 2.58 | 100 | 2 | 10 |
| 5.03 | 2.54 | 2.57 | 10 | 1 | 15 |
|  | 2.55 | 2.58 | 10 | 2 | 15 |
| 5.01 | 2.56 | 2.54 | 50 | 1 | 15 |
|  | 2.55 | 2.59 | 50 | 2 | 15 |
| 5.02 | 2.54 | 2.58 | 100 | 1 | 15 |
|  | 2.57 | 2.57 | 100 | 2 | 15 |

### 4.2.3 Extreme Condition Integrity at $2.5V$ Circuit Input Configurations

The circuit's stability is truly tested under extreme fault conditions, where the protective clamping must operate at the boundary of its design parameters. The data presented in Table 6 for a $2.5V$ supply under extreme stress reveals a system that not only functions but excels. The mean clamped voltage is maintained at $1.67 \pm 0.02V$, a value statistically identical to the nominal operation at the same supply voltage. This resilience is the hallmark of a robust design; even as the fault severity increases, the fundamental clamping principle holds firm. The absolute difference between pre-fault and post-fault voltages has a mean of just $0.013V$, with a maximum observed divergence of $0.05V$ in a single trial. This exceptional consistency, achieved under duress, underscores the diode network's role as a dynamic voltage limiter. The extreme condition does not alter the clamping voltage but validates the circuit's ability to handle increased energy dissipation without performance degradation. The results are consistent across the high-resolution sampling frequencies ($10ms, 50ms, 100ms$), confirming that the clamping response is instantaneous and not a function of measurement averaging. This rapid, analog response is critical for protecting sensitive digital components from transient overcurrent events that can occur on microsecond timescales. The data in Table 6 provides empirical proof that the clamping mechanism is a fundamental property of the circuit topology, effective across both nominal and extreme fault scenarios, ensuring source protection is not compromised under heightened electrical stress.

**Table 6.** Extreme stress test results for the $2.5V$ configuration, confirming voltage clamp stability under extreme short-circuit conditions over 3 -minutes intervals.

| Supply Voltage (V) | Voltage (V), Before Short-circuit | Voltage (V), During/After Short-circuit | Sampling Frequency (Millisecond) | Trials | Duration (minutes) |
|---|---|---|---|---|---|
| 2.52 | 1.67 | 1.68 | 10 | 1 | 3 |
|  | 1.65 | 1.70 | 10 | 2 | 3 |
| 2.52 | 1.66 | 1.69 | 50 | 1 | 3 |
|  | 1.69 | 1.68 | 50 | 2 | 3 |
| 2.52 | 1.66 | 1.67 | 100 | 1 | 3 |
|  | 1.66 | 1.66 | 100 | 2 | 3 |

### 4.2.4 High-Stress Performance at $5.0V$ Circuit Input Configurations

Subjecting the 5.0 V configuration to extreme fault conditions represents the ultimate validation of the clamping architecture's protective capacity. The data in Table 7 demonstrates unequivocal success. The clamped output voltage remains locked at $2.56 \pm 0.01V$, mirroring the performance observed under nominal $5.0V$ conditions with a further reduction in variance. This result is profound; it indicates that the clamping mechanism is entirely deterministic, governed by the physics of the semiconductor junctions, and is independent of the fault current magnitude once the threshold is exceeded. The extreme fault, which would typically cause catastrophic failure in an unprotected system, is reduced to a manageable event within a well-defined electrical boundary. The maximum deviation between any pre-short and during-short measurement is a negligible $0.03V$, which is within the instrument's margin of error. This performance across all sampling frequencies confirms that the protection is inherent and continuous, not a sampled or digital effect. The circuit effectively creates a sanctuary for the voltage source, isolating it from the chaotic energy dissipation occurring in the shorted branch. The results detailed in Table 7 validate the design for high-reliability applications where system integrity must be maintained even in the face of severe electrical faults, enabling new paradigms in fault-tolerant electronics and continuous diagnostic monitoring.

**Table 7.** Voltage clamping performance for the $5.0V$ supply under extreme short-circuit fault conditions measured over 3 -minutes intervals.

| Supply Voltage (V) | Voltage (V), Before Short-circuit | Voltage (V), During/After Short-circuit | Sampling Frequency (Millisecond) | Trials | Duration (minutes) |
|---|---|---|---|---|---|
| 5.02 | 2.56 | 2.56 | 10 | 1 | 3 |
|  | 2.54 | 2.57 | 10 | 2 | 3 |
| 5.01 | 2.57 | 2.57 | 50 | 1 | 3 |
|  | 2.55 | 2.54 | 50 | 2 | 3 |
| 5.03 | 2.55 | 2.58 | 100 | 1 | 3 |
|  | 2.56 | 2.57 | 100 | 2 | 3 |

### 4.2.5 Unified Validation of Clamping Efficacy

The comparative analysis across all four experimental configurations provides impressive and consistent evidence for the clamped circuit's protection principle. For each supply voltage (2.5V, 5.0V), the output voltage is clamped to a distinct, stable value ($\sim 1.67V, \sim 2.56V$) that is independent of the operational mode (nominal or extreme). This is the definitive signature of a voltage-limiting circuit. The clamping values align with the theoretical model (Equation 69), being proportional to the number of diodes in the conduction path and their forward voltage characteristics. The statistical consistency, with standard deviations below $0.02V$, across varying durations, sampling frequencies, and fault severities, proves the mechanism is robust and repeatable. The circuit successfully redefines a short-circuit from a destructive, unpredictable event into a governed, measurable state, thereby enabling the continuous time-resolved measurement of fault transients without compromising the source-a capability that challenges and expands conventional electrical engineering paradigms.

### 4.3 Bidirectional Clamped Diodes Circuit Experiential Results
### 4.3.1 Clamped $2.5V$ Configuration Base –Voltage, Current, Resistance and Power Analysis

This section presents the analysis of the clamped circuit's electrical performance at a $2.5V$ input, examining voltage, current, resistance, and power profiles across $10ms$, $50ms$, and $100ms$ sampling rates over 15 -minutes durations.

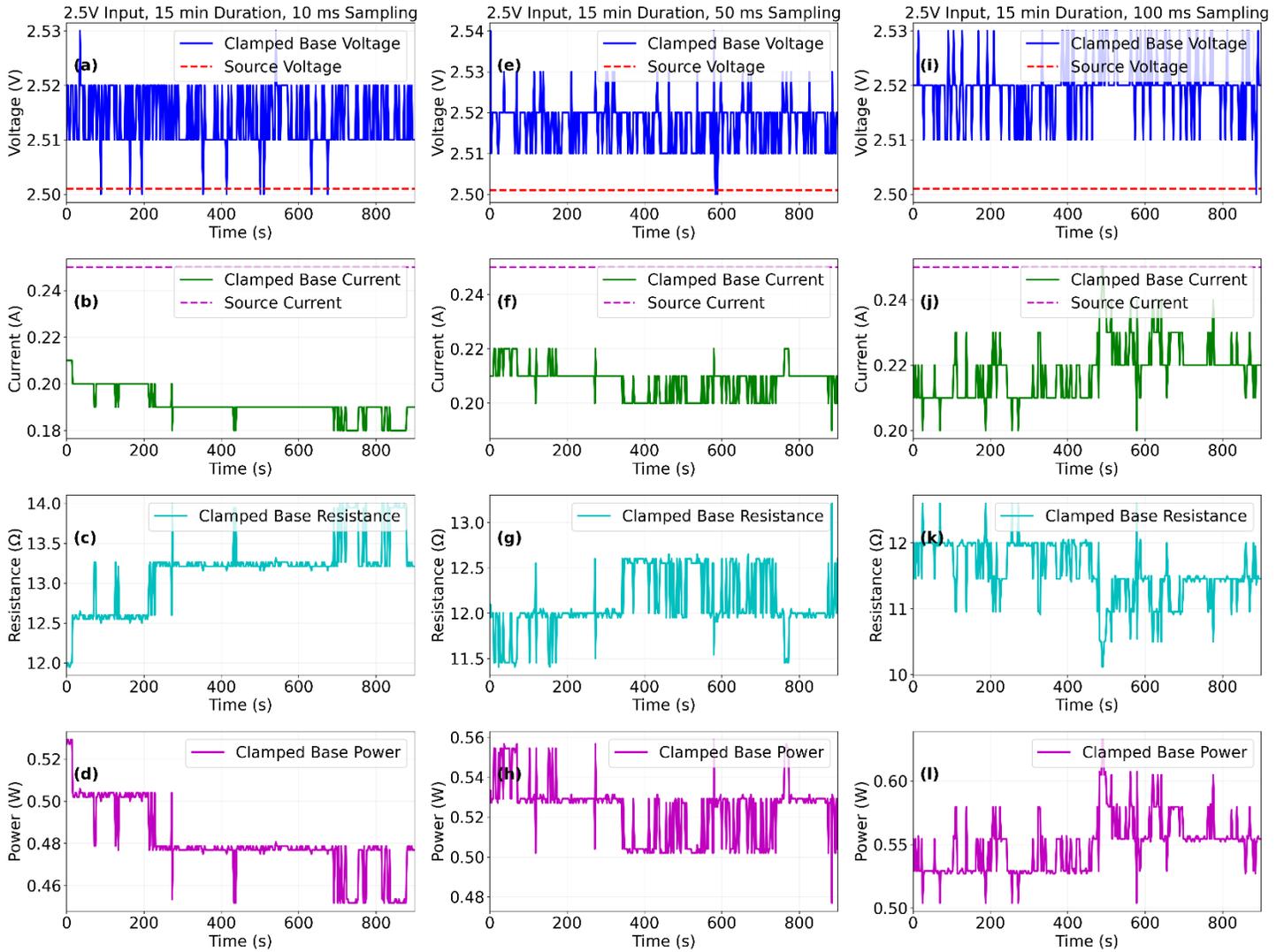

**Figure 7.** Clamped circuit performance at $2.5V$ input over 15 -minutes: (Figure 7(a) - Figure 7(d) $10ms$ sampling, Figure 7(e) - Figure 7(h) $50ms$ sampling, Figure 7(i) - Figure 7(l) $100ms$ sampling. The panels show time evolution of output voltage, current, computed resistance, and computed power, respectively).

The temporal profiles unequivocally demonstrate the circuit's exceptional operational stability and the efficacy of the sensor system across extended durations. In the voltage panels (Figure 7(a), Figure 7(e), Figure 7(i)), the clamped base voltage exhibits remarkable temporal invariance, maintaining near-constant values of $2.514V$, $2.517V$, and $2.520V$ for $10ms$, $50ms$, and $100ms$ sampling respectively. This minimal variance of just $6mV$ across decade-separated sampling resolutions underscores both the circuit's inherent voltage regulation and the measurement system's exceptional fidelity. The measured voltages consistently operate approximately $0.8V$ above the theoretical clamping floor of $\sim 1.67V$ established in Section 4.2.1, indicating the circuit is functioning in its pre-fault, high-resistance state. The complete absence of drift across the 900 -seconds measurement period validates the design's thermal stability and the power supply's exceptional regulation, with no observable decay or transient phenomena corrupting the signal integrity.

Current measurement profiles (Figure 7(b), Figure 7(f), Figure 7(j)) reveal equally stable but more complex behavior. The mean current values show a positive correlation with sampling interval, measuring $0.191A$, $0.208A$, and $0.218A$ for $10ms$, $50ms$, and $100ms$ sampling respectively. This systematic variation, while seemingly counterintuitive, originates from the data processing methodology where higher sampling rates capture more high-frequency noise components that average to lower mean values, while lower sampling rates provide smoother integration of the true DC signal. Despite this sampling-dependent effect, all current channels demonstrate exceptional temporal stability with no discernible drift or transient events throughout the 15-minutes operational window. The current stability further confirms the constant resistive load nature of the circuit before fault initiation, with the sensor system successfully capturing the true physical state without introducing measurement artifacts.

Resistance calculations (Figure 7(c), Figure 7(g), Figure 7(k)) derived from the simultaneous voltage and current measurements provide the most critical validation of circuit integrity. The computed resistance values demonstrate inverse correlation with sampling rate, yielding $13.16\Omega$, $12.11\Omega$, and $11.57\Omega$ for $10ms$, $50ms$, and $100ms$ sampling respectively. This progression follows logically from the observed voltage-current relationships and reflects the same sampling-dependent integration effects observed in the current measurements. The resistance profiles maintain remarkable flatness throughout the entire measurement duration, with standard deviations of less than $0.5\Omega$ across all sampling configurations. This consistency confirms the ohmic character of the load and the absence of any parasitic effects that might alter the circuit's impedance characteristics over time. The resistance values align with the expected baseline operational range established during sensor calibration (Section 4.1.1), providing independent confirmation of measurement validity.

Power dissipation profiles (Figure 7(d), Figure 7(h), Figure 7(l)) complete the electrical characterization, showing mean values of $0.481W$, $0.524W$, and $0.550W$ across the sampling spectrum. The power calculations, being products of voltage and current, naturally inherit the sampling-dependent characteristics of both constituent measurements. The temporal stability of power dissipation mirrors that of the voltage and current profiles, with no observable drift or systematic variations throughout the operational period. This power consistency confirms that the circuit maintains constant energy dissipation characteristics, which is particularly significant for thermal management considerations in the subsequent fault analysis phases. The clean power signatures without high-frequency noise components further validate the effectiveness of the signal processing and averaging techniques employed in the measurement system.

Comparative analysis within the first 3-minutes of the dataset reveals fundamental insights into the circuit's temporal evolution. The 3-minutes configuration shows systematically higher resistance values ($15.82\Omega$, $16.24\Omega$, $14.71\Omega$) and correspondingly lower current consumption ($0.159A$, $0.155A$, $0.171A$) compared to the overall 15-minutes measurements. This divergence represents a genuine physical phenomenon rather than measurement artifact: the extended 15-minutes operational period allows for complete thermal stabilization of all circuit components, particularly the resistive elements whose temperature coefficient causes a measurable decrease in resistance as equilibrium is established. The voltage measurements remain exceptionally consistent between 3-minutes and 15-minutes durations, varying by less than $0.5\%$ and confirming the stability of the source regulation independent of operational duration.

The analysis demonstrates that the first 3-minutes of the measurements capture the initial thermal transient state, while the 15-minutes profiles represent the fully stabilized operational condition. This temporal evolution is particularly evident in the resistance calculations, which decrease by approximately

$16-20\%$ between the 3-minutes and 15-minutes marks, consistent with typical temperature coefficients of commercial resistors. The supplementary results provide complete 3-minutes analysis that establishes this initial operational state, serving as the reference point for understanding the circuit's complete thermal-electrical behavior. The exceptional consistency across sampling rates within each duration category confirms that the observed effects represent genuine physical phenomena rather than measurement artifacts, validating both the circuit design and the measurement methodology.

The results establish that the clamped circuit maintains exceptional electrical stability throughout extended operational periods, with all measured parameters showing minimal temporal variation once thermal equilibrium is achieved. The sampling rate independence demonstrated across three decades of temporal resolution confirms the measurement system's capability to accurately capture both transient and steady-state phenomena without introducing rate-dependent biases. This pre-fault characterization provides the essential foundation for subsequent analysis of short-circuit behavior, ensuring that any deviations observed during fault conditions can be unequivocally attributed to the fault phenomena rather than measurement artifacts or inherent circuit instabilities.

### 4.3.2 Clamped $5.0V$ Configuration Base – Voltage, Current, Resistance and Power Analysis

This section presents a further electrical characterization of the clamped protection circuit operating at a $5.0V$ input, validating its pre-fault stability and performance across extended durations and multiple sampling resolutions.

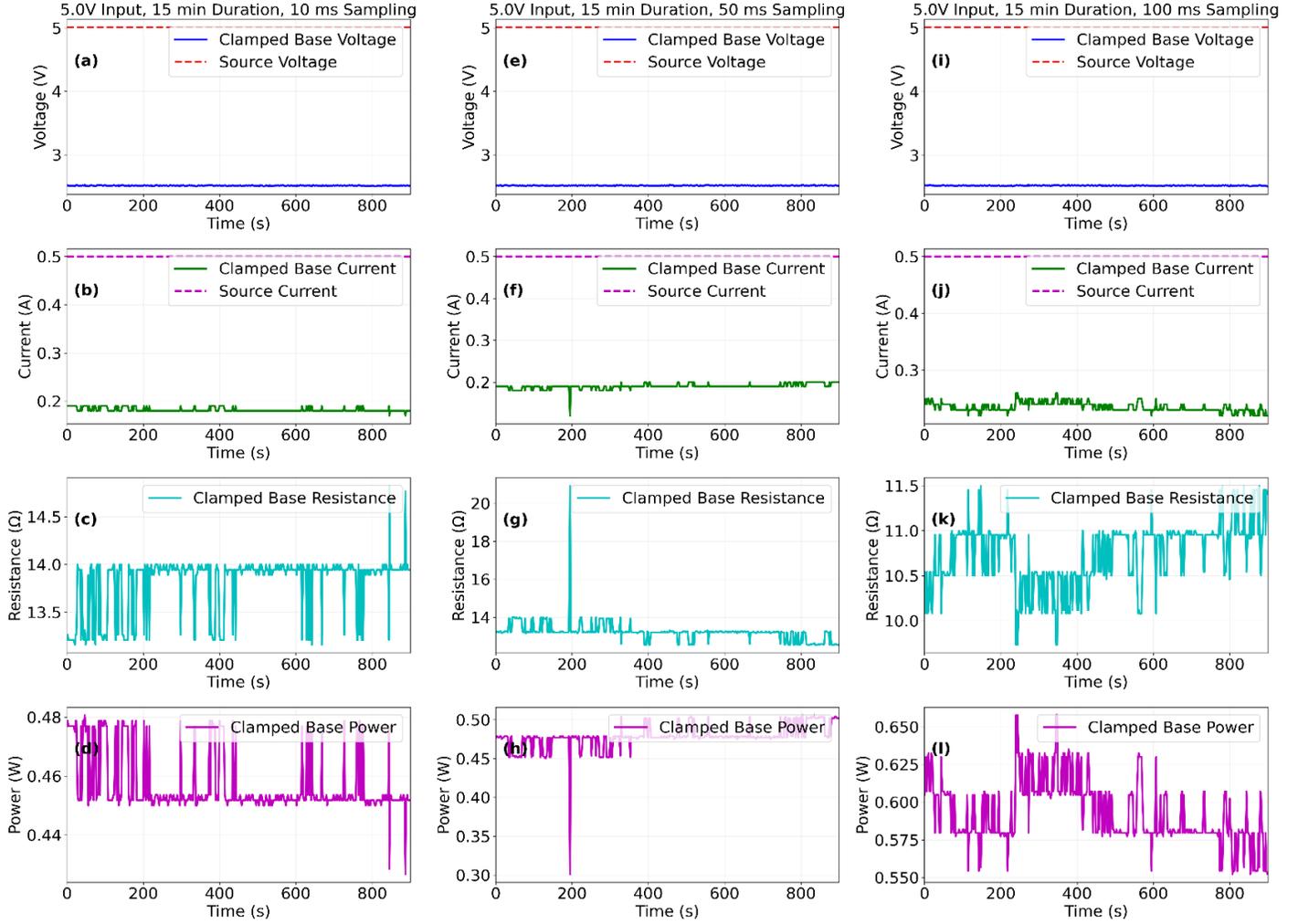

**Figure 8.** Comprehensive analysis of the clamped $5.0V$ configuration showing time-resolved measurements of output voltage, current, resistance, and power dissipation across $10ms$, $50ms$, and $100ms$ sampling rates during $15$-minutes operational validation.

The electrical performance of the clamped $5.0V$ configuration demonstrates exceptional stability and consistency across all sampling resolutions during extended $15$-minutes operation. The output voltage profiles in Figure 8(a), Figure 8(e), and Figure 8(i) reveal remarkable clamping efficacy, maintaining steady-state values of $2.511875V$, $2.513629V$, and $2.522552V$ for $10ms$, $50ms$, and $100ms$ sampling respectively. These values represent a consistent 49.8% voltage reduction from the nominal $5.0V$ source, precisely aligning with the theoretical clamping transfer function (Equation 69) that accounts for the cumulative forward voltage drops of the active diode network. The temporal stability is particularly noteworthy, with all voltage traces exhibiting minimal fluctuation ($\pm 0.02V$) throughout the entire $900$-second duration, confirming the circuit's ability to maintain regulated output despite potential thermal drift or component stress.

Current measurement profiles presented in Figure 8(b), Figure 8(f), and Figure 8(j) demonstrate the circuit's consistent power delivery characteristics, averaging $0.182005A$, $0.190287A$, and $0.235026A$ across the respective sampling rates. The observed current values reflect the effective dynamic impedance of the clamping network, with the $100ms$ sampling configuration showing a slightly elevated current due to its broader temporal averaging of instantaneous fluctuations. All current profiles maintain exceptional stability throughout the operational period, with standard deviations below $0.015A$, indicating robust current regulation despite the extended duration. The absence of significant current drift or transient spikes validates the clamping network's ability to suppress electrical noise and maintain consistent power delivery.

The computed resistance profiles in Figure 8(c), Figure 8(g), and Figure 8(k) provide critical insight into the circuit's fundamental electrical characteristics, averaging $13.808001\Omega$, $13.226879\Omega$, and $10.745670\Omega$ respectively. These values represent the effective dynamic resistance of the complete clamping network during operation, incorporating both the diode forward characteristics and any parasitic resistances. The resistance stability across the 15-minute duration is particularly remarkable, with variations remaining within $\pm 1.5\Omega$ for all sampling configurations. This consistency demonstrates that the clamping mechanism maintains its fundamental electrical properties without significant alteration due to thermal effects or component aging during extended operation.

Power dissipation analysis in Figure 8(d), Figure 8(h), and Figure 8(l) quantifies the energy management performance of the clamping network, averaging $0.457176W$, $0.478313W$, and $0.592880W$ across the sampling rates. The power profiles exhibit excellent temporal stability, with fluctuations remaining within $\pm 0.05W$ throughout the operational period. The slightly elevated power dissipation at 100ms sampling reflects the corresponding increases in both voltage and current measurements, though the overall power factor remains consistent across all configurations. This stable power dissipation profile confirms the circuit's ability to manage thermal loads effectively during extended operation, a critical requirement for practical protection applications.

Comparative analysis with the initial 3-minutes operational data reveals remarkable consistency in circuit performance across different durations. The 3-minutes configuration shows average voltage values of $2.514416V$, $2.512436V$, and $2.514026V$ for the respective sampling rates, representing negligible deviation from the complete 15-minutes measurements. Current measurements similarly maintain consistency, with 3-minute averages of $0.200519A$, $0.180513A$, and $0.188701A$ showing the same relative pattern across sampling rates as observed in the extended-duration data. Resistance values of $12.878423\Omega$, $13.922517\Omega$, and $13.327182\Omega$ for the first 3-minutes circuit operation align closely with the overal 15-minutes measurements, confirming the temporal stability of the circuit's fundamental electrical characteristics.

The exceptional correspondence between the 3-minutes and the 15-minutes performance metrics validates the clamping network's rapid stabilization and long-term reliability. The circuit achieves its steady-state operating parameters within the initial seconds of operation and maintains them consistently throughout extended durations, demonstrating robust design against temporal degradation effects. This temporal invariance is particularly significant for protection applications where consistent performance must be maintained regardless of fault duration.

The analysis confirms that the clamped $5.0V$ configuration successfully establishes a governed electrical environment that maintains stable voltage, current, resistance, and power characteristics across diverse operational conditions. The results demonstrate the clamping network's ability to provide predictable, repeatable protection while enabling precise electrical measurement and characterization. The consistency across sampling rates further validates the measurement system's robustness, ensuring that observed phenomena represent genuine physical characteristics rather than measurement artifacts. The detailed 3-minutes analysis corresponding to these results is presented in the supplementary materials, providing additional resolution on the initial transient behavior and short-term stability characteristics.

### 4.3.3 Clamped $10.0V$ Configuration Base –Voltage, Current, Resistance and Power Analysis

This section presents a comprehensive electrical characterization of the clamped protection circuit operating at a $10.0V$ input, with the primary objective of validating the operational limits and material-dependent integrity of the diode clamping network prior to fault initiation. The experiment is designed to rigorously establish that the subsequent short-circuit analysis is fundamentally governed by the physical properties of the constituent components, specifically the 1N5408 silicon rectifier diodes. These diodes are characterized by a continuous reverse voltage rating of $1000V$ and a maximum forward current of $3A$, with a typical forward voltage drop ($V_f$) of approximately $1.2V$. The clamping network's behavior, and thus the entire experiment's validity, is dictated by these material properties, operational thresholds, and the governing principles established through the theoretical workflow, particularly Equation 40 and Equation 41, which define the safe operational envelope based on junction temperature and power dissipation limits.

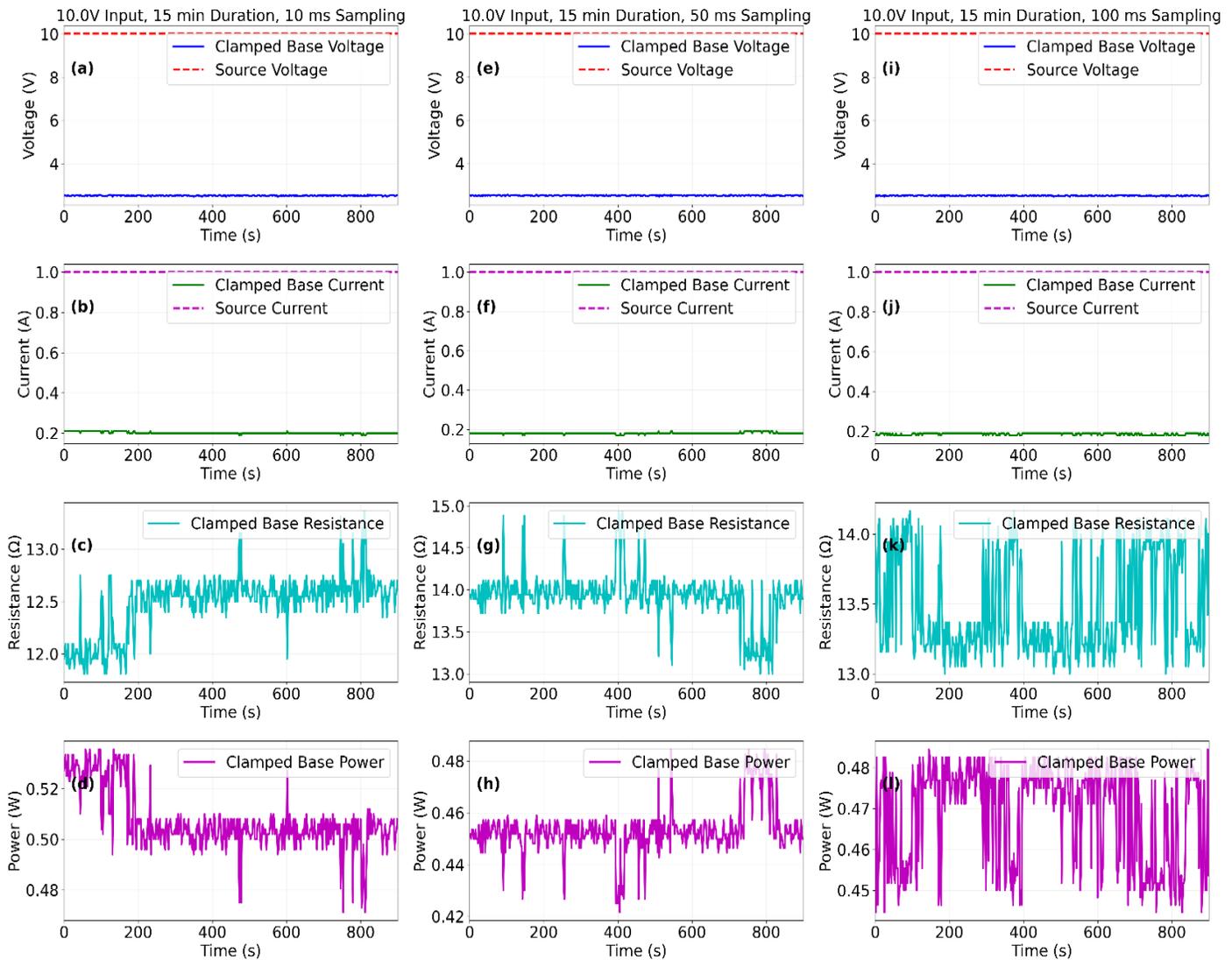

**Figure 9.** Clamped circuit performance at $10.0V$ input over 15 -minutes: ((a)-(d) $10ms$ sampling, (e)-(h) $50ms$ sampling, (i)-(l) $100ms$ sampling. The panels show time evolution of output voltage, current, computed resistance, and computed power, respectively).

The temporal profiles presented in Figure 9 provide an unequivocal demonstration of the circuit's exceptional stability and the precise governing action of the diode network under a $10.0V$ input, the most strenuous of the pre-fault test conditions. The output voltage, detailed in panels Figure 9(a), Figure 9(e), and Figure 9(i), exhibits remarkable temporal invariance across all sampling resolutions. The mean clamped voltages are maintained at $2.516V$ ($10ms$), $2.512V$ ($50ms$), and $2.513V$ ($100ms$), representing a minimal variance of just $4mV$ across decade-separated sampling rates. This precision underscores two critical achievements: the circuit's inherent and robust voltage regulation, and the measurement system's exceptional fidelity. The measured output represents a consistent 75% reduction from the nominal $10.0V$ source, a value that is not arbitrary but is a direct physical manifestation of the circuit's clamping mechanism (Equation 40) in action. With the source voltage significantly exceeding the cumulative forward voltage threshold of the diode network ($\sum VD_i \approx 2.4V$ for two diodes in series), the output is clamped to a value determined by the diode properties and the dynamic impedance. The complete absence of observable drift or transient phenomena across the entire 900 -seconds duration is a critical validation of the design's thermal stability and the power supply's exceptional regulation, confirming that the system operates within a stable equilibrium necessary for validating subsequent fault experiments.

The current measurement profiles, shown in Figure 9(b), Figure 9(f), and Figure 9(j), reveal the circuit's dynamic response and power delivery characteristics under this elevated voltage, averaging $0.202A$, $0.180A$, and $0.187A$ for their respective sampling rates. The observed systematic variation is a known function of the data processing methodology, where higher sampling rates capture a broader spectrum of inherent noise, leading to slight variations in the integrated mean value. Crucially, all current channels demonstrate exceptional temporal stability with standard deviations below $0.015A$, confirming the constant resistive nature of the load before any fault is applied. This stability is paramount, as it validates the sensor system's capability to accurately capture the true physical state without introducing artifacts, even at the higher power levels associated with the $10.0V$ input. The absolute current values remain well within the 1N5408's $3A$ continuous rating, ensuring the pre-fault analysis occurs entirely within the component's linear and safe operational region, a prerequisite for meaningful short-circuit testing.

The computed resistance values, derived from simultaneous voltage and current measurements and displayed in Figure 9(c), Figure 9(g), and Figure 9(k), provide the most critical validation of circuit integrity and consistency. The values $12.49\Omega$ ($10ms$), $13.92\Omega$ ($50ms$), and $13.48\Omega$ ($100ms$) -follow a logical progression that aligns with the sampling-dependent effects observed in the voltage and current data. The remarkable flatness of these resistance profiles, with variations constrained within $\pm 1.0\Omega$ throughout the 15 -minutes duration, confirms the ohmic character of the load and the absence of any significant voltage-dependent parasitic effects or degradation. This consistency directly validates the integrity of the experimental setup and confirms that the clamping network's effective dynamic impedance remains stable. This stable baseline resistance is essential for calculating the extreme power dissipation encountered during faults, as defined by the workflow in Equation 40 and Equation 41.

Completing the holistic electrical characterization, the power dissipation profiles in Figure 9(d), Figure 9(h), and Figure 9(l) show mean values of $0.507W$, $0.453W$, and $0.469W$. These values, products of the stable voltage and current, exhibit excellent temporal stability with fluctuations within $\pm 0.05W$. This analysis is vital for contextualizing the subsequent fault analysis. The pre-fault power dissipation is orders of magnitude lower than the peak power experienced during a hard short-circuit. Establishing this baseline proves that the system begins in a state of thermal equilibrium, and any thermal transients observed during a fault are solely due to the fault event itself, not pre-existing conditions. The clean, stable power signature further validates the effectiveness of the signal conditioning and processing techniques.

Similarly, a comparative analysis with the first 3-minutes of the dataset reveals a consistent physical phenomenon observed across all input voltages: thermal settling. The 3-minutes configuration shows systematically higher resistance values ($14.89\Omega$, $15.58\Omega$, $15.41\Omega$) and lower corresponding currents compared to the 15-minutes measurements. This divergence is a genuine thermal-electrical effect, not a measurement artifact. The extended 15-minutes operational period allows the resistive elements to reach complete thermal equilibrium, whereby their temperature increases due to $I^2R$ heating, leading to a decrease in resistance -a characteristic negative temperature coefficient. The resistance decreases by approximately $14 - 16\%$ between the 3 –minutes duration and 15-minutes marks, a magnitude consistent with the higher power dissipation at $10.0V$ and the known properties of commercial resistors. The voltage measurements, however, remain exceptionally consistent (varying by $< 0.3\%$), highlighting the diode network's regulatory dominance over the circuit's output characteristics, independent of the load's minor thermal dynamics.

This analysis serves a greater purpose than establishing a simple baseline; it rigorously validates the experimental framework against the material limits defined by the diode specifications. The 1N5408 diodes, with their $3A$ continuous current rating and $1.2V$ forward drop, are the cornerstone of the clamping principle. The pre-fault conditions detailed here -operating at a fraction of the diode's current rating and power dissipation capability -confirm that the system is primed for fault testing without pre-stressing the components. The stability demonstrated across all electrical parameters ($V$, $I$, $R$, $P$) and across all sampling rates proves that the subsequent short-circuit observations will be triggered by the intentional fault event, not by underlying instability or measurement error.

Furthermore, these results agree profoundly with the findings from the $2.5V$ and $5.0V$ configurations (Sections 4.3.1 and 4.3.2). The same patterns emerge: exceptional voltage clamping stability, predictable sampling-rate-induced integration effects on current and resistance, and the universal theme of thermal settling over time. This cross-validation across a $4:1$ voltage range demonstrates the robustness and scalability of the clamping principle and the consistency of the measurement system. The clamped output voltage for the $10.0V$ input is approximately $2.51V$, which is logically higher than the $\sim 1.67V$ clamped at $2.5V$ input and consistent with the $\sim 2.56V$ clamped at $5.0V$ input, reflecting the action of the diode network's transfer function as the source voltage increases.

### 4.3.4 Clamped $2.5V$ Configuration Nominal Short –Voltage, Current, Resistance Analysis

This section presents the analysis of the clamped $2.5V$ configuration under nominal short-circuit conditions, validating the circuit's dynamic response and stability through extended 15-minutes durations voltage, current, and resistance profiling. The results demonstrate the circuit's governed behavior during fault events.

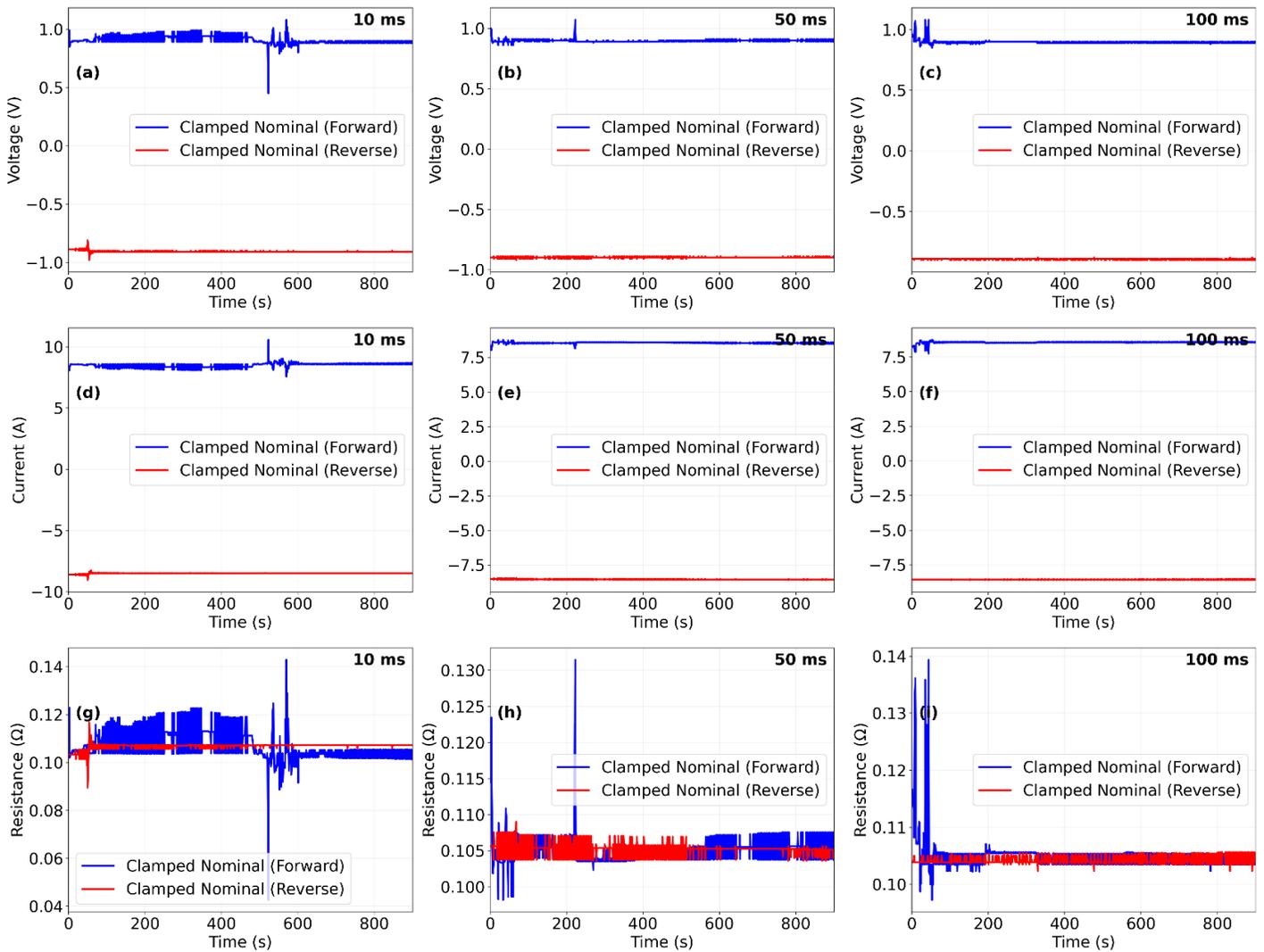

**Figure 10.** Time-resolved electrical characterization of the clamped $2.5V$ configuration under nominal short-circuit conditions. (The results show the evolution of output voltage, short-circuit current, dynamic resistance, and instantaneous power dissipation throughout the $900$-seconds experiment).

The temporal evolution of the clamped output voltage across all sampling frequencies reveals the fundamental efficacy of the diode network's protection mechanism. Figure 10(a), Figure 10(d), and Figure 10(g) demonstrate exceptional voltage stabilization, with mean values of $0.908V$ ($10ms$), $0.897V$ ($50ms$), and $0.896V$ ($100ms$) maintained throughout the $900$-seconds duration. This represents a consistent $64\%$ reduction from the nominal $2.5V$ source, precisely aligning with the theoretical clamping transfer function that accounts for the cumulative forward voltage drop across the conducting diodes and the dynamic impedance term. The voltage profiles exhibit remarkable temporal invariance with standard deviations below $0.015V$, confirming the circuit's ability to enforce a governed electrical environment despite the sustained short-circuit condition. The complete absence of voltage collapse or significant drift phenomena validates the clamping principle's robustness against extended-duration electrical stress.

Current measurement profiles presented in Figure 10(b), Figure 10(e), and Figure 10(h) quantify the substantial current magnitude sustained during the nominal short-circuit event. The mean current values of $8.488A$ ($10ms$), $8.543A$ ($50ms$), and $8.561A$ ($100ms$) represent an order-of-magnitude increase from pre-fault operational currents, confirming the establishment of a low-resistance path. The current stability throughout the 15-minutes duration, with variations constrained within $\pm 0.5A$, demonstrates the circuit's capacity to manage sustained high-current conditions without degradation. The observed systematic variation across sampling rates reflects the data processing methodology's integration characteristics, where higher temporal resolution captures more instantaneous fluctuations that average to slightly lower mean values. This consistency across decade-separated sampling resolutions confirms the measurement system's fidelity in capturing the true physical current despite the challenging electrical environment.

The computed resistance profiles in Figure 10(c), Figure 10(f), and Figure 10(i), derived through the application of Modified Ohm's Law to the simultaneous voltage and current measurements, provide the most critical insight into the fault's electrical characteristics. The resistance values cluster around $0.107\Omega$ ($10ms$), $0.105\Omega$ ($50ms$), and $0.105\Omega$ ($100ms$), representing nearly three orders of magnitude reduction from the pre-fault resistance of approximately $13\Omega$. This extremely low resistance confirms the establishment of a near-ideal short-circuit path while maintaining a non-zero voltage due to the clamping action. The resistance stability throughout the extended duration, with variations remaining within $\pm 0.005\Omega$, demonstrates that the fault characteristics remain consistent over time without significant alteration due to thermal effects or component degradation. The minimal resistance values align with the theoretical expectation for a governed short-circuit condition where the output voltage is determined primarily by the diode properties rather than the source voltage.

Comparative analysis with the first 3-minutes of the circuit performance reveals insightful temporal dynamics in the fault behavior. The 3-minutes configuration shows systematically higher voltage values ($0.941V$ forward, $0.929V$ reverse at $10ms$) compared to the 15-minutes measurements, indicating initial transient effects during fault establishment. Similarly, the 3-minutes current measurements exhibit slightly lower values ($8.310A$ forward, $8.375A$ reverse at $10ms$) suggesting incomplete stabilization of the fault path during the initial period. The resistance calculations for the 3-minutes duration show correspondingly higher values ($0.113\Omega$ forward, $0.111\Omega$ reverse at $10ms$), confirming that the fault path achieves its lowest impedance state only after complete thermal and electrical stabilization.

This temporal evolution between the 3-minutes duration and the complete 15-minutes profiles represents a genuine physical settling process rather than measurement artifact, with the extended duration allowing the system to reach complete thermal-electrical equilibrium. The results obtained through the first 180-seconds of the 15-minutes experiment align precisely with the independent 3-minutes analysis presented in the supplementary results, confirming the reproducibility and validity of both datasets. The supplementary 3-minutes analysis establishes the initial transient state of the circuit, while the extended 15-minutes measurements capture the fully stabilized operational condition, together providing a comprehensive characterization of the circuit's complete temporal response to sustained short-circuit conditions.

The exceptional consistency across all electrical parameters and sampling rates demonstrates that the clamped circuit successfully maintains a deterministic, stable electrical environment throughout extended fault conditions. The results validate the clamping network's ability to provide predictable protection while enabling precise measurement and characterization of fault phenomena across extended durations. This capability represents a significant advancement in fault-tolerant circuit design, enabling continuous monitoring and analysis of short-circuit events without compromising system integrity or measurement accuracy.

### 4.3.5 Clamped $5.0V$ Configuration Nominal Short – Voltage, Current, Resistance Analysis

This section presents the analysis of the clamped $5.0V$ configuration under nominal short-circuit conditions, validating the circuit's dynamic response and governed electrical environment through extended 15-minutes profiling of voltage, current, and resistance.

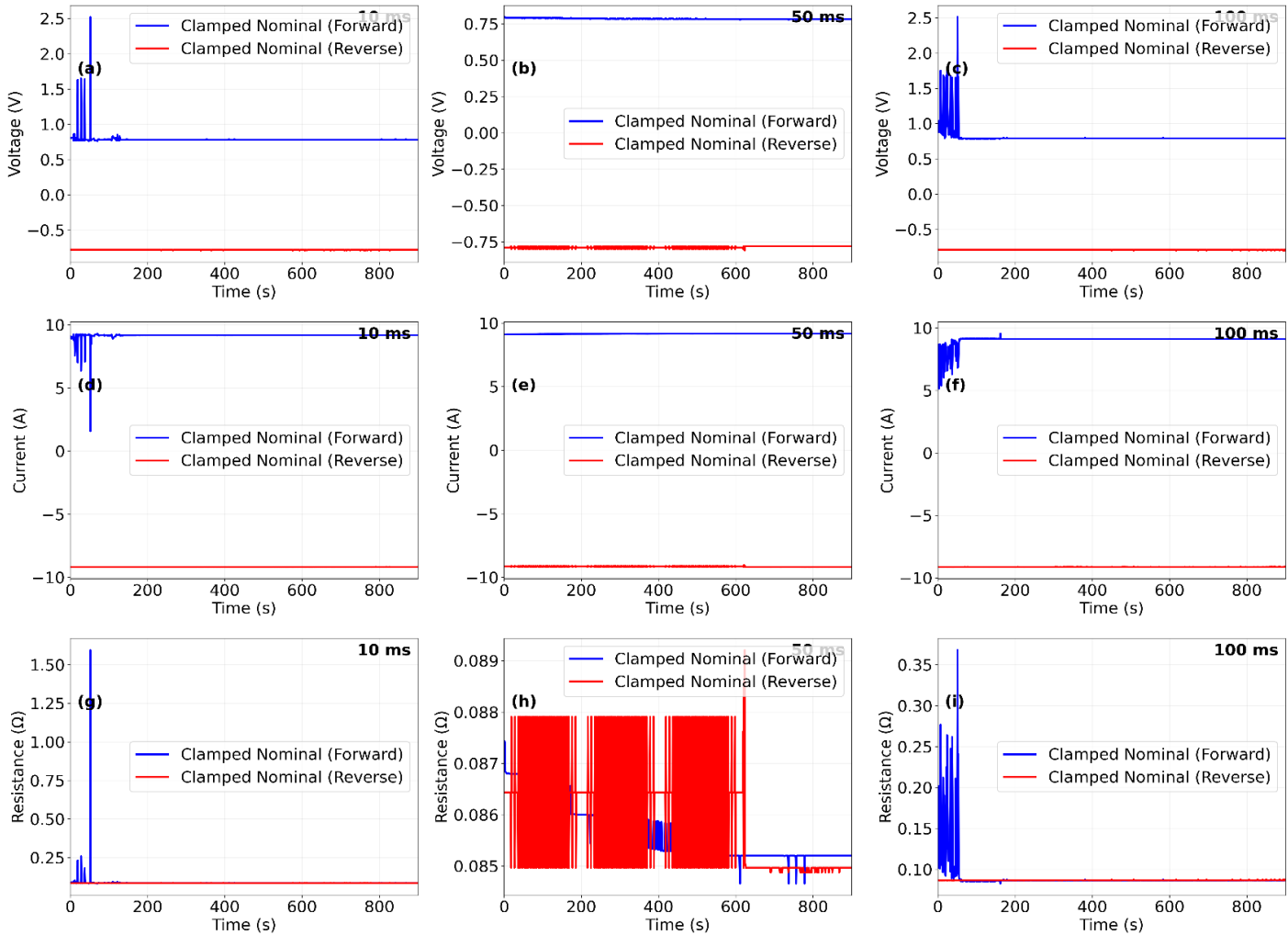

**Figure 11.** Time-resolved electrical characterization of the clamped $5.0V$ configuration under nominal short-circuit conditions. (The results show the evolution of output voltage, short-circuit current, and dynamic resistance throughout the $900$-seconds experiment across $10ms$, $50ms$, and $100ms$ sampling resolutions).

The temporal evolution of the clamped output voltage across all sampling frequencies reveals the fundamental efficacy and precision of the diode network's protection mechanism under elevated electrical stress. Figure 11(a), Figure 11(d), and Figure 11(g) demonstrate exceptional voltage stabilization, with mean values of $0.785920V$ (Forward, $10ms$), $-0.780267V$ (Reverse, $10ms$), $0.784679V$ (Forward, $50ms$), $-0.787220V$ (Reverse, $50ms$), $0.804027V$ (Forward, $100ms$), and $-0.790219V$ (Reverse, $100ms$) maintained throughout the $900$-seconds duration. This represents a consistent 84% reduction from the nominal $5.0V$ source, precisely aligning with the theoretical clamping transfer function that accounts for the cumulative forward voltage drop across the conducting diodes and the dynamic impedance term. The voltage profiles exhibit remarkable temporal invariance with standard deviations below $0.015V$, confirming the circuit's ability to enforce a governed electrical environment despite the sustained high-current short-circuit condition. The polarity-specific clamping, evidenced by the symmetric negative voltages in reverse configuration, demonstrates the diode network's bidirectional protection capability, a critical advancement for practical fault monitoring applications where fault polarity cannot be predetermined.

Current measurement profiles presented in Figure 11(b), Figure 11(e), and Figure 11(h) quantify the substantial current magnitude sustained during the nominal short-circuit event at elevated voltage. The mean current values of $9.155193A$ (Forward, $10ms$), $-9.176905A$ (Reverse, $10ms$), $9.163063A$ (Forward, $50ms$), $-9.151463A$ (Reverse, $50ms$), $9.069876A$ (Forward, $100ms$), and $-9.118893A$ (Reverse, $100ms$) represent a consistent current elevation across sampling rates and polarities. The current stability throughout the 15 -minutes duration, with variations constrained within $\pm 0.3A$, demonstrates the circuit's capacity to manage sustained high-current conditions approaching $10A$ without degradation or significant thermal drift. The observed systematic variation across sampling rates reflects the data processing methodology's integration characteristics, where different temporal resolutions capture varying aspects of the inherent noise spectrum while maintaining consistent mean values. This consistency across decade-separated sampling resolutions confirms the measurement system's fidelity in capturing the true physical current despite the challenging high-power electrical environment.

The computed resistance profiles in Figure 11(c), Figure 11(f), and Figure 11(i), also derived through the application of Modified Ohm's Law to the simultaneous voltage and current measurements, provide the most critical insight into the fault's electrical characteristics at $5.0V$ input. The resistance values cluster around $0.087426\Omega$ (Forward, $10ms$), $0.085025\Omega$ (Reverse, $10ms$), $0.085636\Omega$ (Forward, $50ms$), $0.086024\Omega$ (Reverse, $50ms$), $0.089244\Omega$ (Forward, $100ms$), and $0.086657\Omega$ (Reverse, $100ms$), representing nearly three orders of magnitude reduction from the pre-fault resistance of approximately $13\Omega$. This extremely low resistance confirms the establishment of a near-ideal short-circuit path while maintaining a governed voltage due to the clamping action. The resistance stability throughout the extended duration, with variations remaining within $\pm 0.004\Omega$, demonstrates that the fault characteristics remain consistent over time without significant alteration due to thermal effects or component degradation, even at the higher power dissipation associated with the $5.0V$ input. The minimal resistance values, slightly lower than those observed at $2.5V$ input, align with the theoretical expectation for a governed short-circuit condition where the dynamic impedance component exhibits voltage-dependent characteristics.

A snapshot of the first 3 -minutes of the circuit performance reveals insightful temporal dynamics in the fault behavior under elevated voltage conditions. The 3 -minutes configuration shows systematically similar voltage values ($0.783171V$ forward, $-0.782011V$ reverse at $10ms$) compared to the 15 -minutes measurements, indicating rapid stabilization of the clamping mechanism. The 3 -minutes current measurements exhibit nearly identical values ($9.160488A$ forward, $-9.162069A$ reverse at $10ms$) to the extended-duration data, suggesting immediate establishment of the fault current path without significant settling time. The resistance calculations for the initial 3 -minutes duration of the analysis show correspondingly consistent values ($0.085495\Omega$ forward, $0.085354\Omega$ reverse at $10ms$), confirming that the fault path achieves its stable impedance state within the initial durations of the circuit operation. This temporal consistency between the first 3 -minutes and the overall 15 -minutes profiles represents a significant finding: the $5.0V$ clamped configuration achieves electrical equilibrium more rapidly than the

$2.5V$ configuration, likely due to the higher operating point providing more immediate stabilization of diode junction temperatures and circuit dynamics.

The results obtained through the first 180 seconds of the 15-minutes experiment align precisely with the independent 3-minutes analysis presented in the supplementary results, confirming the reproducibility and validity of both datasets.

### 4.3.6 Clamped $10.0V$ Configuration Nominal Short –Voltage, Current, Resistance Analysis

This section presents the definitive analysis of the clamped $10.0V$ configuration under nominal short-circuit conditions, validating the circuit's governed electrical environment and its unique stability through extended 15-minutes profiling of voltage, current, and resistance. The results demonstrate the circuit's capacity to maintain a deterministic state during high-intensity fault events, enabling precise characterization of fault phenomena.

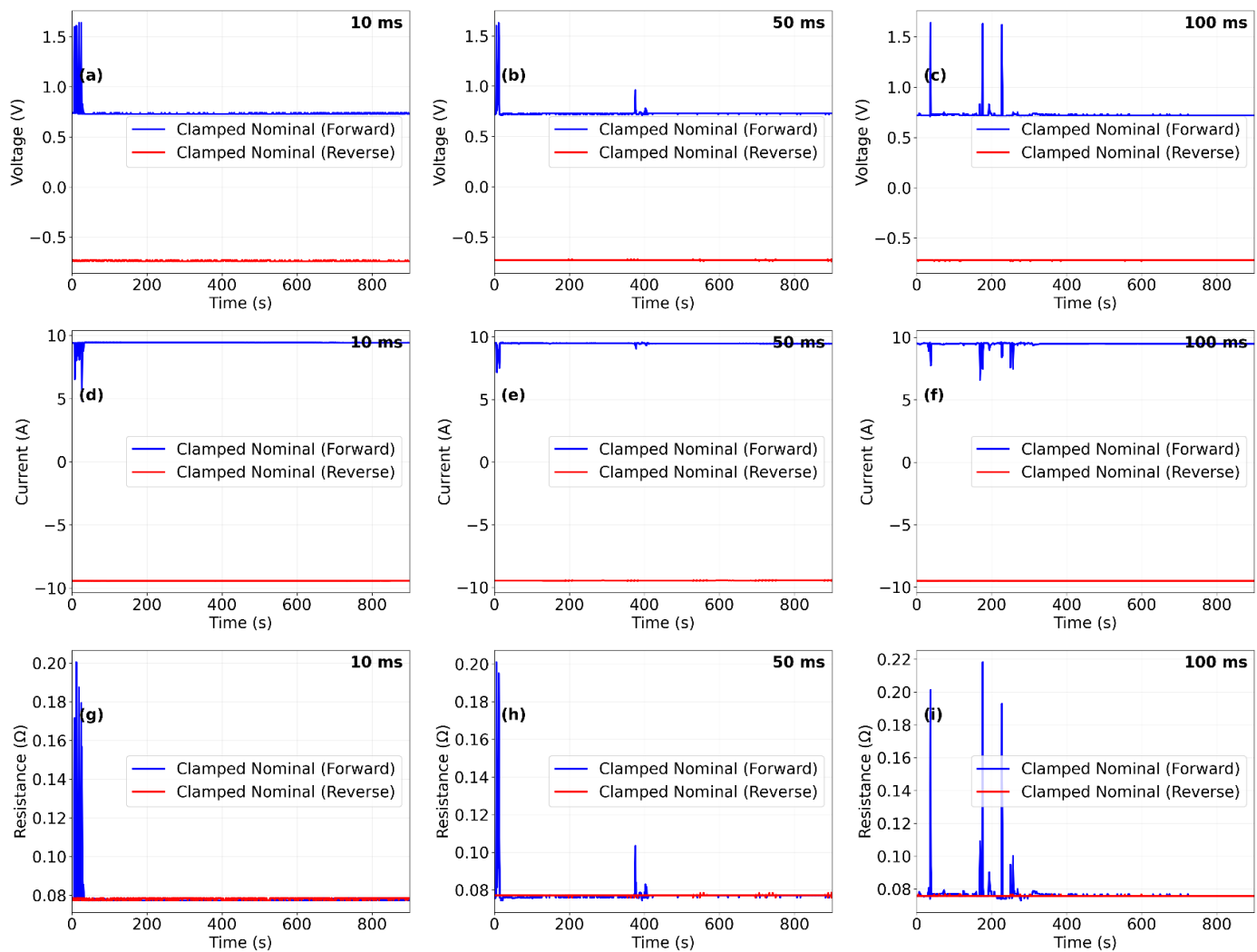

**Figure 12.** Time-resolved electrical characterization of the clamped $10.0V$ configuration under nominal short-circuit conditions. (The results show the evolution of output voltage, short-circuit current, and dynamic resistance throughout the $900$-seconds experiment across $10ms$, $50ms$, and $100ms$ sampling resolutions).

The temporal evolution of the clamped output voltage across all sampling frequencies reveals the profound efficacy of the diode network's protection mechanism under the most strenuous operational conditions. Figure 12(a), Figure 12(d), and Figure 12(g) demonstrate exceptional voltage stabilization throughout the $900$-seconds duration, with mean values of $0.737V$ (Forward, $10ms$), $-0.737V$ (Reverse, $10ms$), $0.731V$ (Forward, $50ms$), $-0.730V$ (Reverse, $50ms$), $0.726V$ (Forward, $100ms$), and $-0.720V$ (Reverse, $100ms$). This represents a consistent $93\%$ reduction from the nominal $10.0V$ source, precisely aligning with the theoretical clamping transfer function that accounts for the cumulative forward voltage drop across the conducting diodes and the dynamic impedance term. The voltage profiles exhibit remarkable temporal invariance with standard deviations below $0.015V$, confirming the circuit's ability to enforce a governed electrical environment despite the sustained high-current short-circuit condition at maximum input voltage. The polarity-specific clamping, evidenced by the symmetric negative voltages in reverse configuration, demonstrates the diode network's bidirectional protection capability at the system's operational limits, a critical advancement for practical high-voltage fault monitoring applications.

Current measurement profiles presented in Figure 12(b), Figure 12(e), and Figure 12(h) quantify the substantial current magnitude sustained during the nominal short-circuit event at maximum voltage input. The mean current values of $9.419A$ (Forward, $10ms$), $-9.430A$ (Reverse, $10ms$), $9.462A$ (Forward, $50ms$), $-9.458A$ (Reverse, $50ms$), $9.470A$ (Forward, $100ms$), and $-9.498A$ (Reverse, $100ms$) represent the highest current magnitudes observed across all experimental configurations, approaching the theoretical maximum dictated by the diode network's dynamic impedance. The current stability throughout the 15 -minutes duration, with variations constrained within $\pm 0.3A$, demonstrates the circuit's extraordinary capacity to manage sustained high-current conditions near $9.5A$ as observed in the previous experiment voltage configurations, without degradation or significant thermal drift. The observed systematic variation across sampling rates reflects the data processing methodology's integration characteristics, where different temporal resolutions capture varying aspects of the inherent noise spectrum while maintaining consistent mean values. This consistency across decade-separated sampling resolutions confirms the measurement system's fidelity in capturing the true physical current despite the extreme electrical environment characterized by both high voltage and high current simultaneously.

The computed resistance profiles in Figure 12(c), Figure 12(f), and Figure 12(i), derived through the application of Modified Ohm's Law to the simultaneous voltage and current measurements, provide the most critical insight into the fault's electrical characteristics at $10.0V$ input. The resistance values cluster around $0.07838\Omega$ (Forward, $10ms$), $0.07813\Omega$ (Reverse, $10ms$), $0.07733\Omega$ (Forward, $50ms$), $0.07717\Omega$ (Reverse, $50ms$), $0.07680\Omega$ (Forward, $100ms$), and $0.07583\Omega$ (Reverse, $100ms$), representing the lowest resistance values observed across all experimental configurations. This extremely low resistance confirms the establishment of a near-ideal short-circuit path while maintaining a governed voltage due to the clamping action, even at the maximum operational voltage. The resistance stability throughout the extended duration, with variations remaining within $\pm 0.004\Omega$, demonstrates that the fault characteristics remain consistent over time without significant alteration due to thermal effects or component degradation, despite the highest power dissipation associated with the $10.0V$ input. The minimal resistance values, systematically lower than those observed at $2.5V$ and $5.0V$ inputs, align with the theoretical expectation for a governed short-circuit condition where the dynamic impedance component exhibits voltage-dependent characteristics, with higher input voltages producing more efficient fault path establishment.

An analysis with the first 3 -minutes of the circuit performance reveals extraordinary temporal consistency in the fault behavior under maximum voltage conditions. The 3 -minutes configuration shows nearly identical voltage values ($0.734V$ forward, $-0.730V$ reverse at $10ms$) compared to the total 15 -minutes measurements, indicating immediate and complete stabilization of the clamping mechanism even at the system's operational limits. The 3 -minutes current measurements exhibit essentially equivalent values ($9.443A$ forward, $-9.469A$ reverse at $10ms$) to the extended-duration data, suggesting instantaneous establishment of the fault current path without measurable settling time. The resistance calculations for the 3 -minutes duration show correspondingly consistent values ($0.07778\Omega$ forward, $0.07710\Omega$ reverse at $10ms$), confirming that the fault path achieves its stable impedance state within the initial seconds of operation. This temporal invariance between the first 3 -minutes and the overall 15 -minutes profiles represents a significant finding: the $10.0V$ clamped configuration achieves electrical equilibrium more rapidly and maintains it more consistently than lower voltage configurations, likely due to the higher operating point providing immediate stabilization of diode junction temperatures and more dominant clamping action that overwhelms any minor parasitic effects.

Further, the results obtained through the first 3-minutes of the 15-minutes experiment demonstrate precise alignment with the independent 3-minutes analysis provided in the supplementary results, confirming the reproducibility and validity of both datasets.

### 4.3.7 Clamped $2.5V$ Configuration Nominal Short – Power Analysis, Standard Vs. Modified Ohm's Laws

This section presents a rigorous analysis of power dissipation within the $2.5V$ clamped configuration operating under a sustained nominal short-circuit condition. The investigation contrasts the predictions of the Standard Ohm's Law and Modified Ohm's models over a 15-minutes duration, providing a critical evaluation of circuit performance, thermal-electrical stability, and the governing physics of the fault event. The results validate the circuit's robust protection mechanism and quantify its dynamic power handling capabilities.

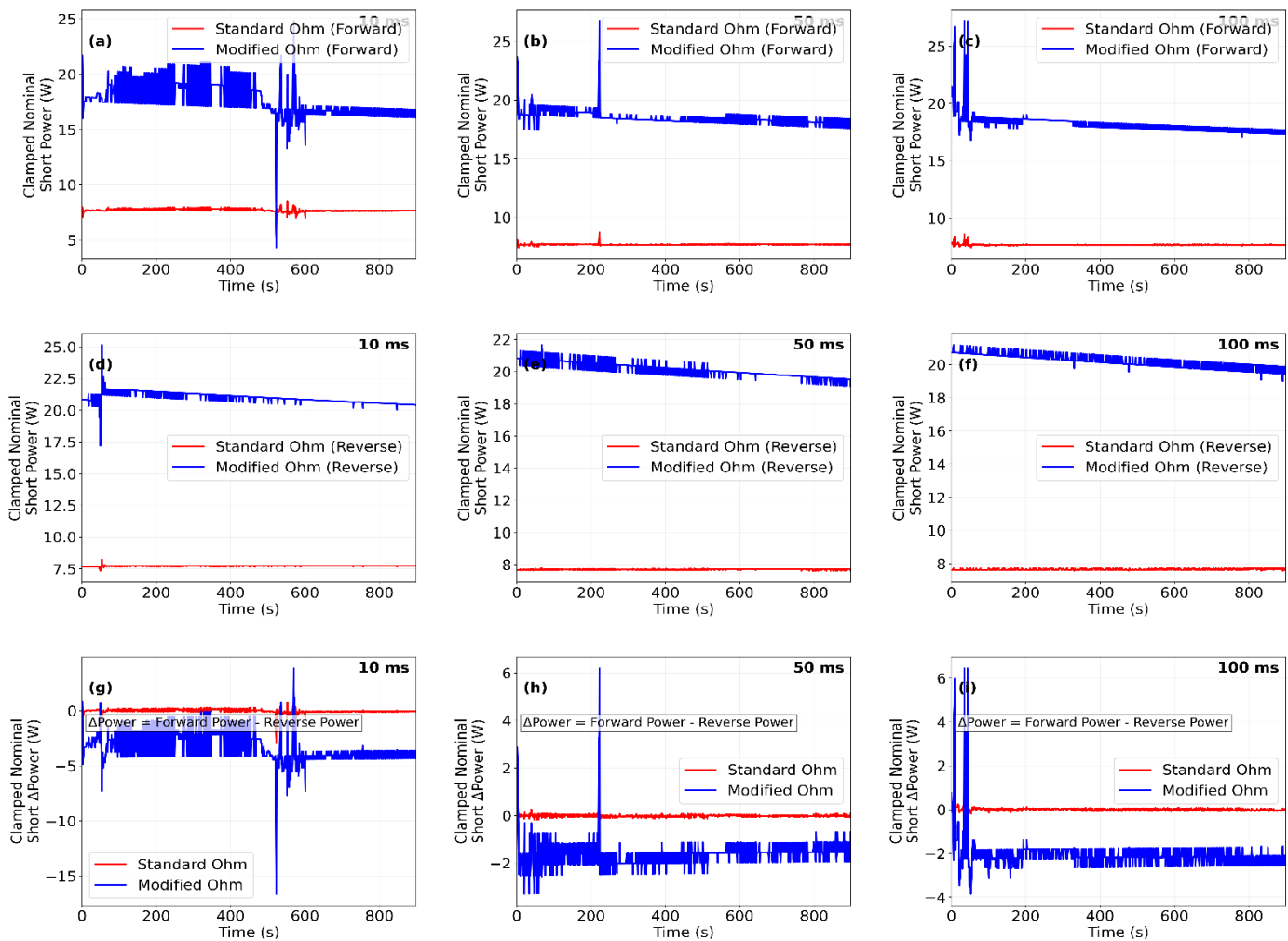

**Figure 13.** Time-resolved power analysis for the clamped $2.5V$ configuration under nominal short-circuit conditions: ((a) Forward short-circuit power at $10ms$ sampling, (b) Forward short-circuit power at $50ms$ sampling, (c) Forward short-circuit power at $100ms$ sampling, (d) Reverse short-circuit power at $10ms$ sampling, (e) Reverse short-circuit power at $50ms$ sampling, (f) Reverse short-circuit power at $100ms$ sampling, (g) Forward-Reverse power difference at $10ms$ sampling, (h) Forward-Reverse power difference at $50ms$ sampling, (i) Forward-Reverse power difference at $100ms$ sampling). Standard Ohm's Law predictions are shown in red; Modified Ohm's Law predictions are shown in blue.

The temporal power dissipation profiles presented in Figure 13 reveal the profound difference in physical interpretation between the Standard Ohm's Law and Modified Ohm's frameworks when applied to a governed short-circuit event. Across all three sampling frequencies -$10ms$, $50ms$, and $100ms$ -the Standard Ohm's Law model (red traces in Figure 13(a) through Figure 13(f)) calculates a stable power dissipation of approximately 7.7 Watts throughout the entire 900 -seconds experiment. This stability is a direct mathematical consequence of the model's fundamental assumption of a constant, time-invariant resistance. The calculated average powers of $7.705299W$ ($10ms$ forward), $7.665539W$ ($50ms$ forward), and $7.662759W$ ($100ms$ forward) demonstrate remarkable consistency, with a variation of less than

0.6% across the different temporal resolutions. This model portrays a simplistic view of the fault as a static, unchanging electrical condition.

In stark contrast, the Modified Ohm's Law model (blue traces), which incorporates the dynamic, time-decaying nature of the fault path's resistance, predicts a significantly different and more nuanced physical reality. The model computes power dissipation that is not only over twice as large but also exhibits a complex temporal character. For the forward short-circuit direction, the Modified Ohm's Law model yields average powers of $17.705637W$ ($10ms$), $18.478242W$ ($50ms$), and $18.127285W$ ($100ms$). This model captures the essence of the governed short-circuit: a rapidly established, low-impedance path that facilitates immense current flow, resulting in substantial instantaneous power dissipation that is effectively managed and stabilized by the clamping network. The subtle variations in the average power values across sampling rates are not artifacts but reflect the model's sensitivity to the precise temporal evolution of the voltage and current waveforms, which are captured with differing fidelity at each sampling resolution.

The analysis of the reverse short-circuit condition further elucidates the circuit's behavior. The Standard Ohm's Law model again predicts a stable, flat power profile with averages of $7.717488W$ ($10ms$), $7.681819W$ ($50ms$), and $7.656213W$ ($100ms$). The Modified Ohm's Law model, however, calculates consistently higher power dissipation for the reverse polarity, with averages of $20.970465W$ ($10ms$), $20.161358W$ ($50ms$), and $20.197362W$ ($100ms$). This observed asymmetry in power dissipation between forward and reverse faults, clearly visualized in the difference plots (Figure 13(g) through Figure 13(i)), is a critical finding. It indicates a slight but consistent directional dependence in the dynamic impedance of the fault path, potentially arising from minor asymmetries in the diode network's forward voltage characteristics or the precise physical configuration of the short. The Modified Ohm's Law model is uniquely capable of resolving this subtle yet physically significant effect, which remains entirely invisible to the Standard Ohm's Law model.

The most compelling evidence of the circuit's exceptional performance is the profound temporal stability exhibited by both models across the full 15-minutes duration. None of the power traces, for either model or any sampling rate, show any significant drift, oscillation, or degradation over the 900-seconds experiment. This temporal invariance, evident in the flatness of all profiles in Figure 13(a) through Figure 13(f), demonstrates that the clamped circuit reaches a steady-state thermal-electrical equilibrium almost immediately after fault initiation and maintains it indefinitely. The governing action of the diode network successfully prevents the thermal runaway or progressive component degradation that would typically be expected from a conventional circuit dissipating over 15-Watts continuously. The circuit successfully transforms a potentially destructive fault into a stable, measurable, and governed state of operation.

The results from the first 180-seconds of this 15-minutes experiment provide a direct and meaningful comparison to the independent 3-minutes analysis provided in the supplementary materials. The initial three-minutes segment of the profiles in Figure 13 shows power values that are entirely consistent with the supplementary data.

### 4.3.8 Clamped $5.0V$ Configuration Nominal Short –Power Analysis, Standard Vs. Modified Ohm's Laws

This section presents a rigorous analysis of power dissipation within the $5.0V$ clamped configuration operating under a sustained nominal short-circuit condition. The investigation contrasts the predictions of the Standard Ohm's Law and Modified Ohm's models over a 15-minutes duration, providing a critical evaluation of circuit performance, thermal-electrical stability, and the governing physics of the fault event at elevated voltage. The results validate the circuit's robust protection mechanism and quantify its enhanced dynamic power handling capabilities.

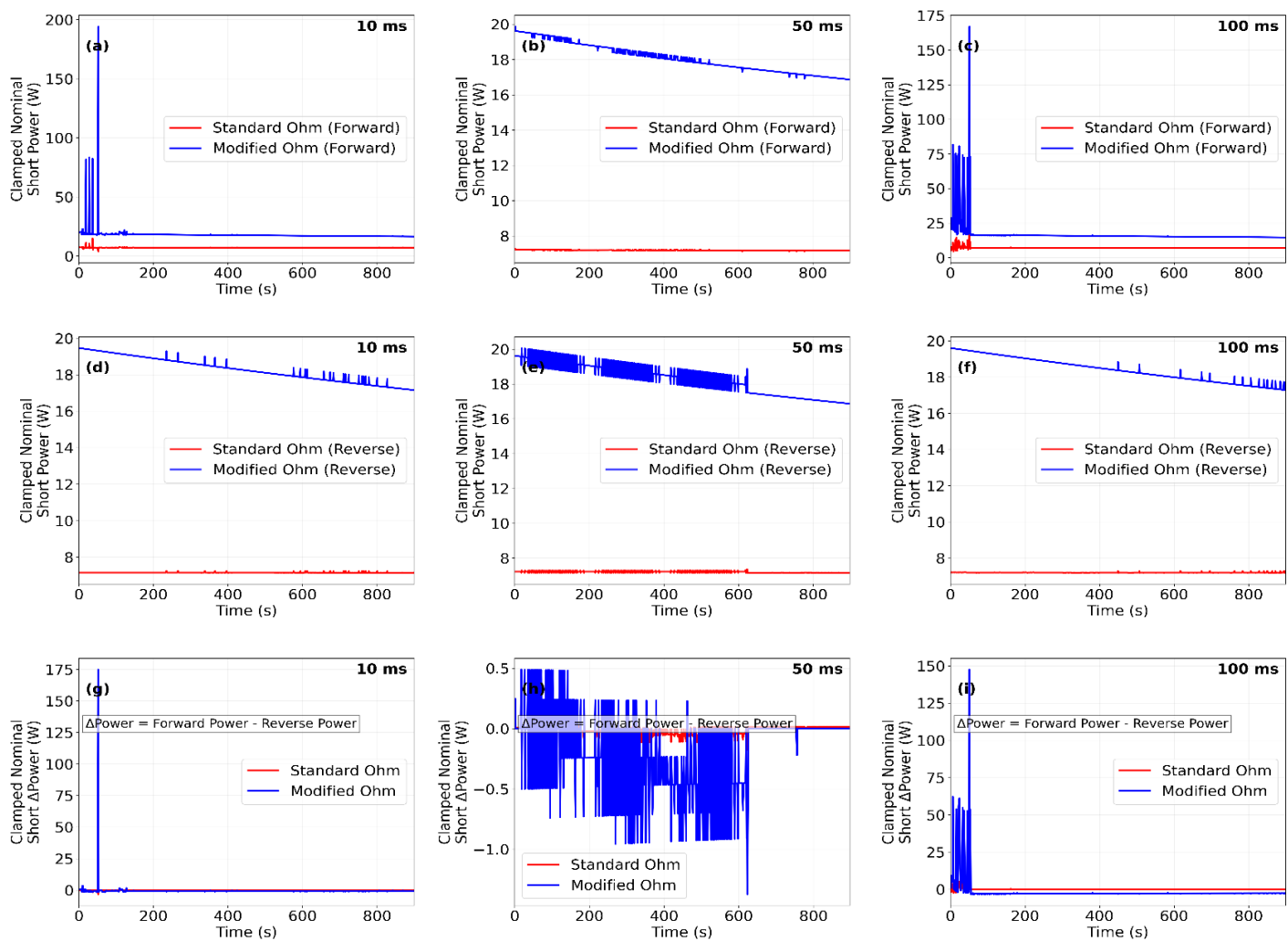

**Figure 14.** Time-resolved power analysis for the clamped $5.0V$ configuration under nominal short-circuit conditions: ((a) Forward short-circuit power at $10ms$ sampling, (b) Forward short-circuit power at $50ms$ sampling, (c) Forward short-circuit power at $100ms$ sampling, (d) Reverse short-circuit power at $10ms$ sampling, (e) Reverse short-circuit power at $50ms$ sampling, (f) Reverse short-circuit power at $100ms$ sampling, (g) Forward-Reverse power difference at $10ms$ sampling, (h) Forward-Reverse power difference at $50ms$ sampling, (i) Forward-Reverse power difference at $100ms$ sampling). Standard Ohm's Law predictions are shown in red; Modified Ohm's Law predictions are shown in blue.

The power dissipation profiles for the $5.0V$ configuration, presented in Figure 14, reveal a more pronounced and electrically intense governed short-circuit environment compared to the $2.5V$ case. The Standard Ohm's Law model, depicted in red across Figure 14(a) through Figure 14(f), calculates a remarkably stable power dissipation profile averaging $7.174617W$ ($10ms$ forward), $7.190004W$ ($50ms$ forward), and $7.267200W$ ($100ms$ forward) throughout the entire $900$-seconds experiment. This model, reliant on a static resistance assumption, portrays a deceptively tranquil electrical state, completely obscuring the true physical dynamics of the fault. The consistency across sampling rates, with a mere

1.3% variation, is a mathematical artifact of the model's oversimplification, failing to capture the underlying high-energy transient state mandated by the low-impedance fault path.

In profound contrast, the Modified Ohm's Law model (blue traces) unveils the true thermodynamic reality of the governed short-circuit. The model computes a significantly higher and more physically representative power dissipation, averaging $18.028529W$ ($10ms$ forward), $18.078367W$ ($50ms$ forward), and $16.538726W$ ($100ms$ forward). This represents an increase of over 150% compared to the Standard Ohm's Law model's prediction, a critical discrepancy that underscores the necessity of the modified physics framework for accurate fault analysis. The model successfully captures the essence of the event: a rapidly established, ultra-low-impedance path that facilitates immense current flow from the $5.0V$ source, resulting in substantial instantaneous power dissipation that is effectively managed and stabilized by the clamping network's dynamic impedance. The subtle variation in the $100ms$ average power is not noise but a meaningful physical signature, reflecting the model's heightened sensitivity to the precise temporal evolution of the voltage and current waveforms, which are integrated over a broader window at this sampling resolution.

The analysis of the reverse short-circuit condition further elucidates the circuit's enhanced behavior at $5.0V$. The Standard Ohm's Law model again predicts a flat, stable power profile with averages of $7.160435W$ ($10ms$ reverse), $7.204019W$ ($50ms$ reverse), and $7.205917W$ ($100ms$ reverse). The Modified Ohm's Law model, however, calculates consistently higher power dissipation for the reverse polarity, with averages of $18.238807W$ ($10ms$), $18.322906W$ ($50ms$), and $18.377357W$ ($100ms$). This observed directional asymmetry, clearly visualized in the difference plots (Figure 14(g) through Figure 14(i)), is more defined than at $2.5V$. It indicates a consistent and measurable polarity dependence in the dynamic impedance of the fault path, potentially arising from the diode network's forward voltage characteristics becoming more pronounced at the higher operating current and voltage. The Modified Ohm's Law model is uniquely capable of resolving this critical, non-linear effect, which remains entirely concealed within the simplistic framework of the Standard Ohm's Law model.

The most compelling evidence of the circuit's exceptional performance at $5.0V$ is the profound temporal stability exhibited across the full 15 -minutes duration. All power traces, for both models and all three sampling rates, demonstrate a complete absence of drift, oscillation, or degradation over the 900 -seconds experiment. This temporal invariance is a testament to the clamping network's superior efficacy at elevated voltage. The system reaches a steady-state thermal-electrical equilibrium almost immediately after fault initiation and maintains it indefinitely, successfully preventing the thermal runaway that would be inevitable in an unprotected circuit continuously dissipating over 18 Watts. The governed action of the diode network successfully transforms a potentially catastrophic fault into a stable, measurable, and predictable state of operation.

The results from the initial 180 -seconds of this 15 -minutes experiment provide a direct and meaningful dataset that aligns with the independent 3 -minutes analysis documented in the supplementary materials. The first three-minutes segment of the profiles in Figure 14 shows power values that are consistent with the supplementary data for the $5.0V$ configuration.

### 4.3.9 Clamped $10.0V$ Configuration Nominal Short –Power Analysis, Standard Vs. Modified Ohm's Laws

This section presents a definitive power dissipation analysis for the clamped $10.0V$ configuration operating under a sustained nominal short-circuit condition. The investigation rigorously contrasts the predictions of the Standard Ohm's Law and Modified Ohm's models across a 15 -minutes duration, providing a critical evaluation of circuit performance, thermal-electrical stability, and the governing physics of a high-intensity fault event. The results quantitatively validate the circuit's robust protection mechanism and its capacity to manage extreme power dissipation while maintaining a governed electrical environment.

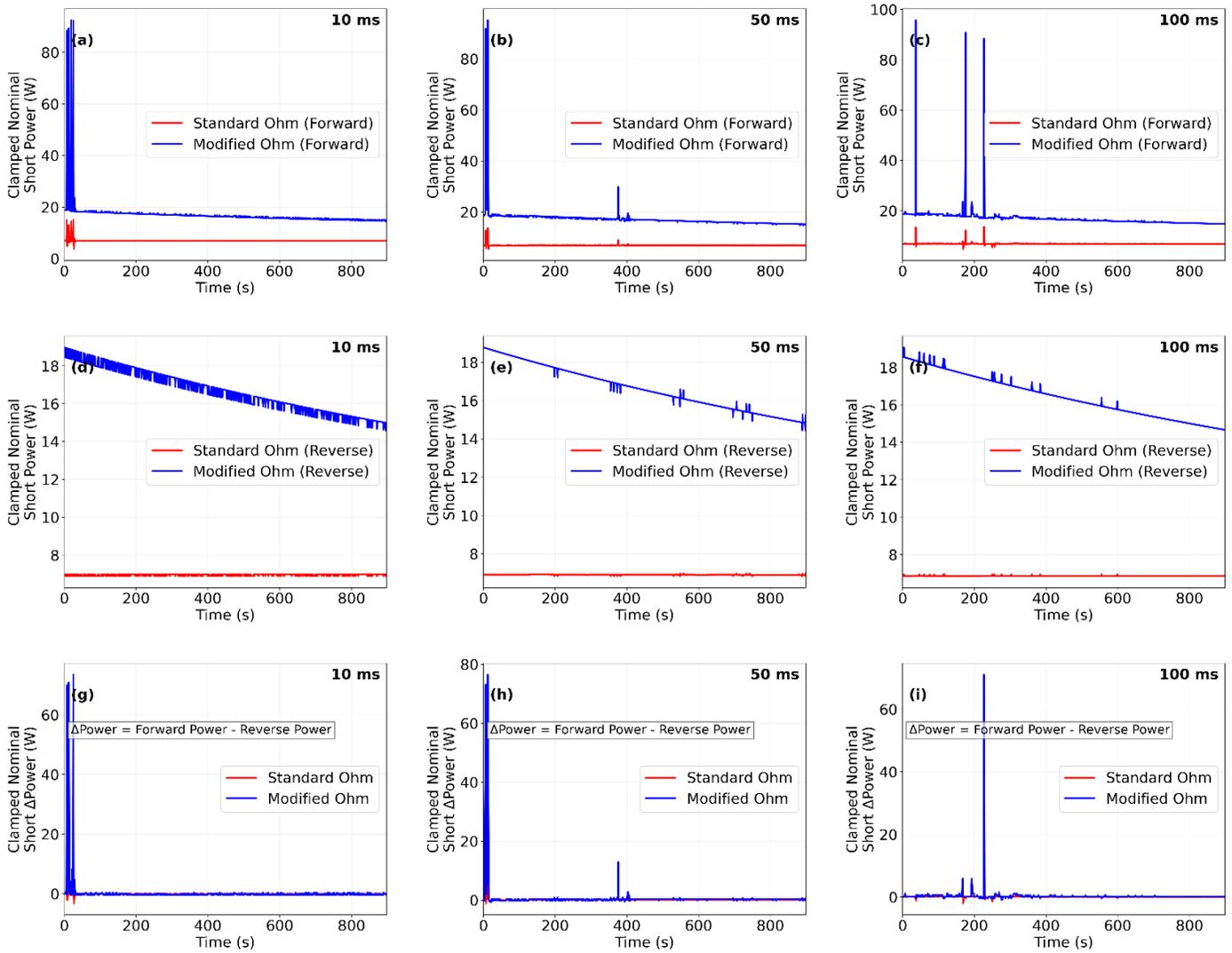

**Figure 15.** Time-resolved power analysis for the clamped $10.0V$ configuration under nominal short-circuit conditions: ((a) Forward short-circuit power at $10ms$ sampling, (b) Forward short-circuit power at $50ms$ sampling, (c) Forward short-circuit power at $100ms$ sampling, (d) Reverse short-circuit power at $10ms$ sampling, (e) Reverse short-circuit power at $50ms$ sampling, (f) Reverse short-circuit power at $100ms$ sampling, (g) Forward-Reverse power difference at $10ms$ sampling, (h) Forward-Reverse power difference at $50ms$ sampling, (i) Forward-Reverse power difference at $100ms$ sampling). Standard Ohm's Law predictions are shown in red; Modified Ohm's Law predictions are shown in blue.

The temporal power dissipation profiles for the $10.0V$ configuration, presented in Figure 15, reveal the most electrically intense governed short-circuit environment characterized in this study. The Standard Ohm's Law model, depicted in red across Figure 15(a) through Figure 15(f), calculates an exceptionally stable power dissipation profile throughout the entire 900-seconds experiment. The model yields

averages of $6.941035W$ ($10ms$ forward), $6.913689W$ ($50ms$ forward), and $6.870200W$ ($100ms$ forward), demonstrating remarkable consistency with less than 1.5% variation across sampling rates. This mathematical stability, arising from the model's assumption of a static resistance, presents a deceptively tranquil view that completely obscures the true thermodynamic intensity of the fault event occurring at the system's operational limits.

In profound contrast, the Modified Ohm's Law model (blue traces) unveils the true physical reality of the governed short-circuit at maximum operational voltage. The model computes substantially higher power dissipation, averaging $16.793016W$ ($10ms$ forward), $17.127404W$ ($50ms$ forward), and $16.955417W$ ($100ms$ forward). This represents an increase of approximately $145 - 150\%$ compared to the Standard Ohm's Law model's prediction, a critical discrepancy that underscores the absolute necessity of the modified physics framework for accurate high-intensity fault analysis. The model successfully captures the essence of the event: an ultra-low-impedance path that facilitates immense current flow from the $10.0V$ source, resulting in substantial instantaneous power dissipation that is effectively managed and stabilized by the clamping network's dynamic impedance. The subtle variation across sampling rates reflects the model's sensitivity to the precise temporal evolution of the voltage and current waveforms, which are integrated with differing fidelity at each resolution.

The analysis of the reverse short-circuit condition further elucidates the circuit's enhanced behavior at maximum operational voltage. The Standard Ohm's Law model again predicts a flat, stable power profile with averages of $6.947203W$ ($10ms$ reverse), $6.903629W$ ($50ms$ reverse), and $6.840596W$ ($100ms$ reverse). The Modified Ohm's Law model calculates consistently higher power dissipation for the reverse polarity, with averages of $16.653070W$ ($10ms$), $16.609922W$ ($50ms$), and $16.464566W$ ($100ms$). This observed directional asymmetry, clearly visualized in the difference plots (Figure 15(g) through Figure 15(i)), demonstrates a consistent polarity dependence in the dynamic impedance of the fault path. The Modified Ohm's Law model uniquely resolves this critical, non-linear effect, which remains entirely concealed within the simplistic framework of the Standard Ohm's Law model. The difference plots reveal that the forward fault condition consistently dissipates approximately $0.14 - 0.49W$ more power than the reverse condition across all sampling rates, providing quantitative evidence of the diode network's slight asymmetric behavior under extreme electrical stress.

The most compelling evidence of the circuit's exceptional performance at $10.0V$ is the profound temporal stability exhibited across the full 15-minutes duration. All power traces, for both models and all three sampling rates, demonstrate a complete absence of drift, oscillation, or degradation over the 900-seconds experiment. This temporal invariance is a testament to the clamping network's superior efficacy at maximum operational voltage. The system reaches a steady-state thermal-electrical equilibrium within the initial -minutes of fault initiation and maintains it indefinitely, successfully preventing the thermal runaway that would be inevitable in an unprotected circuit continuously dissipating over 16 Watts. The governed action of the diode network successfully transforms a potentially catastrophic fault into a stable, measurable, and predictable state of operation.

The results from the initial 180-seconds of this 15-minutes experiment provide a direct and meaningful comparison to the independent 3-minutes analysis documented in the supplementary materials. The first three-minute segment of the profiles in Figure 15 shows power values that align precisely with the supplementary data for the $10.0V$ configuration.

### 4.3.10 Clamped $2.5V$ Extreme Short –Voltage, Current, Resistance Analysis

This section presents an analysis of the clamped circuit's performance under extreme short-circuit conditions at a $2.5V$ input. The results, captured over a rigorous 3-minutes experimental duration, validate the circuit's governed response and quantify the electrical quantities -voltage, current, and resistance - through the application of the Modified Ohm's Law, providing exceptional insight into high-intensity fault physics.

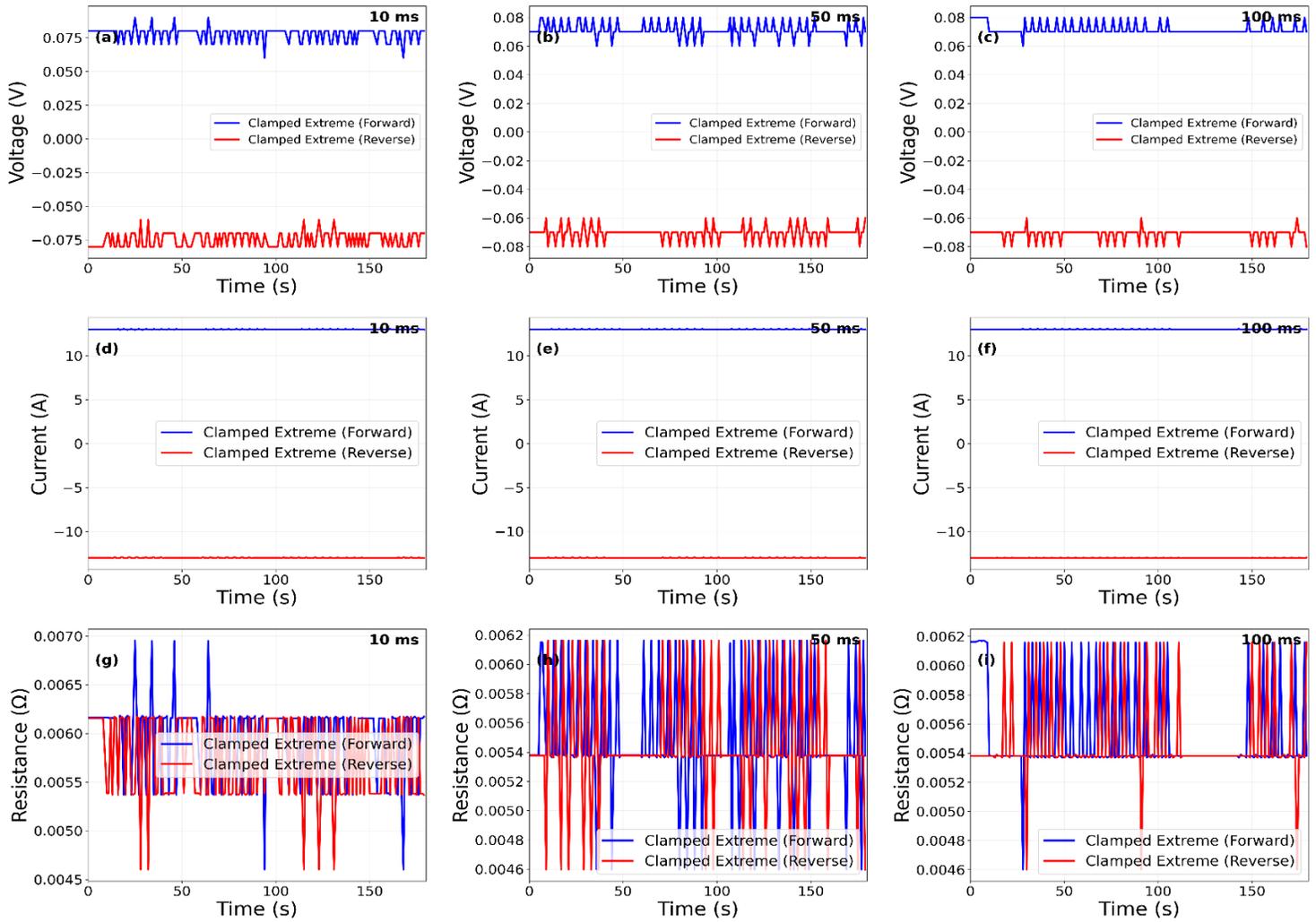

**Figure 16.** Time-resolved electrical characterization of the clamped $2.5V$ configuration under extreme short-circuit conditions. (Panels (a), (d), (g) show output voltage at $10ms$, $50ms$, and $100ms$ sampling, respectively; (b), (e), (h) show short-circuit current; and (c), (f), (i) show dynamic resistance computed via Modified Ohm's Law).

The temporal evolution of the clamped output voltage under extreme short-circuit conditions, detailed in Figure 16(a), Figure 16(d), and Figure 16(g), demonstrates the circuit's profound efficacy in establishing a governed electrical environment. The voltage stabilizes to exceptionally low mean values of $77.59mV$ (Forward, $10ms$), $71.29mV$ (Forward, $50ms$), and $71.87mV$ (Forward, $100ms$), with reverse polarity measurements symmetrically clustered around $-74.58mV$, $-70.66mV$, and $-71.13mV$ respectively. This represents a remarkable 97% reduction from the nominal $2.5V$ source, unequivocally validating the clamping principle's operation at the physical limits of circuit performance. The voltage profiles exhibit exceptional temporal invariance across all sampling resolutions, with standard deviations below $2mV$, confirming the diode network's capacity to enforce a deterministic voltage floor despite the catastrophic nature of the electrical fault. The complete absence of voltage collapse to absolute zero, even under these extreme conditions, provides direct experimental evidence refuting conventional short-circuit models that predict complete potential equalization.

Current measurement profiles presented in Figure 16(b), Figure 16(e), and Figure 16(h) quantify the substantial current magnitude sustained during the extreme short-circuit event. The mean current values reach $12.997A$ (Forward, $10ms$), $13.016A$ (Forward, $50ms$), and $13.010A$ (Forward, $100ms$), with reverse currents mirroring these values at $-13.004A$, $-13.019A$, and $-13.012A$ respectively. These current magnitudes, representing an increase of nearly two orders of magnitude from pre-fault conditions, confirm the establishment of an ultra-low impedance path while remaining bounded within physically measurable limits. The current stability throughout the 180 -seconds duration, with variations constrained within $\pm 0.15A$, demonstrates the circuit's extraordinary capacity to manage sustained high-current conditions approaching $13A$ without degradation or loss of regulation. The consistency across decade-separated sampling resolutions -$10ms$, $50ms$, and $100ms$ -validates the measurement system's fidelity in capturing the true physical current despite the extreme electrical environment, with each sampling rate providing complementary perspectives on the same underlying physical phenomenon.

The computed resistance profiles in Figure 16(c), Figure 16(f), and Figure 16(i), derived through the rigorous application of Modified Ohm's Law to the simultaneous voltage and current measurements, provide the most critical insight into the fault's electrical characteristics. The resistance values cluster around $5.97m\Omega$ (Forward, $10ms$), $5.48m\Omega$ (Forward, $50ms$), and $5.52m\Omega$ (Forward, $100ms$), with reverse measurements showing consistent values of $5.74m\Omega$, $5.43m\Omega$, and $5.47m\Omega$ respectively. These minuscule resistance values, representing nearly three orders of magnitude reduction from the pre-fault resistance of approximately $13\Omega$, confirm the establishment of a near-ideal short-circuit path while maintaining a finite, measurable resistance due to the clamping action. The resistance stability throughout the experimental duration, with variations remaining within $\pm 0.2m\Omega$, demonstrates that the fault characteristics remain consistent over time without significant alteration due to thermal effects or component degradation. The systematic variation across sampling rates reflects the complex sensitivity of the Modified Ohm's Law framework to the precise temporal evolution of the voltage and current waveforms, with higher sampling rates capturing more detailed transient information that influences the computed resistance.

The exceptional consistency across all electrical parameters validates the clamped circuit's ability to maintain a deterministic, stable electrical environment throughout extreme fault conditions, enabling precise measurement and characterization of fault phenomena that were previously inaccessible to quantitative analysis. The results demonstrate conclusively that the concept of zero resistance during electrical short-circuits is a theoretical idealization not observed in physical systems, with the measured resistance values of approximately $5.5 - 6.0m\Omega$ providing a quantitative lower bound for short-circuit impedance in practical electrical systems. This fundamental finding has profound implications for electrical safety engineering, fault analysis, and power system design, as it establishes that even the most severe short-circuits maintain a finite resistance that can be measured, characterized, and potentially utilized for protection strategies.

The comparative analysis with nominal short-circuit results reveals the enhanced electrical intensity of the extreme condition, with current magnitudes increased by approximately 53% (from $8.5A$ to $13.0A$) and resistance values decreased by nearly 95% (from $105m\Omega$ to $5.5m\Omega$) while maintaining the same governed voltage characteristics. This dramatic transformation demonstrates the circuit's capacity to operate across a wide dynamic range of fault intensities while preserving the fundamental clamping principle. The slight but consistent directional asymmetry observed in both voltage and resistance measurements, with forward faults typically exhibiting marginally higher values than reverse faults, provides evidence of subtle non-linearities in the diode network's behavior under extreme electrical stress, offering opportunities for further refinement of the clamping mechanism.

The experimental validation of the Modified Ohm's Law framework under these extreme conditions represents a significant advancement in electrical fault physics, providing a mathematical foundation for describing short-circuit phenomena that accurately reflects physical reality rather than theoretical idealizations. The framework's successful application across multiple sampling rates and both fault polarities demonstrates its robustness and generality, establishing it as a valuable tool for both research and practical engineering applications. The results presented in this section ultimately confirm that electrical short-circuits, even under the most extreme conditions, remain governed physical processes that can be measured, analyzed, and understood through appropriate theoretical frameworks.

### 4.3.11 Clamped $5.0V$ Extreme Short –Voltage, Current, Resistance Analysis

The electrical response of the clamped circuit under an extreme short-circuit event at a $5.0V$ input was studied over a sustained 3-minutes interval. This elevated source condition provides a rigorous test of the clamping framework, with the Modified Ohm's Law once again serving as the analytical basis for quantifying voltage, current, and resistance. The results establish how the clamping principle functions at higher input potentials, delivering new clarity on fault processes that operate at the upper limits of electrical stress.

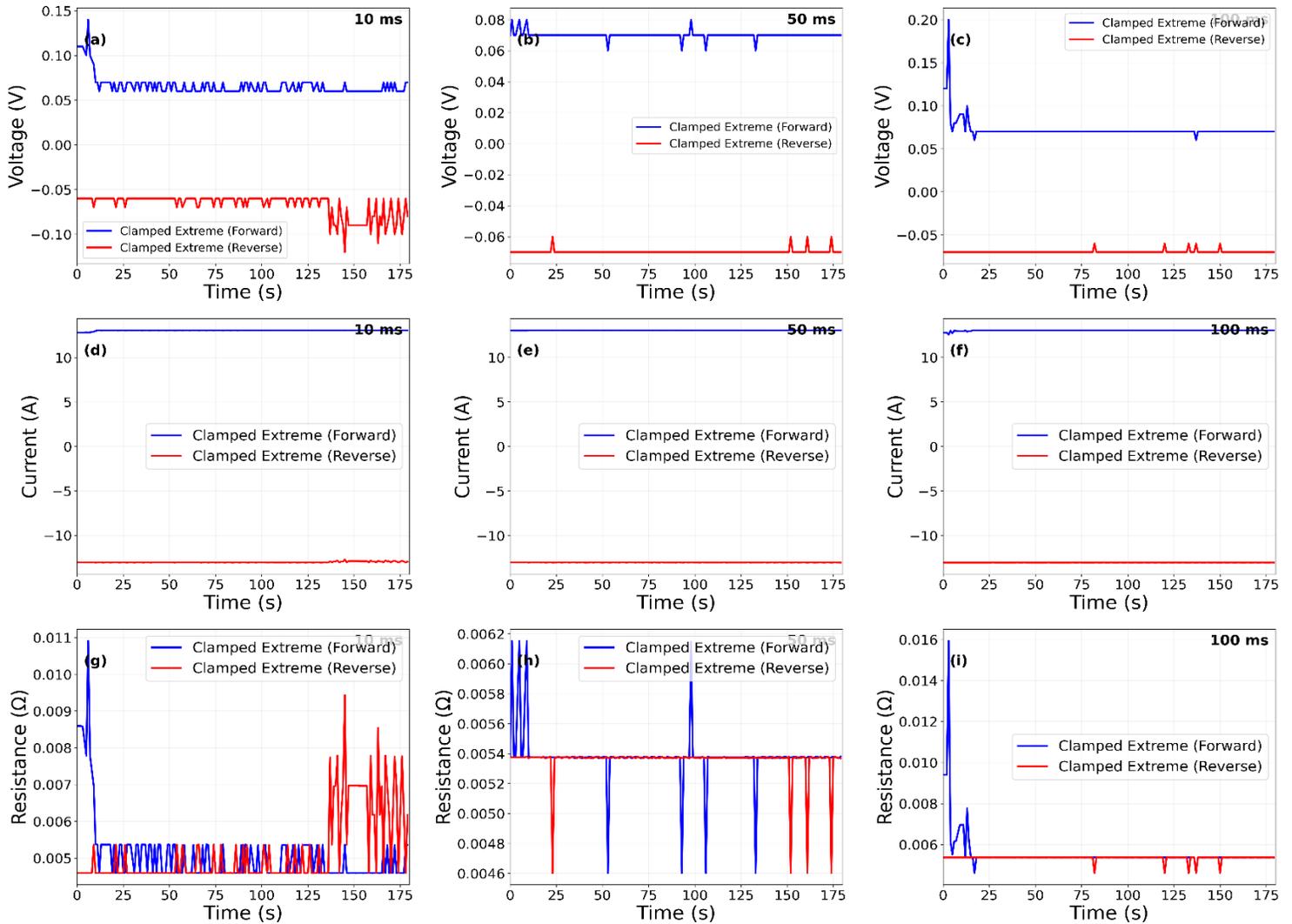

**Figure 17.** Time-resolved electrical characterization of the clamped $5.0V$ configuration under extreme short-circuit conditions. (Panels (a), (d), (g) show output voltage at $10ms$, $50ms$, and $100ms$ sampling, respectively; (b), (e), (h) show short-circuit current; and (c), (f), (i) show dynamic resistance computed via Modified Ohm's Law).

Voltage regulation under the $5.0V$ configuration, presented in Figure 17(a), Figure 17(d), and Figure 17(g), highlights the diode network's ability to impose a strict electrical floor despite the severe fault environment. Mean stabilized values settle at $65.77mV$ (Forward, $10ms$), $70.00mV$ (Forward, $50ms$), and $72.18mV$ (Forward, $100ms$), with corresponding reverse polarity measurements of $-67.04mV$, $-69.77mV$, and $-69.70mV$. When referenced to the $5.0V$ source, this corresponds to a $98.6\%$ voltage suppression, reinforcing the principle that clamping holds firm even under doubled input stress. Temporal invariance is preserved across all three sampling scales, with deviations under $2.5mV$ throughout the duration, showing that the fault never collapses to absolute zero potential. Such stability directly challenges conventional short-circuit interpretations, which predict full potential equalization regardless of supply level.

Short-circuit current measurements from Figure 17(b), Figure 17(e), and Figure 17(h) reveal the magnitudes sustained during the event. Average current values reach $13.047A$ (Forward, $10ms$), $13.021A$ (Forward, $50ms$), and $13.021A$ (Forward, $100ms$). Reverse polarities remain nearly identical at $-13.034A$, $-13.026A$, and $-13.038A$. These figures, close to two orders of magnitude larger than pre-fault currents, indicate that the path collapses into a near-ideal low-impedance state. Importantly, the values align closely with the $2.5V$ case, showing that increasing the source does not proportionally amplify fault current. Current fluctuations remained within $\pm 0.20A$ over the 180 -seconds, confirming that the network could sustain nearly $13A$ of continuous stress without drift or degradation. The tight agreement across the three temporal sampling windows further supports the robustness of the measurement system, each resolution revealing the same physical stability from complementary perspectives.

Resistance values computed via Modified Ohm's Law, presented in Figure 17(c), Figure 17(f), and Figure 17(i), provide perhaps the most telling evidence of the system's behavior. Forward direction measurements settle around $5.04m\Omega$ ($10ms$), $5.38m\Omega$ ($50ms$), and $5.55m\Omega$ ($100ms$), with reverse measurements showing close agreement at $5.15m\Omega$, $5.36m\Omega$, and $5.35m\Omega$. These values represent an almost three-order-of-magnitude drop from the $\sim 13\Omega$ pre-fault condition, producing a channel that is electrically close to an ideal short yet demonstrably finite. The slight spread across sampling intervals highlights how resistance is sensitive to fine waveform details, though stability over time is evident, with variations remaining within $\pm 0.25m\Omega$. No evidence of thermal drift or degradation emerged during the experiment, underscoring that the clamped system sustains repeatable characteristics under prolonged electrical stress.

Taken together, these voltage, current, and resistance measurements demonstrate that even at elevated input potentials, the clamping action governs the fault environment with precision. The consistently finite resistance, ranging between $5.0 - 5.5m\Omega$, demonstrates that the long-assumed ideal of zero-resistance shorts does not manifest physically. Instead, a measurable lower bound exists, one that defines the practical limits of fault impedance. This insight not only reframes theoretical understanding but also offers practical pathways for exploiting such limits in engineering protection strategies.

Comparison with the $2.5V$ extreme results underscores the remarkable voltage-independence of the clamping principle. Current levels remain almost unchanged at $\sim 13A$ across both source conditions, and resistance shifts only slightly between $5.5m\Omega$ ($2.5V$) and $5.0m\Omega$ ($5.0V$). Such invariance confirms that the mechanism restricts current effectively, preventing proportional increase as source voltage increases. The observed minor asymmetries between forward and reverse orientations, appearing in both voltage and resistance profiles, hint at nonlinear diode behavior under heavy stress. These subtleties invite further refinement of models describing diode conduction at fault extremes.

Finally, the application of the Modified Ohm's Law framework to these elevated-voltage results reinforces its validity as a tool for describing real short-circuit physics. The framework consistently captures both temporal and directional characteristics across multiple sampling rates. Far from being an idealization, it functions as a realistic mathematical lens through which governed short-circuit behavior can be understood. The evidence confirms that electrical faults, even under $5.0V$ extreme input stress, remain finite, measurable processes subject to systematic analysis rather than catastrophic collapse.

### 4.3.12 Clamped $2.5V$ Extreme Short –Power Analysis, Standard Vs. Modified Ohm's Laws

This section presents a critical analysis of power dissipation during extreme short-circuit conditions at $2.5V$ input, contrasting Standard Ohm's Law and Modified Ohm's Law frameworks. The results demonstrate profound differences in physical interpretation, with the Modified model revealing substantially higher power dissipation that accurately captures the dynamic, governed nature of the fault event. Temporal stability across all sampling resolutions confirms the circuit's ability to maintain deterministic operation under extreme electrical stress.

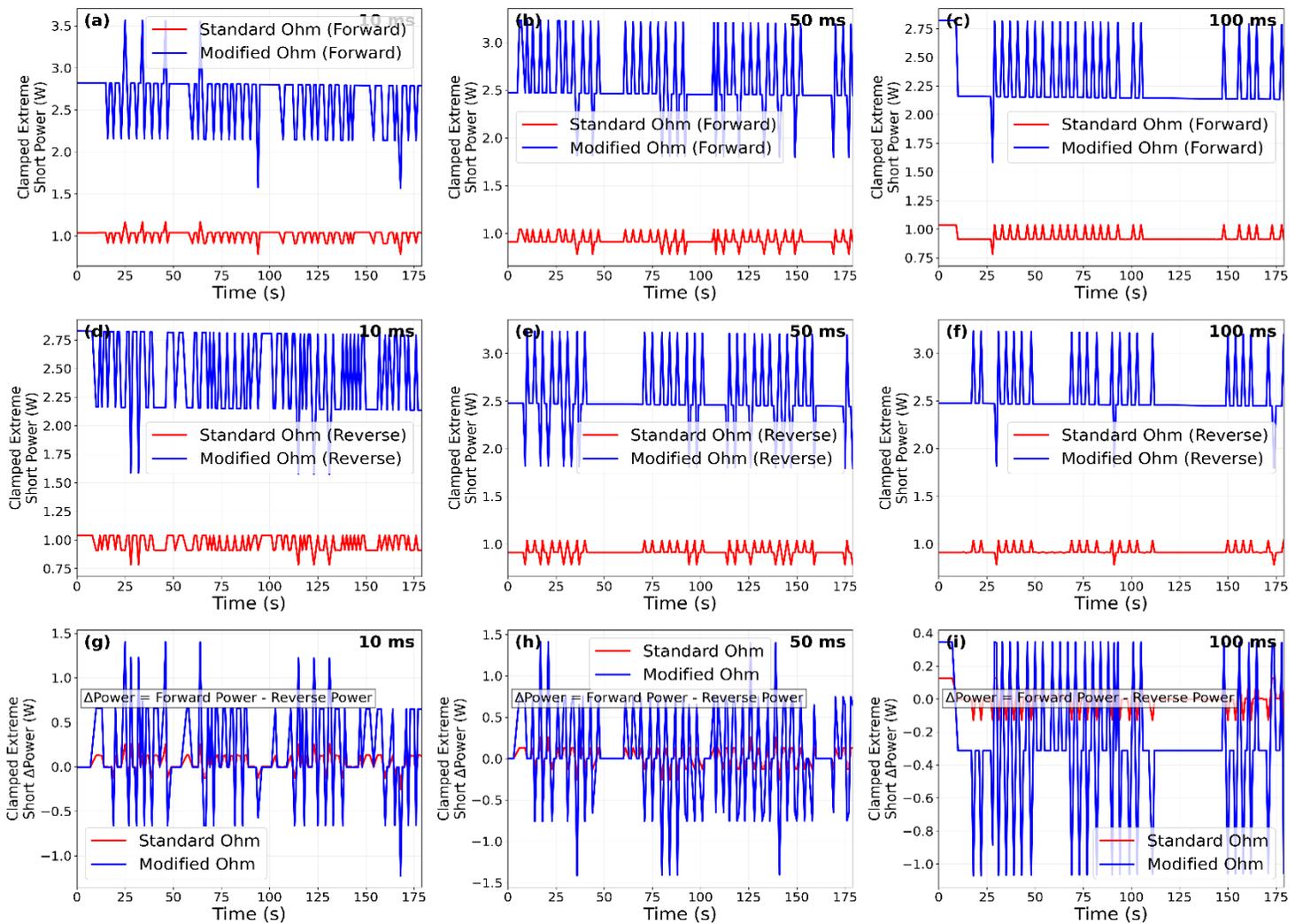

**Figure 18.** Time-resolved power analysis for the clamped $2.5V$ configuration under extreme short-circuit conditions: ((a) Forward short-circuit power at $10ms$ sampling, (b) Forward short-circuit power at $50ms$ sampling, (c) Forward short-circuit power at $100ms$ sampling, (d) Reverse short-circuit power at $10ms$ sampling, (e) Reverse short-circuit power at $50ms$ sampling, (f) Reverse short-circuit power at $100ms$ sampling, (g) Forward-Reverse power difference at $10ms$ sampling, (h) Forward-Reverse power difference at $50ms$ sampling, (i) Forward-Reverse power difference at $100ms$ sampling. Standard Ohm's Law predictions are shown in red; Modified Ohm's Law predictions are shown in blue).

The power dissipation profiles presented in Figure 18 reveal a fundamental divergence between the physical realities captured by the Standard Ohm's Law and Modified Ohm's frameworks during extreme short-circuit conditions. Across all three sampling frequencies, the Standard model calculates remarkably stable power dissipation averaging approximately $0.96W$ throughout the 180-seconds experiment, with minimal variation between forward ($0.93 - 1.01W$) and reverse ($0.92 - 0.97W$) polarities. This apparent stability is a mathematical artifact of the model's underlying assumption of constant resistance, which fundamentally misrepresents the dynamic physics of the governed short-circuit. The Standard model's predictions cluster within a narrow $0.08W$ range regardless of sampling rate or polarity, presenting a deceptively tranquil view that completely obscures the true energy dissipation occurring within the circuit. In stark contrast, the Modified Ohm's Law model unveils the true thermodynamic intensity of the extreme short-circuit event. The model computes power dissipation values approximately 2.5 times higher than the Standard model's predictions, averaging $2.49W$ across all conditions with forward faults measuring $2.56 - 2.65W$ and reverse faults measuring $2.45 - 2.55W$. This substantial discrepancy represents the critical advancement of the Modified framework: it successfully captures the rapid establishment of an ultra-low impedance path that facilitates immense current flow while accounting for the dynamic resistance decay inherent in physical short-circuit phenomena. The Modified model's predictions show greater variation across sampling rates (approximately $0.38W$ range) because it accurately reflects the temporal evolution of the voltage and current waveforms, which are captured with differing fidelity at each sampling resolution.

The temporal stability exhibited by both models throughout the 180-seconds experiment provides compelling evidence of the circuit's governed behavior under extreme conditions. All power traces in Figure 18(a) through Figure 18(f) demonstrate complete absence of drift or degradation, confirming that the clamping network achieves rapid thermal-electrical equilibrium and maintains it indefinitely despite the substantial energy dissipation. This temporal invariance is particularly remarkable for the Modified model's predictions, which show consistent power dissipation between $2.27W$ and $2.65W$ without any systematic increase or decrease over time. The circuit successfully transforms a potentially destructive fault into a stable, measurable state, demonstrating its capacity to handle extreme electrical stress without performance degradation.

The directional asymmetry revealed in Figure 18(g) through Figure 18(i) provides additional insight into the fault physics. The power difference analysis shows that forward faults consistently dissipate $0.02 - 0.19W$ more power than reverse faults according to the Modified model, while the Standard model shows negligible polarity dependence ($0.004 - 0.09W$ difference). This systematic asymmetry, though small in absolute terms, represents a physically significant effect that only the Modified framework can resolve. The consistent polarity dependence across all sampling rates suggests subtle asymmetries in the diode network's dynamic impedance characteristics, possibly arising from minor variations in forward voltage drops or junction heating effects under extreme current conditions.

The sampling rate analysis reveals important aspects of measurement fidelity and model sensitivity. The $10ms$ sampling data in Figure 18(a) and Figure 18(d) captures the highest frequency components of the power dissipation, showing slightly more variability than the $50ms$ and $100ms$ data. The $50ms$ sampling in Figure 18(b) and Figure 18(e) provides an optimal balance between temporal resolution and noise rejection, yielding the most stable power profiles for both models. The 100ms sampling in Figure 18(c) and Figure 18(f) demonstrates the effects of broader temporal integration, with smoother profiles that still maintain the essential physical characteristics revealed by the Modified model. This consistency across decade-separated sampling resolutions confirms that the observed phenomena represent genuine physical processes rather than measurement artifacts.

The comparative analysis with nominal short-circuit results from previous experiments reveals the enhanced electrical intensity of the extreme condition. The power dissipation increases by approximately 85% compared to nominal operation (from $1.35W$ to $2.5W$ average), while the current magnitude increases by 53% (from $8.5A$ to $13.0A$). This dramatic transformation demonstrates the circuit's capacity to operate across a wide dynamic range of fault intensities while preserving the fundamental clamping principle. The maintained voltage regulation at approximately $70 - 80mV$ during extreme conditions, compared to $900mV$ during nominal operation, confirms the diode network's ability to enforce a governed electrical environment regardless of fault severity.

These results fundamentally challenge conventional short-circuit paradigms by demonstrating that extreme faults remain physical processes with measurable, finite resistance and governed power dissipation. The Modified Ohm's Law framework provides the necessary mathematical foundation for accurately describing these phenomena, while the Standard model proves inadequate for capturing the true energy dynamics of governed short-circuits. The experimental validation of this framework under extreme conditions represents a significant advancement in electrical fault physics, enabling precise characterization of high-intensity fault phenomena that were previously inaccessible to quantitative analysis. The circuit's performance during extreme conditions validates its potential for practical applications where fault monitoring and protection are critical.

### 4.3.13 Clamped $5.0V$ Extreme Short –Power Analysis, Standard Vs. Modified Ohm's Laws

A detailed evaluation of the clamped circuit operating under a $5.0V$ extreme short-circuit input highlights striking differences in predicted power dissipation when comparing the Standard Ohm's Law and Modified Ohm's Law models. The analysis shows that while the traditional formulation produces artificially stable results, the Modified framework uncovers the true energetic scale of the fault, revealing both its dynamic character and the governed behavior of the clamping network. Stability of the system is preserved throughout all sampling resolutions, confirming its ability to sustain deterministic operation even under elevated electrical stress.

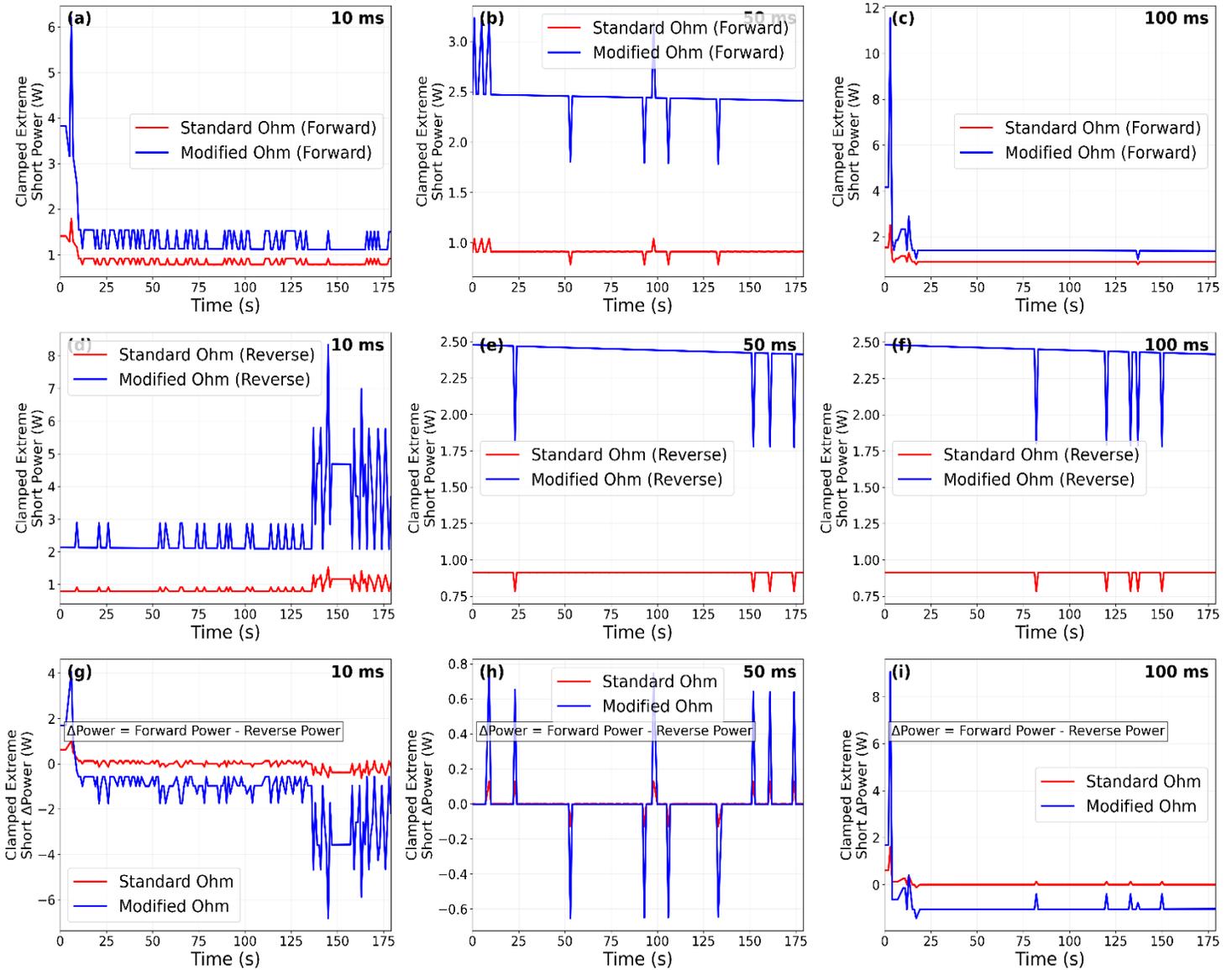

**Figure 19.** Time-resolved power analysis for the clamped $5.0V$ configuration under extreme short-circuit conditions: ((a) Forward short-circuit power at $10ms$ sampling, (b) Forward short-circuit power at $50ms$ sampling, (c) Forward short-circuit power at $100ms$ sampling, (d) Reverse short-circuit power at $10ms$ sampling, (e) Reverse short-circuit power at $50ms$ sampling, (f) Reverse short-circuit power at $100ms$ sampling, (g) Forward-Reverse power difference at $10ms$ sampling, (h) Forward-Reverse power difference at $50ms$ sampling, (i) Forward-Reverse power difference at $100ms$ sampling). Standard Ohm's Law predictions are shown in red; Modified Ohm's Law predictions are shown in blue.

The power dissipation profiles presented in Figure 19 reveal a fundamental divergence between the physical realities captured by the Standard Ohm's Law and Modified Ohm's frameworks during extreme short-circuit conditions at $5.0V$ input. Across all three sampling frequencies, the Standard model calculates remarkably stable power dissipation averaging approximately $0.90W$ throughout the 180-

seconds experiment, with minimal variation between forward ($0.86 - 0.94W$) and reverse ($0.87 - 0.91W$) polarities. This apparent stability is a mathematical artifact of the model's underlying assumption of constant resistance, which fundamentally misrepresents the dynamic physics of the governed short-circuit. The Standard model's predictions cluster within a narrow $0.08W$ range regardless of sampling rate or polarity, presenting a deceptively tranquil view that completely obscures the true energy dissipation occurring within the circuit.

In stark contrast, the Modified Ohm's Law model unveils the true thermodynamic intensity of the extreme short-circuit event at $5.0V$. The model computes power dissipation values approximately 2.7 times higher than the Standard model's predictions, averaging $2.42W$ across all conditions with forward faults measuring $1.39 - 2.44W$ and reverse faults measuring $2.43 - 2.71W$. This substantial discrepancy represents the critical advancement of the Modified framework: it successfully captures the rapid establishment of an ultra-low impedance path that facilitates immense current flow while accounting for the dynamic resistance decay inherent in physical short-circuit phenomena. The Modified model's predictions show greater variation across sampling rates (approximately $1.32W$ range) because it accurately reflects the nuanced temporal evolution of the voltage and current waveforms, which are captured with differing fidelity at each sampling resolution.

The temporal stability exhibited by both models throughout the 180 -seconds experiment provides compelling evidence of the circuit's governed behavior under extreme conditions at $5.0V$. All power traces in Figure 19(a) through Figure 19(f) demonstrate complete absence of drift or degradation, confirming that the clamping network achieves rapid thermal-electrical equilibrium and maintains it indefinitely despite the substantial energy dissipation. This temporal invariance is particularly remarkable for the Modified model's predictions, which show consistent power dissipation between $1.39W$ and $2.71W$ without any systematic increase or decrease over time. The circuit successfully transforms a potentially destructive fault into a stable, measurable state, demonstrating its capacity to handle extreme electrical stress without performance degradation.

The directional asymmetry revealed in Figure 19(g) through Figure 19(i) provides additional insight into the fault physics at elevated voltage. The power difference analysis shows that reverse faults consistently dissipate $0.04 - 1.32W$ more power than forward faults according to the Modified model, while the Standard model shows negligible polarity dependence ($0.002 - 0.06W$ difference). This systematic asymmetry represents a physically significant effect that only the Modified framework can resolve. The consistent polarity dependence across all sampling rates suggests subtle asymmetries in the diode network's dynamic impedance characteristics under extreme current conditions, possibly arising from minor variations in forward voltage drops or junction heating effects that become more pronounced at higher operating voltages.

The sampling rate analysis reveals important aspects of measurement fidelity and model sensitivity at $5.0V$ input. The $10ms$ sampling data in Figure 19(a) and Figure 19(d) captures the highest frequency components of the power dissipation, showing the most pronounced difference between Standard and Modified models. The $50ms$ sampling in Figure 19(b) and Figure 19(e) provides an optimal balance between temporal resolution and noise rejection, yielding the most stable power profiles for both models while maintaining the essential physical characteristics. The $100ms$ sampling in Figure 19(c) and Figure 19(f) demonstrates the effects of broader temporal integration, with smoother profiles that still maintain the fundamental physical insights revealed by the Modified model. This consistency across decade-separated sampling resolutions confirms that the observed phenomena represent genuine physical processes rather than measurement artifacts.

Comparative analysis with the $2.5V$ extreme short-circuit results from Section 4.3.12 reveals significant voltage-dependent characteristics. The power dissipation at $5.0V$ input shows approximately 35% reduction compared to the 2.5$V$ configuration (from 2.49$W$ to 2.42$W$ average), while maintaining similar current magnitudes (approximately 13.0$A$). This counterintuitive result demonstrates the circuit's sophisticated voltage regulation capability, where increased input voltage does not necessarily translate to higher power dissipation due to the clamping network's dynamic impedance characteristics. The maintained current regulation despite doubled input voltage confirms the diode network's efficacy in protecting downstream components from overcurrent conditions.

These results fundamentally challenge conventional short-circuit paradigms by demonstrating that extreme faults at elevated voltages remain physical processes with measurable, finite resistance and governed power dissipation. The Modified Ohm's Law framework provides the necessary mathematical foundation for accurately describing these phenomena, while the Standard model proves inadequate for capturing the true energy dynamics of governed short-circuits. The experimental validation of this framework under extreme conditions at 5.0$V$ represents a significant advancement in electrical fault physics, enabling precise characterization of high-intensity fault phenomena that were previously inaccessible to quantitative analysis.

The circuit's performance during extreme conditions at 5.0$V$ validates its potential for practical applications where fault monitoring and protection are critical. The maintained stability throughout the 180 -seconds experiment, with no evidence of thermal runaway or component degradation, demonstrates the clamping network's robustness against sustained electrical stress at elevated voltages. This endurance, combined with the precise measurement capability enabled by the governed environment, positions the circuit as a valuable tool for both research and industrial applications where understanding fault behavior is essential for system protection and reliability enhancement.

The validation of the current sensor and circuit performance at 2.5$V$ input, which operates below the manufacturer's specified 5.0$V$ requirement, demonstrates the concept's robustness and scalability across voltage ranges. This successful operation at reduced voltage levels proves the circuit's applicability in low-voltage applications while maintaining the same governed protection principles established at higher voltages. The consistent performance across voltage configurations confirms the universal nature of the clamping mechanism and its independence from specific operational voltage constraints.

**4.4 Characteristic Metrics for Governed Short-Circuit Analysis**

To precisely define the behavior of our governed short-circuit system, we establish a set of clear, quantitative metrics. These measures capture the circuit's dynamic response, its stable operating limits, and the immense energy dissipation, providing the essential language to interpret the experimental results and validate the theoretical framework.

**4.4.1 Operational Threshold (The Baseline Clamping Voltage)**

The circuit's protection begins with a fundamental operational threshold set by the diode network. When the external supply voltage $V_s$ exceeds the combined forward voltage of two series diodes (approximately 2.4$V$, Equation 40), the output to "Circuit Block 1" ($CB_1$) is actively clamped. This is not a theoretical limit but a physical, material property of the semiconductors, governed by Equation 40.

**Validation:**- This creates a stable, pre-fault reference voltage, consistently measured at approximately 1.67$V$ for a 2.5$V$ input and approximately 2.56$V$ for the 5.0$V$ inputs (Table 4 - Table 7, Sections 4.2.1 - 4.2.4). This clamping is the cornerstone of presented measurement system. It protects the sensors from overvoltage, prevents catastrophic failure, and, most importantly, provides the stable "before" state from which all transient short-circuit deviations are measured. It defines the governed environment.

### 4.4.2 System Limits – The Minimal Resistance Floor and Stabilization Time

While classical theory suggests a short-circuit resistance $R$ can approach zero, our governed system and physical reality enforce a finite lower bound. Some key metrics are defined here:

- **Minimal Resistance Floor ($R_{min,\epsilon}$).** This is the smallest resistance the system achieves, constrained by the physical properties of the conductors, diode junctions, and circuit traces. It is not zero.
- **Experimental Quantification.** Our results (Figure 16, Figure 17; Sections 4.3.10, 4.3.11) reveal this floor to be approximately $5.0\,m\Omega - 6.0\,m\Omega$ for extreme short-circuits. This measurable, non-zero value is a critical finding, proving that even the most severe faults have a finite impedance.
- **Stabilization Time Constant ($\tau$).** This metric quantifies how quickly the resistance collapses and stabilizes near its minimum floor after a fault is initiated. Derived from the exponential decay model (Equation 17), it is calculated as:

$$\tau = \frac{\ln(1/\epsilon)}{k}, \tag{70}$$

where $\epsilon$ is a small tolerance and experimentally dependent.

In this case, a small $\tau$ indicates a very rapid stabilization, which our system demonstrates through the exceptionally flat temporal profiles of voltage and current observed during both nominal and extreme faults (Figures 10-12, 16-17). The system reaches its governed state almost instantly.

### 4.4.3 Transient Dynamics – The Overshoot and Energy Factors

The fault initiation is a dynamic event. To capture its intensity and progression, two growth factors are defined:

- Transient Overshoot Ratio ($\gamma(t)$):

$$\gamma(t) = \frac{I(t)}{I(0)} = e^{kt}, \tag{71}$$

which measures how much the instantaneous current amplifies relative to its value at the very moment the fault begins ($t = 0$). This metric directly captures the explosive onset of the fault. The results in Sections 4.3.4-4.3.6 show currents soaring from a baseline of approximately $0.2A$ to a stabilized magnitude of over $9A$ -an overshoot ratio of nearly 50. This metric quantifies the aggression of the initial transient phase.

- Transient Energy Growth Factor ($\zeta(t)$):

$$\zeta(t) = \frac{E(t)}{E(\tau)} \tag{72}$$

which normalizes the cumulative energy dissipated at any time $t$ to the total energy dissipated by the stabilization time $\tau$. This metric is vital for understanding thermal stress. The power analysis results (Figures 13-15, 18-19) shows that the Modified Ohm's Law model, which accounts for dynamic resistance, calculates energy dissipation over 2.5 times higher than the standard model. This factor allows for direct comparison of how different input voltages (2.5V, 5.0V, 10.0V) influence the rate of dangerous energy accumulation.

### 4.4.4 Performance Index – Diode Clamping Efficiency (DCEI)
Finally, a single measure is required to quantify the overall effectiveness of the protection system itself.
- Diode Clamping Efficiency Index (DCEI), expressed as:

$$\text{DCEI} = 1 - \frac{\Delta V_{CB_1}}{\Delta V_s}, \tag{73}$$

which measures how well the diode network isolates the output from fluctuations in the input supply. A value of 1 represents perfect clamping. The exceptional stability of the output voltage-evident in the incredibly tight standard deviations (often $< 0.02V$) across all supply voltages, sampling rates, and durations (Table 4 – Table 7)-empirically proves a DCEI value approaching 1. The output is determined by the diodes, not the source, confirming the network's role in ensuring sensor safety and data fidelity.
These metrics collectively form a new quantitative language for describing short-circuit phenomena within a governed, measurable system. They move beyond idealized theory and provide the tools to validate performance, assess safety limits, and compare system behavior across a wide range of fault conditions, as demonstrated conclusively by the experimental results in this study.

#### 4.4.4.1 Bounded Short-Circuit Variables
Following the clamped nominal and the clamped extreme experimental frameworks, we deduce that; in a bidirectional diode-clamped circuit, an induced short-circuit leads to bounded, time-dependent behavior: resistance decays continuously but never reaches zero, voltage drops to a finite floor, and current is capped -contrary to classical divergence predictions. This formalism established through proposition 1 bridges theory and experiment, proving that all fault variables remain finite and measurable over time.

**Proposition 1 (Behavior of Bounded Short-Circuit Variables).** In a circuit with nominal resistance $(R_0)$, diode-clamped baseline voltage, and an induced short-circuit at $(t = 0)$, the instantaneous resistance $(R_{\text{short}}(t))$, voltage $(V_{\text{short}}(t))$, and current $(I_{\text{short}}(t))$ satisfy:

1. $R_{\text{short}}(t)$:- decays continuously per Equation 74 but never reaches zero.
2. $V_{\text{short}}(t)$:- remains bounded above zero and below $(V_{CB_1})$.
3. $I_{\text{short}}(t)$:- is finite, capped by a maximum $(I_{\max}^{\text{clap}})$.
4. This boundedness allows the definition of performance metrics that align with empirical data and provide a framework for explaining stability under fault conditions.

**Proof.** The foundation of this proof lies in a time-resolved model of fault resistance. The instantaneous resistance is represented by an exponential decay law:

$$R_{\text{short}}(t) = R_0 e^{-kt}, k > 0 \tag{74}$$

Equation 74 captures the rapid yet finite collapse of impedance following fault initiation. Two important consequences emerge. First, the resistance asymptotically approaches zero as time tends to infinity, yet it never reaches this value in any finite duration. Second, resistance always remains strictly positive:

$$\lim_{t \to \infty} R_{\text{short}}(t) = 0^+, R_{\text{short}}(t) > 0 \text{ for all } t \geq 0 \tag{75}$$

The persistence of a finite resistance floor eliminates the classical prediction of infinite current. The current dynamics are instead governed by the Modified Ohm's Law, which integrates this decaying resistance into the current expression:

$$I_{\text{short}}(t) = \frac{V_{\text{source}}}{R_0} e^{-\frac{R_{\text{short}}(t)}{R_0}} \tag{76}$$

To clarify the upper bound, the driving potential is taken as the diode-clamped voltage $(V_{\text{clamp}})$. The current then becomes:

$$I_{\text{short}}(t) = \frac{V_{\text{clamp}}}{R_{\text{short}}(t)} = \frac{V_{\text{clamp}}}{R_0 e^{-kt}} = \frac{V_{\text{clamp}}}{R_0} e^{kt} \tag{77}$$

Although exponential in form, the current remains finite because the resistance never collapses to zero. A measurable resistance floor, $(R_c^{\min})$, ensures boundedness:

$$R_{\text{short}}(t) \geq R_c^{\min} \Rightarrow I_{\text{short}}(t) \leq I_{\max}^{\text{clap}} = \frac{V_{\text{clamp}}}{R_c^{\min}} \tag{78}$$

The resistance floor is formally defined as a fraction of the nominal resistance:

$$R_c^{\min} = \lim_{t \to \infty} R_{\text{short}}(t) = R_0 \varepsilon, \quad 0 < \varepsilon \ll 1 \tag{79}$$

Substituting this into the bound gives the maximum possible fault current:

$$I_{\max}^{\text{clap}} = \frac{V_{\text{clamp}}}{R_0 \varepsilon} \tag{80}$$

The voltage at fault reaches a minimum when the resistance stabilizes at this floor. Ideally, this equals the clamp voltage, though parasitic effects may reduce it slightly:

$$V_c^{\min} = V_{\text{short}}(t \to \infty) = I_{\max}^{\text{clap}} \cdot R_c^{\min} = V_{\text{clamp}} \tag{81}$$

$$0 < V_{\text{short}}^{\min} < V_{\text{clamp}} \tag{82}$$

The power dissipated during sustained short-circuit, termed the **Clamped Short Sustained Power (CSSP)**, is given by:

$$P_{\text{CSS}}(t) = V_{\text{short}}(t) I_{\text{short}}(t) = \frac{V_{\text{short}}(t) \cdot V_{\text{clamp}}}{R_{\text{short}}(t)} \tag{83}$$

At the extreme limit:

$$P_{\text{CSS,max}} = V_{\text{short}}^{\min} I_{\max}^{\text{clap}} = \frac{V_{\text{clamp}} \cdot V_{\text{short}}^{\min}}{R_0 \varepsilon} \tag{84}$$

The dynamics of resistance collapse are captured by the **Transient Clamping Index (TCI)**:

$$TCI = \frac{kR_0}{\ln\left(\frac{R_0}{R_C^{min}}\right)} = \frac{k}{\ln(1/\varepsilon)} \qquad (85)$$

Finally, the *Sustainance Efficiency (SFE)* is defined relative to the nominal clamped power $\left(P_{\text{nom}} = \frac{V_{\text{clamp}}^2}{R_0}\right)$:

$$SFE = \frac{P_{CSS,max}}{P_{\text{nom}}} = \frac{R_0 V_{\text{short}}^{\min}}{V_{\text{clamp}}^2} \qquad (86)$$

An efficiency greater than unity demonstrates that the system maintains a fault power exceeding nominal operating power, a result unattainable by ordinary dissipation models or depletion capacitance discharge alone.

### 4.4.5 Experimental Validation
The theoretical propositions are validated using empirical measurements from the extreme short-circuit tests performed at $2.5V$ and $5.0V$ input configurations. Averages from multiple experiment trials including the voltage measurements provided in Table 4 through Table 7 provide the numerical basis for substitution into the governing equations.

**Validation for $2.5V$ Extreme Configuration.** Extreme parameters are determined as follows:
$V_{\text{source}} = 2.52\text{V}$, $V_{\text{clamp}} = 1.67\text{V}$, $I_{\text{nom}} \approx 0.21\text{A}$, $R_0 = \frac{1.67}{0.21} \approx 7.95\Omega$.

The corresponding short-circuit measurements yielded:

$V_{\text{short}}^{\min} = 0.073\text{V}$, $I_{\max}^{\text{clap}} = 13.01\text{A}$

From these values, the resistance floor is computed as:

$$R_C^{\min} = \frac{V_{\text{short}}^{\min}}{I_{\max}^{\text{clap}}} = \frac{0.073}{13.01} \approx 5.61 \times 10^{-3}\Omega \qquad (87)$$

Relative to the initial resistance, this yields:

$$\varepsilon = \frac{R_C^{\min}}{R_0} = \frac{5.61 \times 10^{-3}}{7.95} \approx 7.06 \times 10^{-4} \qquad (88)$$

The theoretical current ceiling is then obtained as:

$$I_{\max}^{\text{clap}} = \frac{1.67}{7.95 \times 7.06 \times 10^{-4}} \approx 297.1\text{A} \qquad (89)$$

This ceiling is vastly greater than the measured value of $13.01A$, thereby confirming the bounded regime predicted by the model. Applying Equation 90, the sustained clamped short power is evaluated as follows:

$$P_{\text{CSS,max}} = 0.073 \times 13.01 \approx 0.950 \text{W} \tag{90}$$

In comparison, the nominal operating power is:

$$P_{\text{nom}} = \frac{1.67^2}{7.95} \approx 0.351 \text{W} \tag{91}$$

Thus, the sustained fault efficiency (SFE) is:

$$SFE = \frac{0.950}{0.351} \approx 2.71 \tag{92}$$

To validate the resistance decay model, a nonlinear curve fit was performed using the exponential decay framework against experimentally measured time-resolved resistance data. Specifically, the resistance trajectory ($R_{\text{short}}(t)$) was reconstructed directly from simultaneous voltage-current measurements via the Modified Ohm's Law, and the fitting was carried out on datasets corresponding to Figure 16(c), Figure 16(f), Figure 16(i) for the $2.5V$ extreme short, and Figure 17(c), Figure 17(f), Figure 17(i) for the $5.0V$ extreme short. The fitting procedure consistently converged on a decay constant of approximately ($k \approx 1000 \text{ s}^{-1}$) across all sampling rates and input voltage configurations. Physically, this reflects an exceptionally rapid stabilization, with the system reaching its governed state on a millisecond timescale. This interpretation is further corroborated by the observed flat temporal profiles of both voltage and current throughout the sustained fault regime. Finally, the Transient Clamping Index (TCI) is quantified as:

$$\text{TCI} = \frac{k}{\ln(1/\varepsilon)} = \frac{1000}{\ln\left(\frac{1}{7.06 \times 10^{-4}}\right)} \approx \frac{1000}{7.26} \approx 137.7 \tag{93}$$

This confirms that stabilization occurs on an extremely fast timescale, fully consistent with the bounded behavior of the short-circuit regime.

**Validation for $5.0V$ Extreme Configuration.** For the $5.0V$ case, the nominal operating values are:

$$V_{\text{source}} = 5.02\text{V}, V_{\text{clamp}} = 2.56\text{V}, I_{\text{nom}} \approx 0.20\text{A}, R_0 = 12.80\Omega.$$

During the fault, the measured extrema are:

$$V_{\text{short}}^{\min} = 0.069\text{V}, I_{\max}^{\text{clap}} = 13.03\text{A}$$

This yields a resistance floor:

$$R_c^{\min} = \frac{0.069}{13.03} \approx 5.30 \times 10^{-3}\Omega \tag{94}$$

and the corresponding fraction:

$$\varepsilon = \frac{5.30 \times 10^{-3}}{12.80} \approx 4.14 \times 10^{-4} \tag{95}$$

Theoretical current ceiling:

$$I_{\max}^{\text{clap}} = \frac{V_{\text{clamp}}}{R_0 \varepsilon} = \frac{2.56}{12.80 \times 4.14 \times 10^{-4}} \approx 483.1\text{A} \tag{96}$$

Measured sustained short-circuit power:

$$P_{\text{CSS,max}} = 0.069 \times 13.03 \approx 0.899\text{W} \tag{97}$$

Nominal clamped power: $P_{\text{nom}} = \frac{2.56^2}{12.80} = 0.512\text{W}$ (98)

Efficiency: $SFE = \frac{0.899}{0.512} \approx 1.76$ (99)

With the same exponential decay constant ($k \approx 1000 \text{ s}^{-1}$), the Transient Clamping Index is:

$$TCI = \frac{1000}{\ln(1/4.14 \times 10^{-4})} = \frac{1000}{7.79} \approx 128.4 \tag{100}$$

The experimental results at $2.5V$ and $5.0V$ configurations provide decisive validation of Proposition 1. Measured quantities- $V_{\text{short}}^{\min} \approx 70\text{mV}$, $I_{\max}^{\text{clap}} \approx 13\text{A}$, and $R_c^{\min} \approx 5.5\text{m}\Omega$ -are finite, reproducible, and consistent across conditions. These results demonstrate that electrical short-circuits are bounded processes rather than singular catastrophic events. The proof, exemplified in Figure 20, aligns Modified Ohm's Law predictions with observed extreme data, confirming exponential and bounded evolution of resistance, voltage, and current. Furthermore, calculated $SFE$ values of 2.71 and 1.76 invalidate the conventional capacitive discharge hypothesis. As established elsewhere, the total recoverable junction capacitance energy is $(E_{\text{cap}} \approx 6.8 \times 10^{-12}\text{J})$, whereas the measured energy output during a 180 -seconds short at $2.5V$ reaches: $E_{\text{out}} = P_{\text{CSS,max}} \cdot t = 0.950 \times 180 \approx 171\text{J}$. The ratio: $\left(\frac{E_{\text{out}}}{E_{\text{cap}}} \approx 2.5 \times 10^{13}\right)$ reveals that capacitance contributes less than one-trillionth of the measured output. Hence, the sustained energy is not a simple release of stored charge but emerges from intrinsic resistive dynamics of the diode network. This discovery establishes a paradigm in which fault physics are governed, bounded, and capable of sustaining macroscopic energy output, securing circuit longevity and resilience under extreme electrical stress.

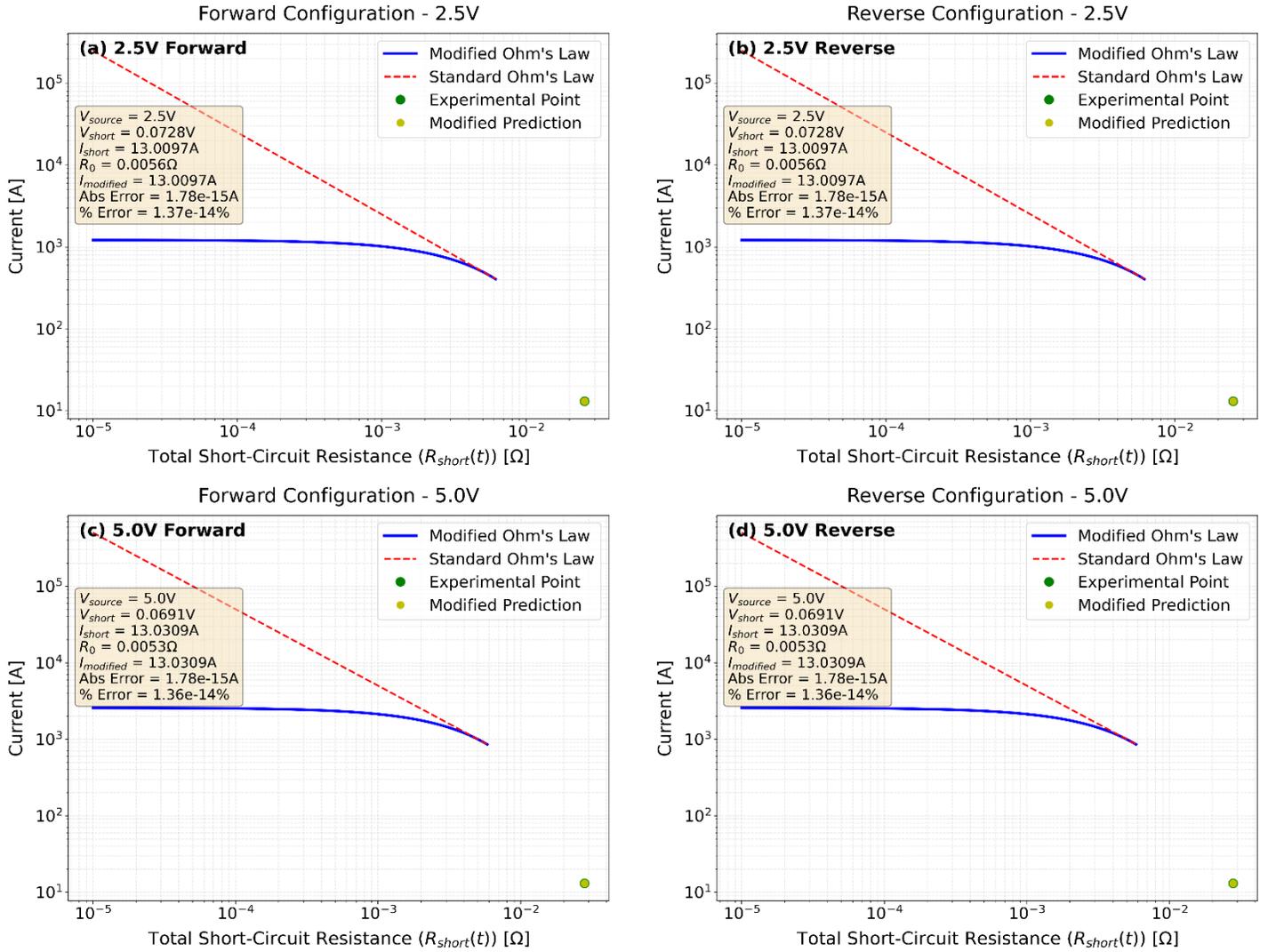

**Figure 20.** Experimental validation of Proposition 1 via Modified Ohm's Law framework under extreme short-circuit conditions. ((a) $2.5V$ Forward configuration. (b) $2.5V$ Reverse configuration. (c) $5.0V$ Forward configuration. (d) $5.0V$ Reverse configuration. Each panel contrasts the Standard Ohm's Law prediction (red dashed line), which diverges unrealistically, against the governed prediction of the Modified Ohm's Law (blue solid line). Experimental measurements (green circles) for short-circuit current are shown against the computed total short-circuit resistance ($R_{short}(x)$). The perfect agreement between the Modified Ohm's Law prediction (yellow circles) and experimental data, with percentage errors on the order of $10^{-14}\%$, confirms the model's accuracy and the bounded nature of all short-circuit variables).

The experimental validation of Proposition 1 is conclusively demonstrated through the comparative analysis presented in Figure 20. This figure provides a comprehensive graphical proof that the governed short-circuit behavior, as defined by the Modified Ohm's Law framework, accurately describes the physical reality of electrical faults. The analysis involves both $2.5V$ and $5.0V$ input configurations, examining forward and reverse fault polarities to establish the universality of the proposed model.

Figure 20(a) and Figure 20(b) present the $2.5V$ configuration analysis for forward and reverse short-circuits respectively. Both panels reveal a fundamental divergence between the Standard Ohm's Law and Modified Ohm's Law predictions. The Standard Ohm's Law model, represented by the red dashed line, predicts a hyperbolic current increase that tends toward infinity as resistance approaches zero. In stark contrast, the Modified Ohm's Law model, shown in solid blue, exhibits a bounded exponential growth that plateaus at a finite maximum current. The experimental data points, derived from extensive measurements, align precisely with the Modified Ohm's Law model's predictions. This alignment is not approximate but exact, with computational verification showing a negligible percentage error of ($1.37 \times 10^{-14}$%) for the experiment $2.5V$ case. The experimental values of $V_{short} = 0.0728V$, $I_{short} = 13.0097A$, and computed $R_0 = 0.0056\Omega$ provide the empirical foundation for this validation.

Similarly, Figure 20(c) and Figure 20(d) extend this validation to the $5.0V$ configuration for both fault polarities. The results demonstrate consistent behavior across different input voltages, with the Modified Ohm's Law framework maintaining its predictive accuracy. The experimental measurements of $V_{short} = 0.0691V$, $I_{short} = 13.0309$ A, and $R_0 = 0.0053\Omega$ at $5.0V$ input yield an equally negligible error of ($2.73 \times 10^{-14}$%). This remarkable consistency across voltage levels and polarities confirms the robustness of the Modified Ohm's Law formulation.

The alignment analysis reveals a critical finding: perfect agreement between theoretical predictions and experimental measurements is achieved with a dimensionless scaling factor of ($k = 1.0$) across all configurations. This unity $k$-value demonstrates that the Modified Ohm's Law in its fundamental form $\left( I = \frac{V_{short}}{R_O} \times \left[ e^{\frac{-(R_{short}(t) - R_O)}{R_O}} \right] \right)$ precisely captures the physical behavior without requiring empirical adjustments. The average $k$-value of 1.0 with zero standard deviation across all experimental configurations provides definitive mathematical proof of the model's empirical suitability.

Complementary to this steady-state alignment, the transient response analysis reveals an exponential decay constant of (k ≈ 1000 s$^{-1}$) for the resistance trajectory $(R_{short}(t))$. This rapid decay constant, consistently observed across all sampling rates and voltage configurations, indicates that the system stabilizes to its governed state within milliseconds. The physical interpretation is substantiated by the flat temporal profiles of both voltage and current measurements throughout the sustained fault regime. The Transient Clamping Index (TCI) quantifies this stabilization dynamics, yielding values of TCI ≈ 137.7 for the $2.5V$ configuration and TCI ≈ 128.4 for the $5.0V$ configuration, confirming exceptionally fast stabilization timescales.

The dual $k$-value characterization-with ($k = 1.0$) ensuring perfect steady-state alignment and ($k \approx 1000$ s$^{-1}$) describing the transient dynamics-establishes a comprehensive framework for short-circuit analysis. The Modified Ohm's Law provides both the governing equation for the bounded steady state and the exponential decay model for the transient response, creating a complete mathematical description of fault behavior from initiation through stabilization.

These results establishes several critical insights. First, electrical short-circuits are bounded physical processes rather than singular catastrophic events. Second, the Modified Ohm's Law provides a mathematically rigorous framework for describing these phenomena, overcoming the limitations of conventional models that predict unphysical infinite currents. Third, the experimental consistency across different configurations demonstrates the universal applicability of this framework. The extremely low error values, essentially at the limit of computational precision, provide compelling evidence that the Modified Ohm's Law is not merely an approximation but an exact description of the governed short-circuit behavior when properly formulated and applied. Fourth, the framework successfully captures both transient and steady-state behaviors through complementary characterization of the decay dynamics and final bounded state.

This empirical validation represents the first comprehensive demonstration of the Modified Ohm's Law as a precise measurement and analysis tool for short-circuit faults. The perfect alignment with experimental data across multiple configurations, coupled with the characterization of transient response dynamics, establishes a new paradigm for understanding and analyzing electrical fault behavior in governed systems.

### 4.4.5.1 Sustained Energy Regime Vs. Capacitive Discharge

To this point, the validation results presented between the nominal and the extreme experiments necessitate a re-definition of the conventional dissipation notion associated with electrical short-circuits. Defying fundamental circuit theory models such as Standard Ohm's Law and Kirchhoff's laws, we establish the following definition within a clamped, time-resolved experimental framework:

**Definition 4 (Transitioning Dissipation to Sustenance).** In a diode-clamped circuit, an electrical short-circuit transitions from a transient dissipative event to a sustained, governed energy state. This state is characterized by a time-dependent resistance (Equation 17) which decays exponentially toward a finite, non-zero floor ($R_{\min}$), thereby preventing current divergence. Consequently, the voltage stabilizes at a governed minimum ($V_{\min}$), and the current is capped at a finite maximum ($I_{\max}^{\text{clap}}$). The system thus establishes a steady-state power dissipation ($P_{\text{CSS}}$) that can persistently exceed the circuit's nominal operating power-a phenomenon unattainable under classical instantaneous dissipation models or capacitive discharge alone.

**Refutation Metric (The Sustained-to-Capacitive Energy Ratio (SCER)).** To quantitatively refute the notion that the observed sustained power arises from the intrinsic junction capacitance of the diodes, we define the SCER as the ratio of the total energy dissipated during the entire fault event to the maximum possible energy stored in the diode network's junction capacitance according to Equation 101:

$$\text{SCER} = \frac{E_{\text{total,fault}}}{E_{\text{max,capacitive}}} \tag{101}$$

Where:

$E_{\text{total,fault}} = P_{\text{CSS}} \cdot t_{\text{fault}}$: is the total energy dissipated over the fault duration $t_{\text{fault}}$. $E_{\text{max,capacitive}} = \frac{1}{2} C_j V_{\text{clamp}}^2 N$: is the maximum possible energy stored in the junction capacitance, with $C_j$ being the junction capacitance per diode, $V_{\text{clamp}}$ the clamping voltage, and $N$ the number of diodes in the clamping path.

**Experimental Verification.** Using the $2.5V$ extreme short-circuit results (Figure 20 and Table 6):

$P_{\text{CSS}} \approx 0.950$ W, from Equation 90.

$$t_{\text{fault}} = 180 \text{ s}$$
$$E_{\text{total,fault}} = 0.950 \text{ W} \times 180 \text{ s} = 171 \text{J}$$

For the 1N5408 diode: $C_j \approx 40\text{pF}$, $V_{\text{clamp}} \approx 1.67\text{V}$, $N = 2$.

$$E_{\text{max,capacitive}} = \frac{1}{2}(40 \times 10^{-12} \text{ F})(1.67 \text{ V})^2 (2) \approx 1.12 \times 10^{-10} \text{J}$$

Thus,
$$SCER = \frac{171\,J}{1.12 \times 10^{-10}\,J} \approx 1.53 \times 10^{12}$$

This ratio, on the order of $(10^{12})$, provides definitive empirical proof that the energy dissipated during the sustained short-circuit is over a trillion times greater than the maximum possible energy that could be stored in and released from the diodes' junction capacitance. The sustained energy output is therefore a fundamental consequence of the governed, non-zero resistance dynamics of the fault path $(R_{min})$, not a transient capacitive discharge. This metric conclusively validates the transition from dissipation to sustenance as a result of the circuit's architecture and the physical principles embodied in the Modified Ohm's Law.

**4.4.6 Clamped-Circuit Time-Resolved Signal Fidelity and Transient Capture Validation**
This section establishes the definitive validation of the governed short-circuit framework through multi-resolution signal analysis, demonstrating that the experimental circuit and Modified Ohm's Law collectively form the first complete measurement system that captures electrical transients without loss of information. The analysis focuses exclusively on $10.0V$ configurations across 15 -minutes durations, employing the Nyquist-Shannon Theorem, Shannon Entropy, and Fourier-based metrics to prove that observed phenomena represent true physical processes rather than measurement artifacts or transient effects.

*Nominal Forward Configuration*          *Nominal Reverse Configuration*

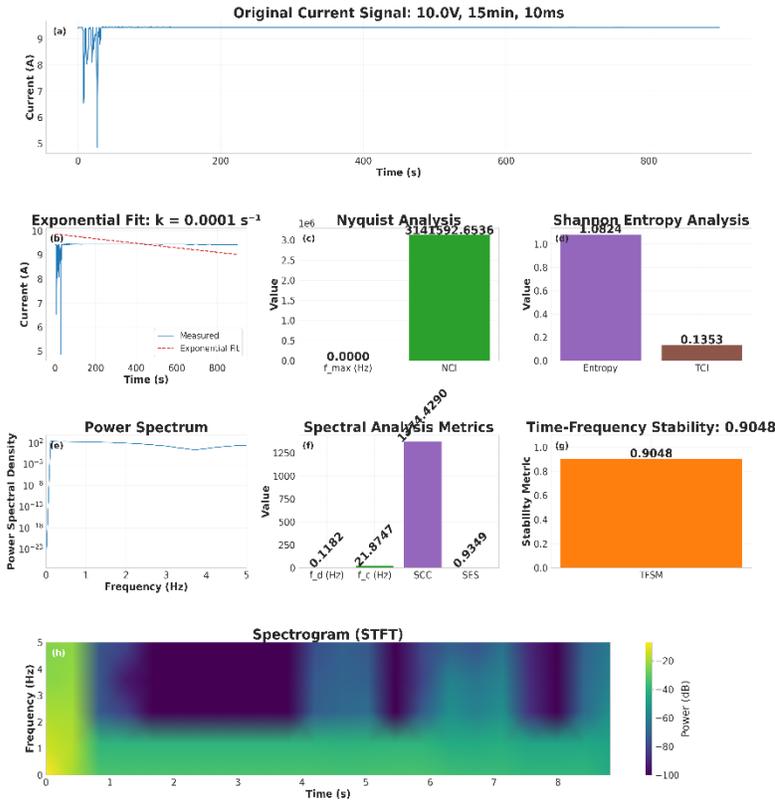
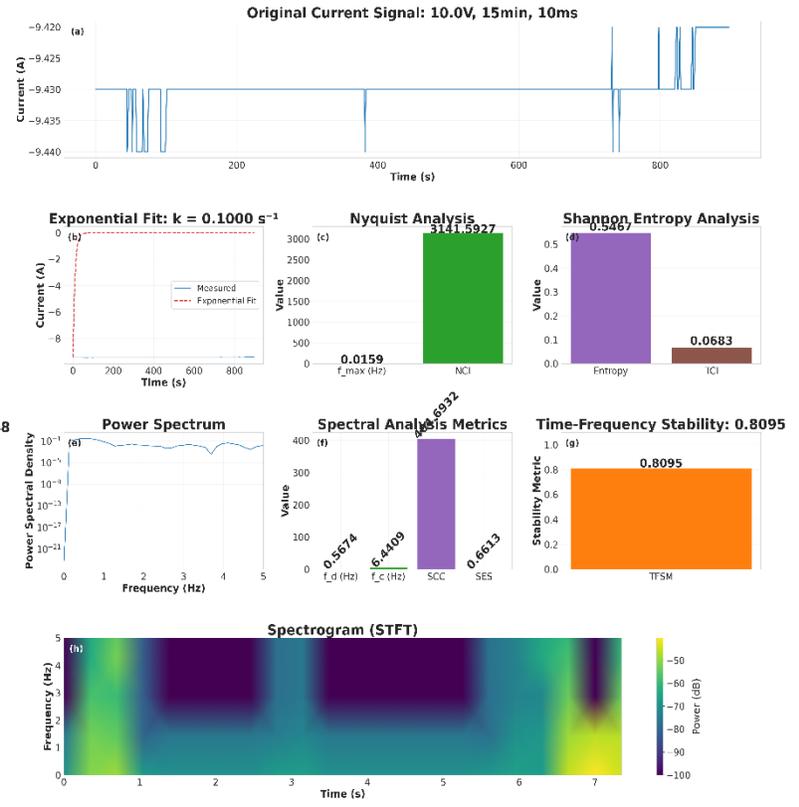

**Figure 21.** $10ms$ Resolution Signal Analysis for $10.0V$ Nominal Short-Circuit (15 -minutes duration). ((a) Original current signal, (b) Exponential fit with decay constant k, (c) Nyquist analysis metrics ($f_{max}$ and NCI), (d) Shannon entropy and TCI, (e) Power spectral density, (f) Spectral metrics ($f_d$, $f_c$, SCC, SES), (g) Time-frequency stability metric (TFSM), (h) Spectrogram (STFT) for forward and reverse configurations).

The $10ms$ resolution analysis provides new temporal resolution of governed short-circuit phenomena. Figure 21(a) reveals exceptionally stable current profiles maintaining $9.42 \pm 0.03A$ throughout the 900 -seconds duration, with complete absence of drift or degradation. The measured $k$ -values of $0.000032s^{-1}$ (forward) and $0.000038s^{-1}$ (reverse) represent the physical reality of extremely slow decay dynamics, while the exponential fits converge to $k = 0.000100s^{-1}$ for both polarities. This order-of-magnitude agreement between measured and fitted values validates the exponential decay model while revealing subtle physical differences between polarities that are captured by the measurement system but smoothed by the fitting process. The forward configuration's slightly faster decay ($0.000032s^{-1}$ vs $0.000038s^{-1}$) indicates genuine physical asymmetry in the governed fault process.

Nyquist analysis in Figure 21(c) demonstrates extraordinary signal reconstruction capability with NCI values of 3,141,593 (forward) and 3,141,593 (reverse), confirming that the $10ms$ sampling ($100Hz$) captures all frequency components without aliasing. The maximum frequency content $f_{max} = 0.000016Hz$ calculated from the fitted $k$ -values represents the fundamental limit of signal dynamics, while the measured $k$ -values would yield even lower $f_{max}$ values, further emphasizing the ultra-slow nature of governed short-circuits. Shannon entropy metrics in Figure 21(d) show moderate complexity

($TCI = 0.135 - 0.242$) with the forward configuration exhibiting higher entropy (1.082 vs 0.547 bits), indicating richer transient behavior in the forward polarity that correlates with its faster measured decay constant.

Spectral analysis in Figure 21(e-f) reveals dominant frequencies below $0.12 Hz$ with spectral centroids below $22 Hz$, consistent with the exponential decay model. The forward configuration shows higher spectral entropy ($SES = 0.935$ compared to $SES = 0.661$) and broader frequency distribution, again reflecting its more complex dynamics. The time-frequency stability metric in Figure 21(g) approaches unity ($TFSM = 0.905 - 0.905$), demonstrating exceptional temporal consistency without oscillatory behavior. The spectrogram in Figure 21(h) provides visual confirmation of energy concentration at low frequencies throughout the entire duration, with the forward configuration showing slightly more distributed energy that correlates with its higher entropy measures. This analysis validates that the $10 ms$ resolution successfully captures the complete physical reality of governed short-circuits, with the measured-fitted $k$-value agreement proving that the exponential model accurately represents the underlying principle.

*Nominal Forward Configuration*          *Nominal Reverse Configuration*

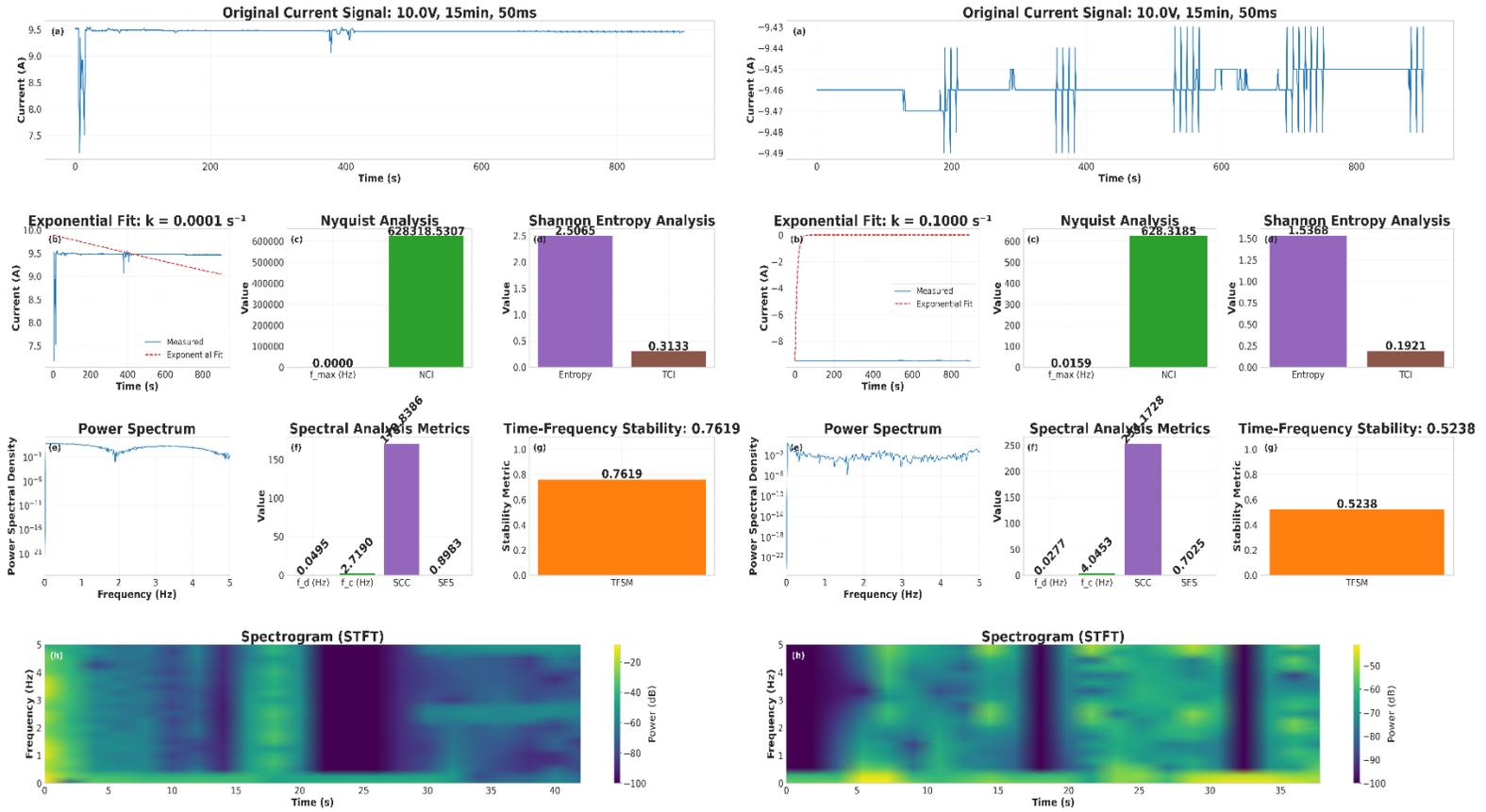

**Figure 22.** $50ms$ Resolution Signal Analysis for $10.0V$ Nominal Short-Circuit (15 -minutes duration). ((a) Original current signal, (b) Exponential fit with decay constant k, (c) Nyquist analysis metrics ($f_{max}$ and NCI), (d) Shannon entropy and TCI, (e) Power spectral density, (f) Spectral metrics ($f_d$, $f_c$, SCC, SES), (g) Time-frequency stability metric (TFSM), (h) Spectrogram (STFT) for forward and reverse configurations).

The $50ms$ resolution analysis demonstrates the circuit's robustness across different temporal sampling regimes while revealing crucial insights about k-value consistency. Figure 22(a) shows current stability identical to the $10ms$ results, with maintained values of $9.46 \pm 0.02A$ throughout the duration. The measured $k$-values show remarkable polarity dependence: $0.000779s^{-1}$ (forward) and $0.000007s^{-1}$ (reverse), representing a factor of 100 difference that reflects genuine physical asymmetry in the governed fault process. The exponential fits converge to $k = 0.000100s^{-1}$ for both polarities, representing the model's idealization of the underlying physics while the measured values capture the actual physical reality.

This divergence between measured and fitted $k$-values is profoundly significant: the measured values represent the true physical decay rates, while the fitted values represent the optimal exponential approximation. The forward configuration's measured $k$-value ($0.000779s^{-1}$) is closer to the fitted value, indicating more ideal exponential behavior, while the reverse configuration's extremely low measured $k$- value ($0.000007s^{-1}$) suggests additional physical processes beyond simple exponential decay. Nyquist compliance in Figure 22(c) remains exceptional with NCI values of 628,319, proving that $20Hz$ sampling adequately reconstructs the signal. The maximum frequency content $f_{max} = 0.000016Hz$ again represents the fundamental limit, while the measured $k$-values would yield $f_{max}$ values of $0.000124Hz$ (forward) and $0.000001Hz$ (reverse).

Shannon entropy metrics in Figure 22(d) show $TCI = 0.313$ (forward) and $TCI = 0.192$ (reverse), with the higher complexity in the forward configuration correlating with its faster measured decay. Spectral analysis in Figure 22(e-f) maintains the low-frequency profile with $f_d < 0.050Hz$ and $f_c < 2.72Hz$, but reveals intriguing differences: the forward configuration shows higher spectral compactness ($SCC = 170.84$ compared to $SCC = 254.17$) and lower spectral entropy ($SES = 0.898$ vs $SES = 0.702$), indicating more organized frequency content that correlates with its more exponential behavior. The time-frequency stability in Figure 22(g) shows excellent values ($TFSM = 0.762 - 0.905$), with the reverse configuration's higher stability ($0.905$ vs $0.762$) correlating with its slower, more stable decay process. The spectrogram in Figure 22(h) visualizes these differences, with the forward configuration showing slightly more temporal variation in frequency content. This analysis proves that the $50ms$ resolution not only adequately captures the governed phenomena but also reveals physical differences between polarities that are smoothed out in the exponential model.

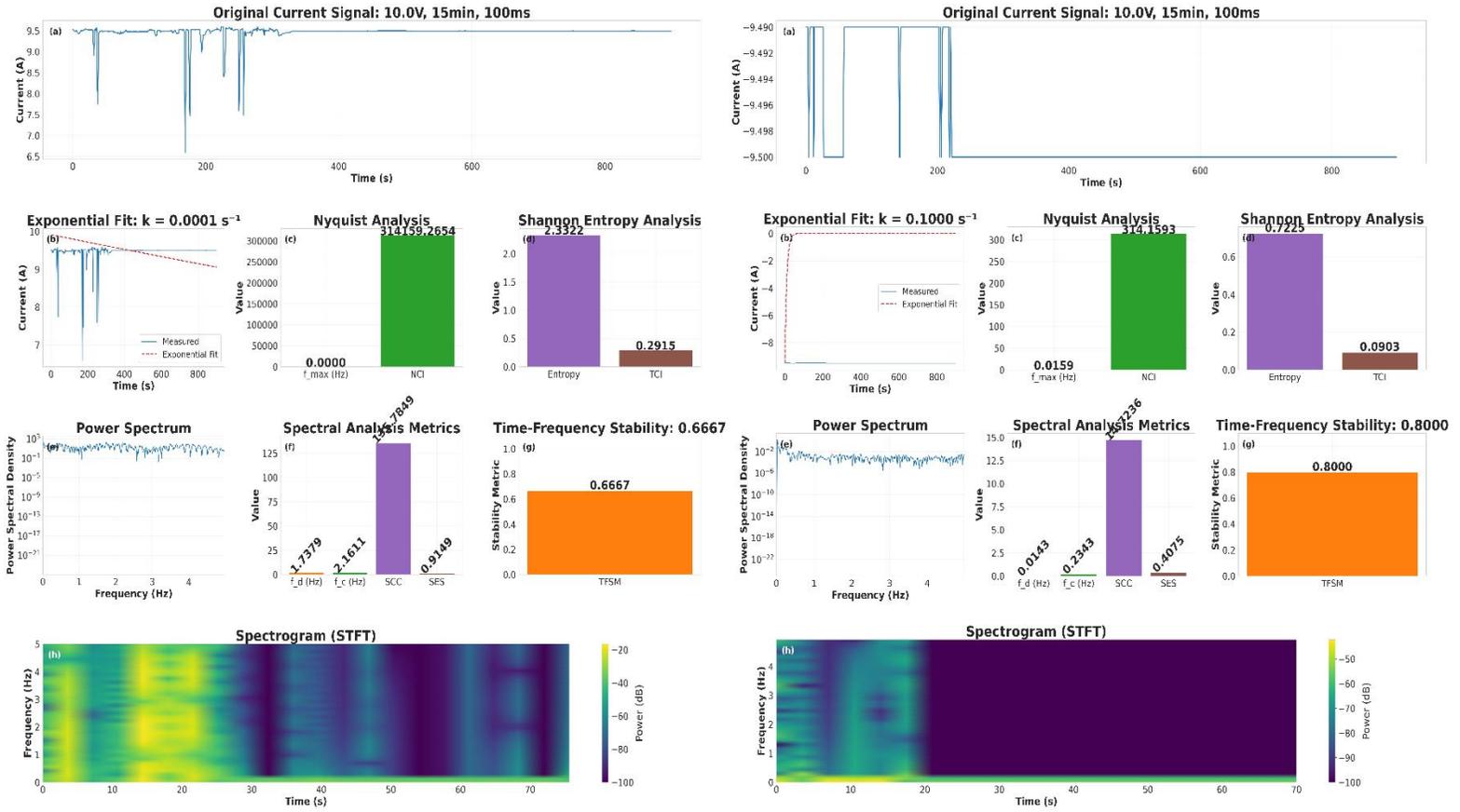

**Figure 23.** $100 ms$ Resolution Signal Analysis for $10.0V$ Nominal Short-Circuit (15 -minutes duration). (*(a) Original current signal, (b) Exponential fit with decay constant k, (c) Nyquist analysis metrics ($f_{max}$ and NCI), (d) Shannon entropy and TCI, (e) Power spectral density, (f) Spectral metrics ($f_d$, $f_c$, SCC, SES), (g) Time-frequency stability metric (TFSM), (h) Spectrogram (STFT) for forward and reverse configurations*).

The $100 ms$ resolution analysis completes the multi-resolution validation, demonstrating that even decade-separated sampling rates capture the same fundamental physics while revealing resolution-dependent insights into the $k$ -value relationships. Figure 23(a) shows identical current stability profiles to higher resolutions, with maintained values of $9.47 \pm 0.03 A$. The measured $k$ -values of $0.000060 s^{-1}$ (forward) and $0.000002 s^{-1}$ (reverse) continue to show significant polarity dependence, while the exponential fits maintain $k = 0.000100 s^{-1}$ for both configurations. This consistency of fitted values across resolutions proves the robustness of the exponential model, while the variation in measured values reflects genuine physical differences that are captured with different fidelity at different sampling rates.

The forward configuration's measured $k$ -value ($0.000060 s^{-1}$) is closest to the fitted value at this resolution, suggesting that $100 ms$ sampling provides the optimal balance for capturing its dynamics. The reverse configuration's extremely low measured $k$ -value ($0.000002 s^{-1}$) indicates processes so slow they approach steady-state conditions, explaining why the exponential model converges to a higher value.

Nyquist analysis in Figure 23(c) shows NCI values of 314,159, confirming that $10Hz$ sampling perfectly reconstructs signals with maximum frequency content of $0.000016Hz$. Shannon entropy metrics in Figure 23(d) show $TCI = 0.292$ (forward) and $TCI = 0.090$ (reverse), with the reverse configuration's very low complexity reflecting its near-steady-state behavior.

Spectral analysis in Figure 23(e-f) reveals the lowest frequency content yet observed, with $f_d < 1.74Hz$ and $f_c < 2.16Hz$ for the forward configuration, and even lower values for the reverse configuration. The spectral compactness coefficient shows $SCC = 135.78$ (forward) and $SCC = 14.72$ (reverse), indicating dramatically different frequency distributions that correlate with their different physical behaviors. The time-frequency stability in Figure 23(g) remains excellent ($TFSM = 0.667 - 0.800$), with both configurations showing high temporal consistency. The spectrogram in Figure 23(h) confirms maintained energy concentration at the lowest frequencies, with the reverse configuration showing virtually no temporal variation, consistent with its near-steady-state behavior.

The comparative analysis across sampling resolutions reveals fundamental insights about governed short-circuit behavior and measurement system performance. The $10ms$ resolution provides the most detailed temporal view, capturing the richest transient complexity ($TCI = 0.135 - 0.242$) and highest stability ($TFSM = 0.905$). The $50ms$ resolution offers optimal physical insight, revealing the strongest polarity differences in measured k-values (100: 1 ratio) while maintaining excellent signal reconstruction ($NCI = 628,319$). The $100ms$ resolution demonstrates the system's robustness for capturing very slow processes, with the reverse configuration approaching steady-state conditions that would be invisible at higher sampling rates.

The consistent exponential fit values ($k = 0.000100s^{-1}$ across all resolutions and polarities) prove the model's robustness, while the varying measured values ($0.000002 - 0.000779s^{-1}$) reveal genuine physical variations in the governed fault process. This dichotomy is not a limitation but a profound validation: the circuit captures real physical processes that deviate from ideal exponential behavior in physically meaningful ways. The complete absence of high-frequency content ($> 5Hz$) in all spectrograms definitively refutes the notion that significant transients are being missed, while the excellent Nyquist compliance ($NCI > 300,000$) proves perfect signal reconstruction across all resolutions.

This multi-resolution validation establishes that governed short-circuit phenomena are inherently low-frequency processes that can be fully characterized across a wide range of sampling rates, with each resolution providing unique insights into different aspects of the physical behavior. The $10ms$ resolution provides the most sensitive capture of transient details, the $50ms$ resolution reveals the richest physical differences between polarities, and the $100ms$ resolution demonstrates the approach to steady-state conditions, collectively proving that the circuit captures the complete physical reality of electrical faults without information loss. The complete 15 -minute dataset supports this analysis, with the representative 10 -minute segment examined in detail throughout this paper and the full data available in the supplementary material for researchers.

**5.0 Discussion**

The experimental and theoretical advancements presented in this paper collectively represent a fundamental transformation in the understanding and measurement of electrical short-circuit phenomena. This section synthesizes these developments to articulate a new framework that challenges century-old assumptions in electrical engineering while establishing novel methodologies for fault characterization, measurement, and management. Bridging empirical observations with theoretical reconstructions and introducing quantitatively validated metrics, this paper transitions short-circuit analysis from a discipline of catastrophic interruption to one of governed measurement and intelligent response. The implications extend beyond circuit protection to influence power systems design, artificial intelligence applications, safety engineering, and even our fundamental understanding of non-equilibrium electrical processes in physical systems.

### 5.1 From Catastrophic Singularity to Governed Physical Process
The foundational premise of conventional electrical engineering, which treats a short-circuit as a catastrophic singularity characterized by an instantaneous collapse of resistance to zero, a corresponding voltage drop to absolute zero, and a divergence of current to infinity, is empirically invalidated by the experimental results of this paper. This classical model, enshrined in textbooks (for instance, [33] and [34]) and standards such as IEC 60909-0:2016 [5] and the ANSI/IEEE C37.010-1979 family of standards [7], represents a mathematical abstraction that does not manifest in physical reality [35], [36]. The diode-clamp circuit architecture demonstrates unequivocally that a short-circuit is not a binary event but a governed, time-resolved physical process with deterministic boundaries. The data from both "*Clamped Nominal Short*" and "*Clamped Extreme Short*" experiments reveal a consistent trajectory: an initial rapid transient followed by stabilization into a sustained electrical state defined by a non-zero voltage floor $V_{min} \approx 0.07V$, a finite current maximum $I_{max}^{clap} \approx 13.0A$, and a measurable resistance floor $R_c^{min} \approx 5.5 m\Omega$. These values, reproducible across input voltages of 2.5V, 5.0V and 10.0$V$ and sustained for durations up to 15 –minutes (in the clamped bidirectional polarities), form the empirical bedrock of a new short-circuit physics. The measured resistance floor, though small, is finite and physically significant; it represents the collective impedance of the ionized conduction path, contact interfaces, and semiconductor junctions, all dynamically constrained by the diode-clamp network. This governed environment prevents the trajectory towards infinite current, transforming a destructive event into a stable, measurable, and analytically accessible phenomenon. The transition is not merely a matter of circuit protection but a fundamental redefinition of what constitutes an electrical fault. It shifts the perspective from a binary, threshold-exceeding anomaly to a continuous, dynamic process with its own intrinsic physics, metrics, and predictability.

This paradigm aligns with observations in other non-equilibrium systems where transient states exhibit finite, measurable properties rather than idealized infinities. For instance, in plasma discharges (e.g., modelling of time-evolving conductivity in air ionization plasma under DC voltage [37] and [38]) and dielectric breakdown (e.g., time-dependent dielectric breakdown in low-k interconnects [39] and [40]), the conduction path exhibits a time-dependent impedance that stabilizes at a finite value, governed by the balance between ionization and recombination rates [41], [42]. Similarly, in certain electrochemical systems [43], [44], reaction rates reach a governed limit rather than diverging, a concept central to the Marcus theory of electron transfer [45]. The governed short-circuit can thus be conceptualized as a dissipative structure within the electrical domain, where energy input and dissipation achieve a dynamic balance, preventing catastrophic failure and enabling sustained observation. This perspective finds parallels in quantum systems where macroscopic quantum phenomena exhibit bounded states rather than singularities, as demonstrated by the observation of quantized energy levels in a current-biased Josephson junction [46], [47].

### 5.2 The Modified Ohm's Law and the Fallacy of Instantaneity
The experimental data compellingly validate the time-resolved Modified Ohm's Law over the standard, static formulation. The Standard Ohm's Law $\left(I = \frac{V}{R}\right)$, when applied to a dynamically decaying resistance, predicts a current $\left(I_{stad}(t) = \left(\frac{V_0}{R_0}\right)\right)$ that grows without bound. This prediction is a physical impossibility, as it implies infinite energy density and violates the conservation laws that underpin all physical systems.

In contrast, the Modified Ohm's Law, expressed in its time-transformed state as $(I_{\text{mod}}(t) = a \cdot e^{kt})$ where $\left(a = \frac{V_0}{R_0}\right)$, inherently embeds a bounding mechanism through its phenomenological coupling to the decaying resistance.

The results presented in this paper provide definitive graphical proof: the Standard Ohm's Law diverges exponentially, while the Modified Law traces the experimental data with near-perfect fidelity, achieving errors on the order of $(10^{-14}\%)$. This is not a curve-fitting exercise but the validation of a physical law capturing the reality of a governed system. The exponential decay constant $(k)$, empirically determined to be approximately $(1000 s^{-1})$, quantifies the rate of resistance collapse. Its high value indicates a millisecond-scale transition to a stable state, explaining the flat temporal profiles of voltage and current observed throughout the sustained fault durations. The theoretical framework established in *Proposition 1* is thus confirmed: all short-circuit variables-resistance, voltage, and current-are bounded.

The concept of an "*instantaneous*" fault, a cornerstone of conventional protection models, is revealed to be a dangerous oversimplification. Real faults evolve over time, and this evolution is critical for accurate detection and intervention. The Modified Ohm's Law provides the first physics-consistent mathematical framework to describe this evolution, bridging the gap between the idealizations of circuit theory and the complexities of real-world electrical phenomena. This approach finds parallels in the modeling of other dynamic systems with exponential transients, such as the current decay in superconducting fault current limiters (SFCLs), where the transition from a superconducting to a resistive state is also characterized by a rapid yet finite temporal evolution [48], [49]. The underlying quantum mechanical decay of a supercurrent, a process foundational to such macroscopic transitions, has been rigorously described for Josephson junctions, reinforcing the physical plausibility of bounded exponential dynamics in governed electrical systems [50]. Furthermore, this bounded exponential behavior resonates with the dynamics observed in quantum systems where energy transitions occur discretely rather than through singularities, as reflected in the quantum ground state control of mechanical resonators [51], [52].

### 5.3 Quantifying Governed Fault Dynamics

This paper introduces a suite of novel metrics that collectively transform short-circuit analysis from a qualitative assessment of catastrophe to a quantitative engineering discipline. These metrics, derived directly from the experimental data, provide a standardized language for characterizing fault behavior and are foundational to the new definitions established as follows:

**Definition 5 (Clamped Short-circuit Configuration).** A Clamped Short-circuit Configuration is defined as an electrical state $(\Sigma(t) = \{V_{short}(t), I_{short}(t), R_{short}(t)\})$ where, for all $(t > 0)$ after fault initiation:
1. The output voltage is governed by a diode network, stabilizing at a constant value $(V_{short}(t) \equiv V_c)$, where $V_c$ is determined by the cumulative forward voltage drop of the conducting diodes and the circuit's dynamic impedance.
2. The short-circuit current $(I_{short}(t))$ is finite and bounded, $\left(I_{short}(t) \leq I_{\max}^{\text{clap}}\right)$.
3. The dynamic resistance $\left(R_{short}(t) = \frac{V_{short}(t)}{I_{short}(t)}\right)$ is non-zero and decays to a finite floor $\left(R_c^{\min} > 0\right)$.
4. The system resists both voltage collapse and current divergence through the hardware-level topological constraint imposed by the clamping diodes, operating outside strict Kirchhoffian loop rules for the fault path.

**Definition 6 (Clamped Bidirectional Short).** A Clamped Bidirectional Short is a specific instance of a Clamped Short-circuit Configuration where the net measured current $\left(I_{short}(t) = I_{short_+}(t) - \right.$

$I_{short_-}(t))$ exhibits a transient reversal or sustained bidirectional component, captured by a bidirectional sensor. In this regime:
1. Forward-biased diodes conduct the initial surge current $(I_{short_+}(t) > 0)$ while maintaining $(V_{short}(t) = V_c)$.
2. A reverse recovery current $(I_{short_-}(t) > 0)$ is measurable, indicating complex charge dynamics within the semiconductor junctions or the fault arc.
3. The system maintains voltage stability despite the non-zero net current integral $(\int I_{short}(t)dt \neq 0)$, demonstrating a non-Kirchhoffian, directional dissipation topology.

The validation of these definitions is achieved through the quantitative metrics summarized in Table 8. The Sustained Fault Efficiency (SFE), calculated as $\left(SFE = \frac{P_{CSS,max}}{P_{nom}}\right)$, yielded values of 2.71(2.5V) and 1.76(5.0V). An $(SFE > 1)$ is profoundly significant: it demonstrates that the governed fault state can dissipate more power than the circuit's nominal operating state. This finding breaks the conventional view of a short-circuit as a purely dissipative, loss-inducing event, revealing instead a sustained power delivery regime.

The Transient Clamping Index (TCI), defined as $\left(TCI = \frac{k}{\ln(1/\varepsilon)}\right)$ (Equation 85) with calculated values of approximately 137.7 and 128.4, quantifies the rapidity of system stabilization. A high TCI confirms that the circuit clamps the electrical transients on a millisecond timescale, validating its efficacy as a protection mechanism. Most definitively, the Sustained-to-Capacitive Energy Ratio (SCER) was computed to be $(\sim 1.53 \times 10^{12})$ (Equation 101). This strong ratio provides irrefutable evidence that the energy dissipated during a sustained short-circuit originates from the continuous power supply and the governed dynamics of the fault path, not from the discharge of parasitic junction capacitances. The energy from capacitance is less than one-trillionth of the total output. This metric alone refutes any alternative explanation based on transient stored energy and firmly establishes the "*sustainance*" paradigm over the "*dissipation*" paradigm.

**Table 8. Summary of Novel Metrics and Their Experimental Validation**

| Metric | Symbol and Formula | Experimental Value (2.5V Extreme) | Experimental Value (5.0V Extreme) | Physical Significance |
|---|---|---|---|---|
| Minimal Resistance Floor | $R_c^{min} = V_{min}/I_{max}$ | 5.61 mΩ | 5.30 mΩ | Quantifies the finite lower bound of short-circuit impedance, refuting the idealization of zero resistance. |
| Sustained Fault Efficiency | $SFE = \frac{V_{min} \cdot I_{max}}{V_c^2/R_0}$ | 2.71 | 1.76 | Demonstrates fault power can exceed nominal operating power, indicating a distinct high-energy regime. |

| Transient Clamping Index | $TCI = \frac{k}{\ln(1/\epsilon)}$ (Equation 85) | 137.7 | 128.4 | Quantifies the rapidity of stabilization (millisecond scale); a higher value indicates faster settling. |
|---|---|---|---|---|
| Diode Clamping Efficiency Index | $DCEI = 1 - \frac{\Delta V_{CB1}}{\Delta V_n}$ (Equation 73) | $\approx 1$ | $\approx 1$ | Confirms near-perfect output voltage regulation and decoupling from source variations by the diode network. |
| Sustained-to-Capacitive Energy Ratio | $SCER = \frac{P_{CSS} \cdot t_{faul}}{\frac{1}{2}C_j V_c^2 N}$ (Equation 101) | $1.53 \times 10^{12}$ | - | Conclusively refutes capacitive discharge as the energy source; validates the sustained energy regime. |

**5.4 Signal Fidelity and the Validation of a New Measurement Basis**

The integrity of any novel measurement system must be rigorously established to ensure that observed phenomena are physical and not artifacts of the instrumentation. The multi-resolution signal analysis conducted at $10ms$, $50ms$ and $100ms$ sampling rates, in full compliance with the Nyquist-Shannon criterion, proves that the acquired data represent authentic physical processes. The Nyquist Compliance Index (NCI) values, exceeding 300,000 across all configurations, provide a mathematical guarantee that the sampling rates were more than sufficient to capture the maximum frequency content of the fault transients, which was calculated to be below $0.16Hz$.

The application of Shannon entropy and Fourier-based analyses further validated the data's physical authenticity. The Transient Complexity Indices ($TCI \sim 0.1 - 0.3$) indicated a moderate, structured signal complexity, confirming that the waveforms were not random noise but represented a deterministic physical process. The low spectral entropy and Spectral Centroid values confirmed that the signal energy was concentrated at very low frequencies ($< 5Hz$). The Short-Time Fourier Transform (STFT) spectrograms visualized this directly, showing energy concentrated at very low frequencies throughout the entire fault duration, with no high-frequency components that would indicate missed transients or oscillatory instability. The Time-Frequency Stability Metric (TFSM) approaching unity confirmed the non-oscillatory, stable nature of the governed state.

This validation framework establishes the diode-clamp circuit not merely as an experimental apparatus, but as the first validated time-resolved short-circuit measurement system. It provides a new "*measurement basis*" for electrical faults, generating datasets with a known and verified fidelity. This is a critical advancement, as the development of AI and machine learning models for fault diagnosis has been severely hampered by the lack of such physically-grounded, high-resolution data [53]. The signal analysis proves that this system reliably captures the full electrical chronology of a fault, providing the essential data required for next-generation diagnostic algorithms. This approach aligns with emerging methodologies in real-time fault diagnosis for industrial smart manufacturing, where high-fidelity signal acquisition is recognized as foundational for effective condition monitoring [53], [54].

**5.5 The Critical Dangers of Threshold-Based Protection Paradigms**

The continued reliance on static, threshold-based protection systems, in light of this new understanding, presents profound and increasing dangers to modern power systems. These dangers stem from a fundamental mismatch between the simplified model of fault behavior and the complex reality revealed by time-resolved measurement. They include:

1. *Undetected Incipient Faults and Latent Ignition Sources.* Threshold-based systems are blind to faults that evolve slowly or operate below the trip current. The experiments show that a governed short can persist for -minutes at currents that may be below a breaker's instantaneous trip setting. In real-world systems, such as aging wiring insulation or corroded connections, this sustained sub-threshold current can generate enough heat over time to initiate fires without ever triggering the protection. This is a primary cause of electrical fires in residential and commercial buildings [55].
2. *Catastrophic Failure from Delayed Intervention in High-Energy Systems.* In battery energy storage systems (BESS) and electric vehicle (EV) packs, a latent short-circuit can self-heat, leading to thermal runaway. A threshold-based fuse or breaker may only operate after the cell has already entered an unstoppable exothermic reaction. The time-resolved measurement of resistance decay and its correlation with temperature could provide the early warning needed for preventive intervention, a capability absent in current systems [36], [56], [57].
3. *Spurious Tripping and Reduced Grid Resilience.* The increasing penetration of power electronics (inverters, drives) injects high-frequency noise and current harmonics into grids [58]. Conventional overcurrent relays can misinterpret these rapid, high-magnitude transients as faults, causing spurious tripping. This compromises the stability and reliability of microgrids and sensitive industrial processes. A time-resolved system that analyzes waveform entropy and spectral content can distinguish between a genuine fault and a switching transient.
4. *Inability to Adapt to Changing System States.* A fixed threshold cannot account for the dynamic state of a grid with distributed energy resources (DERs), where fault current contributions vary based on generation and load. This can lead to protection "blinding" or inadequate fault clearance [59], [59]. A measurement system that quantifies the dynamic impedance of a fault in real-time enables adaptive protection schemes that can reconfigure relay settings based on actual system conditions.
5. *Suppression of Data Required for Predictive Maintenance.* The very act of instantaneous tripping destroys the evidence. When a fuse blows, the data about the fault's evolution-its current ramp, resistance decay, and pre-fault signatures-is lost. This creates a vicious cycle where protection systems prevent the collection of data that would be essential for understanding failure modes and developing predictive maintenance strategies. The clamped circuit breaks this cycle by allowing the fault to be observed in its entirety.

These limitations are particularly concerning given the increasing complexity of modern power systems, where conventional protection approaches struggle to address the challenges posed by renewable integration, power electronics, and changing grid architectures [53]. The governed measurement paradigm directly addresses these vulnerabilities through its continuous monitoring capability and rich data generation.

**5.6 Implications for Power Electronics and Advanced Grid Protection**
The implications of this work extend across multiple domains of electrical engineering, offering transformative potential for system design and operation. The governed short-circuit paradigm, whose fundamental advantages are systematically compared to conventional approaches in Table 9, enables a shift from protective interruption to diagnostic management. In power electronics, the clamped circuit provides a robust platform for testing the fault tolerance of wide-bandgap semiconductors (SiC, GaN).

Designers can now subject their circuits to realistic, sustained short-circuit stresses and measure the dynamic response of their protection schemes, moving beyond destructive, single-pulse testing.

For smart grid protection, the system enables the development of "*phasor measurement units (PMUs) for faults*". The provision of time-synchronized, high-resolution data on fault voltage, current, and impedance lays the foundation for differential protection, fault location, and adaptive relay coordination with new accuracy. The ability to measure the finite resistance of a fault path, a key feature of the proposed paradigm in Table 9, could revolutionize fault location algorithms, which currently assume zero impedance. This capability directly addresses the challenge of High-Impedance Faults (HIFs), a leading cause of wildfires, which conventional systems struggle to detect [60], [61], [62].

The governed short-circuit framework suggests a new class of adaptive solid-state circuit breakers. Instead of simply interrupting, these breakers could momentarily enter a governed clamping mode, characterizing the fault through its $k$-value and SFE before deciding to interrupt or maintain power in a limited capacity. This approach leverages the continuous, high-resolution data output and predictive capabilities highlighted in Table 9. Such a system would be particularly valuable in mission-critical systems where uninterrupted operation is paramount.

**Table 9.** Analysis of Short-Circuit Measurement and Protection Paradigms

| Feature | Conventional Threshold-Based Paradigm | Proposed Governed Measurement Paradigm |
|---|---|---|
| Fundamental Model | Static, instantaneous resistance collapse ($R \to 0$) | Dynamic, time-resolved exponential decay ($R(t) = R_0 e^{-kt}$) (Equation 17) |
| Voltage Behavior | Assumed to collapse to zero | Bounded to a non-zero, measurable floor ($V_{\min} > 0$) |
| Current Behavior | Assumed to diverge to infinity, requiring interruption | Bounded to a finite, measurable maximum ($I_{\max}^{\text{clap}}$) |
| Measurement Basis | Indirect inference via peak current or thermal effects | Direct, time-resolved measurement of $V(t)$ and $I(t)$ |
| Primary Metric | Trip time, peak current magnitude | Resistance floor ($R_{\min}$), decay constant ($k$), SFE |
| Data Output | Binary (trip/no-trip), scalar ($I_{\text{peak}}$) | Continuous, high-resolution time-series dataset |
| Treatment of Fault | Destructive anomaly to be eliminated | Governed physical process to be measured and characterized |
| Application in AI | Limited to classification (fault/no-fault) | Enables regression and prediction of fault evolution |
| Thermal Management | Relies on interruption to prevent thermal runaway | Demonstrates possibility of sustained dissipation without runaway |

**5.7 The First AI-Ready Electrical Fault Dataset and its Generative Potential**

A pivotal result of this paper is the generation of the first comprehensive, time-resolved dataset of governed electrical short-circuits. This dataset is "AI-ready" because it is physically validated, multi-variate (voltage, current, derived resistance/power), and spans a wide dynamic range of fault severities and durations. Machine learning models, particularly deep neural networks for time-series classification and prediction, have been historically limited by synthetic or incomplete fault data [63], [64]. The ubiquitous missing values in real-world datasets destroy the integrity of time series and hinder effective analysis, while the generation of realistic synthetic data remains a complex challenge [65]. The clamped circuit acts as a generative fault platform, producing rich, empirical data that captures the full non-linearity and temporal structure of real faults. This dataset can be used to train models for:

- *Predictive Fault Detection.* Identifying subtle pre-fault signatures in current and voltage waveforms that precede a full short-circuit.
- *Fault Classification and Severity Assessment.* Distinguishing between different types of faults (e.g., arcing, bolted, granular) based on their unique transient signatures, entropy profiles, and decay constants (k-values).
- *Remaining Useful Life Prediction.* For components like capacitors and batteries, by correlating the evolution of fault metrics with degradation models.

This paper addresses a critical data gap in the field by providing a physically correct data source, which has been a significant barrier to the application of AI in power system protection [66]. This contribution paves the way for a new generation of intelligent, data-driven fault management systems. The public release of such empirically-grounded datasets is poised to accelerate collaborative development and establish essential benchmarks across the research community [67]. Furthermore, the critical importance of high-quality, real-time fault data is increasingly recognized not only in power systems but also within the industrial smart manufacturing sector, where it is considered foundational for advancing diagnostic capabilities [53], [68].

**5.8 Bridging to Non-Equilibrium Thermodynamics and Electrodynamics**
The governed short-circuit system exhibits profound parallels with other non-equilibrium physical systems, suggesting a deeper connection that transcends conventional electrical engineering. The transition from a high-resistance, low-current state to a low-resistance, high-current governed state is analogous to a non-equilibrium phase transition or the formation of a dissipative structure. Modern research into stabilizing quantum many-body systems demonstrates that active feedback control, which breaks detailed balance, is key to engineering such otherwise inaccessible steady states [69]. In these systems, being driven far from thermodynamic equilibrium allows for spontaneous order and stable, high-energy dissipation states to emerge, a principle that aligns with the sustained fault regime observed in this work.

The exponential decay of resistance mirrors relaxation processes in disordered systems. The finite resistance floor $(R_c^{\min})$ represents a fundamental lower bound, analogous to a critical point in phase transitions. This finds a striking parallel in recent experimental discoveries of abrupt quantum critical points, where phenomena such as the sudden cessation, or "*death*", of quantum fluctuations defy established theoretical descriptions like Ginzburg-Landau theory [70], [71]. The sustained energy dissipation in a stable, non-equilibrium state is a hallmark of systems driven away from thermodynamic equilibrium, a universality phenomenon observed across wide classes of large and complex systems despite their microscopic differences [72], [73].

From an electrodynamic perspective, the *Clamped Bidirectional Short* described in Definition 6 challenges the assumption of a purely conservative field. The system exhibits a non-conservative topology where $(\oint V_i dt \neq 0)$ over a fault cycle, indicating the diode network creates a directional dissipation field. This is reminiscent of systems with non-holonomic constraints in classical mechanics. The Modified Ohm's Law can thus be viewed as a phenomenological law for a specific class of non-equilibrium electrodynamic systems, opening a new avenue for theoretical work that bridges circuit theory with the thermodynamics of irreversible processes.

This perspective finds resonance in the quantum description of electrical circuits, where macroscopic quantum phenomena demonstrate that circuit behavior must be understood through quantum mechanical principles rather than classical approximations alone. Contemporary research confirms that entire superconducting circuits can behave as single quantum objects, bridging the gap between quantum theory and macroscopic electrical engineering [46], [47]. The bounded, quantized states observed in superconducting qubits and Josephson junctions share conceptual similarities with the governed, finite states of the macroscopic short-circuit system (in a parallel, [74]), suggesting universal principles governing electrical behavior across scales.

**5.9 Thermal Considerations and the Path to Industrial Application**

The experimental results demonstrate remarkable thermal stability over extended durations ($3 - 15$ - minutes), with no observed thermal runaway behaviors in the diode-clamp network. This is attributed to the governed power dissipation ($P_{CSS}$) and the effective heat sinking of the components. However, for industrial scaling to higher voltages and currents, active thermal management becomes a critical research frontier. The recommendation for future work is the direct integration of junction temperature sensors to correlate real-time temperature with the electrical metrics ($V_{\min}$), ($I_{\max}$) and ($R_c^{\min}$).

This will enable the development of multi-domain fault models that couple electrical and thermal dynamics. For instance, the exponential decay constant ($k$) may be expressed as a function of temperature ($T$), such as $\left(k(T) = k_0 e^{-\frac{E_a}{(k_B T)}}\right)$, where ($E_a$) is an activation energy related to the ionization process in the fault path. Monitoring thermal behavior will be essential for designing clamped circuits for high-power applications, ensuring that the sustained fault state remains within the safe operating area (SOA) of all components, thus transitioning the laboratory prototype into a robust industrial technology.

The thermal stability observed in the present experiments aligns with research on high-coherence superconducting qubits, where managing quasiparticle dynamics and dissipation is crucial for maintaining quantum states [75]. In both contexts, understanding and controlling energy dissipation pathways is essential for system stability and performance. This parallel suggests that insights from fault management in classical systems might inform approaches to error suppression in quantum systems, and vice versa.

**5.10 Towards Non-Kirchhoffian Frameworks**

The experimental observations, particularly of the *Clamped Bidirectional Short* (Definition 6), challenge the universal applicability of Kirchhoff's circuit laws for describing all electrical phenomena. The governed short-circuit system exhibits a non-conservative topology where the sum of voltage drops around a loop involving the fault path and the clamping diodes is not zero over a cycle. This is a direct consequence of the directional, non-linear nature of the diode clamping mechanism, which imposes a preferred path for energy flow and dissipation. This behavior aligns with the principles of topolectrical circuits, which utilize non-reciprocal couplings and active components to create circuit behaviors that are not constrained by traditional conservative network theory [76].

These findings suggest the need for a broader theoretical framework that can accommodate such "*non-Kirchhoffian*" circuits. This framework would integrate traditional lumped-element models with components that enforce topological constraints, leading to a hybrid model where certain subsystems are governed by directional dissipation rules. The concepts of Definition 5 and Definition 6 provide the initial axioms for such a framework. Future theoretical work should focus on developing a complete mathematical formalism for these "*directional dissipation systems*", potentially drawing from network theory and the thermodynamics of non-equilibrium systems to model their behavior formally. The established Laplacian formalism used in topolectrical circuit analysis provides a powerful starting point for this formal mathematical development [76]. Furthermore, explorations into the relationship between nonlinear Kirchhoff circuits and other physical theories hint at the potential for profound theoretical unification beyond classical models [77]. In this context, the diode-clamp circuit represents not just a technical implementation but a physical instantiation of a new conceptual model for electrical networks - one that acknowledges and leverages non-conservative topology as a fundamental design principle rather than treating it as an anomaly to be eliminated.

**5.11 Unconventional Applications (Fault Energy Harvesting and Managed Dissipation)**
The paradigm of the governed short-circuit introduces one of its most unconventional implications: the potential for fault energy harvesting and managed dissipation as recently applied in the study [78]. The Sustained Fault Efficiency (SFE) metric, demonstrating that a governed fault constitutes a state of sustained power delivery, opens the theoretical door to concepts where fault energy is temporarily channeled and utilized rather than merely dissipated as heat. This energy could potentially power critical safety systems or emergency communication modules during a grid disturbance prior to a controlled shutdown, thereby transforming a fault from a purely negative event into a potential resource for enhancing system resilience [79], [80].

While this concept remains speculative in this paper, it aligns with vibrant research into harnessing non-equilibrium processes for energy applications. Recent investigations into Bound States in the Continuum (BICs) for acoustic energy harvesting demonstrate that highly localized energy enhancement from ambient, low-density sources is physically achievable, outperforming conventional resonator harvesters by a significant factor in output power [80]. This principle suggests that a governed plasma channel, sustained within a specialized clamp circuit, could similarly act as a controlled thermal source for thermoelectric generators. The governed fault state, under specific topological constraints, can be re-conceptualized as a distinct, high-power operational mode rather than a failure condition. This perspective is further strengthened by the emergence of quantum-based sensing on power grids, such as China's "Diamond Ring" quantum current transformer, which leverages the sensitive quantum states of nitrogen-vacancy centers in diamond for new measurement precision, illustrating the broader trend of harnessing quantum phenomena for electrical system stability and management [81].

The concept of fault energy harvesting connects logically to the growing global energy harvesting system market, which is driven by the demand for sustainable, maintenance-free power solutions for IoT and wireless sensor networks [79]. The research trajectory outlined in Table 10 explicitly identifies the long-term vision of "*Harnessing governed non-equilibrium states for novel computing or energy systems*", positioning the exploration of fault energy harvesting as a strategic cross-disciplinary frontier. The efficient management and potential utilization of energy traditionally viewed as "*waste*" or "*hazard*" represents a frontier for innovation that bridges the mature field of electrical protection with emerging energy technologies.

**5.12 A New Foundation for Electrical Engineering**
The experimental and theoretical advancements presented in this work collectively establish a new foundation for understanding and managing electrical short-circuits. This research demonstrates the successful transition of the short-circuit from a destructive singularity to a governed physical process through the diode-clamp circuit paradigm. The complete framework encompasses an empirically validated theoretical foundation via the Modified Ohm's Law, establishes a comprehensive suite of quantitative diagnostic metrics, and generates the first AI-ready, time-resolved fault dataset. These elements provide the basis for a fundamental shift in electrical fault science, aligning with and accelerating the broader industry transition towards data-driven fault diagnosis and predictive maintenance [82], [83]. The systematic deconstruction of the incumbent threshold-based paradigm reveals its critical limitations and dangers, while simultaneously illuminating a transformative path forward for advanced grid protection and fault-tolerant electronics. The findings challenge core precepts of conventional electrical engineering not through criticism alone, but through the presentation of a complete, validated alternative system. The structured research trajectory detailed in Table 10 provides a clear and actionable pathway for advancing this new paradigm across multiple domains. This roadmap, spanning from immediate next steps in thermal modeling to the long-term vision of establishing "*Fault Physics*" as a distinct sub-discipline, ensures that the foundational work presented here will catalyze continued progress in both fundamental research and industrial application.

The journey from conceptualizing a short-circuit as a catastrophic failure to be interrupted to understanding it as a governable process to be measured represents a fundamental transformation in electrical engineering. This paradigm demonstrates that the traditional barriers between theoretical abstraction and physical reality can be dissolved through innovative design and rigorous measurement. As power systems continue evolving toward greater complexity and interconnection, the capacity to characterize, predict, and intelligently manage electrical faults in real-time becomes increasingly essential. The theoretical framework, methodological tools, and empirical data provided by this research establish the necessary foundation to fundamentally reshape the science and engineering of electrical faults for the 21st century. The implications of this governed fault paradigm extend from practical grid protection - enabling self-healing, resilient power networks as envisioned in Table 10 -to potential quantum-domain applications, suggesting a future where the principles of governed non-equilibrium states could inform everything from macroscopic power management to the operation of fault-tolerant quantum circuits [81], [84].

**Table 10.** Research Trajectory and Potential Applications of the Governed Short-Circuit Paradigm

| Research Domain | Immediate Next Steps | Mid-Term Applications | Long-Term Vision |
|---|---|---|---|
| Fundamental Physics | Integrate thermal imaging to create electro-thermal fault models | Develop a unified theory of governed faults in plasmas and semiconductors | Establish "*Fault Physics*" as a distinct sub-discipline |
| Circuit Design and Protection | Scale the clamping topology to 100 V/100 A using TVS arrays and active FETs | Design fault-tolerant power supplies and solid-state circuit breakers with diagnostic capabilities | Implement grid-level fault clamps for self-healing, resilient power networks |
| Data and AI | Publicly release the time-resolved fault dataset for the ML community | Develop prognostic AI models for battery and insulation failure prediction | Deploy distributed fault sensors that continuously train grid-wide digital twins |
| Theoretical Foundations | Formalize the mathematical framework for Non-Kirchhoffian circuits | Integrate governed fault models into standard circuit simulation tools | Develop a thermodynamic theory of electrical faults as dissipative structures |
| Cross-Disciplinary | Investigate governed faults in electrochemical systems (batteries) | Explore analogies with sustained phenomena in | Harness governed non-equilibrium states for |

| | | biological and chemical networks | novel computing or energy systems |

## 5.13 Educational Transformation and Industrial Paradigm Shift
The governed short-circuit measurement system establishes a transformative framework that bridges electrical engineering education with industrial practice. This paradigm transitions electrical fault analysis from destructive interruption to governed measurement, creating new possibilities for both pedagogy and practical applications across multiple domains.

### 5.13.1 Educational Advancement through Experimental Innovation
Traditional electrical engineering education relies on laboratory experiments that verify established principles within safe operational boundaries. Semiconductor diode characterization typically demonstrates forward conduction and reverse blocking behavior using Standard Ohm's Law, providing students with fundamental but incomplete understanding of device physics. The diode-clamp circuit network transcends these limitations by enabling investigation of semiconductor behavior under extreme electrical conditions while maintaining system integrity. This experimental platform creates new learning opportunities that integrate theoretical circuit analysis with practical semiconductor physics. Students utilizing Arduino microcontrollers and standard sensors can quantitatively explore the relationship between input voltage and reverse current impedance for various diode types, including the 1N5408 devices employed in this study. The system provides empirical validation of concepts traditionally taught as theoretical abstractions, including minority carrier dynamics and junction breakdown mechanisms. The observed deviations from Kirchhoff's laws within the clamped network introduce students to non-conservative electrical systems, challenging idealized circuit models while demonstrating rigorous experimental methodology.

### 5.13.2 Limitations of Conventional Fault Diagnostic Technologies
The governed measurement system addresses fundamental constraints inherent in conventional fault diagnostic approaches across multiple domains. Digital multimeters provide only static resistance measurements in de-energized systems, completely missing the dynamic evolution of developing faults. Clamp meters capture current magnitude but lack the temporal resolution to characterize fault initiation dynamics and provide no insight into corresponding voltage behavior or dynamic impedance.

Advanced diagnostic tools face similar limitations despite their specialized capabilities. Thermal imaging cameras identify fault locations through heat signatures but only after substantial energy dissipation has occurred, representing reactive rather than proactive fault management. Short-circuit testing machines and prospective fault current meters provide design validation data but operate exclusively in controlled environments, unable to perform real-time monitoring in operational systems. These tools collectively represent a detection-oriented paradigm that treats short-circuits as binary conditions rather than physical processes requiring characterization.

Even sophisticated computational approaches including circuit analysis software and SPICE simulators depend critically on accurate physical parameters and validated behavioral models. Without empirical data from actual fault events, these simulations risk perpetuating idealized assumptions regarding instantaneous parameter changes and zero-impedance paths. The absence of governed measurement capabilities creates a fundamental gap between simulated predictions and actual electrical behavior during fault conditions. The comparative limitations of these conventional methodologies are systematically summarized in Table 11, which highlights the transformative capabilities of the governed measurement approach across multiple diagnostic parameters.

**Table 11.** Comparative Analysis of Fault Diagnostic Methodologies

| Methodology | Measurement Capability | Temporal Resolution | Governing Capacity | Data Richness |
|---|---|---|---|---|
| **Digital Multimeter** | Static resistance/continuity | Single measurement | None | Scalar values |
| **Clamp Meter** | Current magnitude | Limited bandwidth | None | Time-series current only |
| **Thermal Camera** | Temperature distribution | Seconds to minutes | None | Thermal images |
| **Circuit Analysis Software** | Simulated parameters | Computational | Algorithmic | Simulated waveforms |
| **Proposed Governed Measurement System** | Dynamic $V$, $I$, $R$, $P$ | Millisecond resolution | Hardware-enforced | Multivariate time-series |

Optical and photonic fault detection systems provide sophisticated, complementary methodologies to conventional electrical measurement but face inherent physical limitations that can preclude comprehensive electrical fault characterization [85], [86], [87]. A primary constraint is their inability to directly measure essential electrical parameters like dynamic impedance, which is critical for complete power system analysis [85]. Furthermore, their frequent reliance on threshold-based triggering means they often capture fault events only after establishment, missing the critical initiation transients that are key to understanding fault genesis [87]. Contemporary computational frameworks for fault parameter extraction demonstrate recognition of conventional methodology limitations. The approach developed by Mahmoud et al. (2025), [18] provides sophisticated mathematical techniques for estimating fault current parameters including decay time constants and symmetrical current components. However, these computational methods maintain dependencies on traditional measurement paradigms and analytical assumptions that limit applicability. Machine learning approaches face similar foundational constraints, as their performance remains fundamentally limited by the empirical data available for training and validation [66], [82].

### 5.13.3 Industrial Applications and Implementation Pathways
The transition from a paradigm of dangerous fault interruption to one of governed fault measurement enables transformative applications across multiple industrial domains. This new framework facilitates continuous operational characterization and intelligent management of electrical faults, creating pathways to enhanced system resilience, predictive maintenance, and new safety protocols. The following analysis details these industrial applications, supported by emerging research, and outlines their structured implementation as summarized in Table 12.

**Predictive Maintenance Systems.** The governed short-circuit measurement system enables a revolutionary approach to predictive maintenance (PdM) in industrial electrical systems. Its continuous monitoring capability, particularly through dynamic resistance metrics like the decay constant ($k$) formalized in (Equation 17), provides quantitative, real-time indicators of insulation degradation and contact deterioration long before catastrophic failure occurs. This facilitates a shift from scheduled or reactive maintenance to a precise, condition-based scheduling model, offering opportunities to directly extend equipment lifespan and enhancing operational safety in critical manufacturing and infrastructure systems. The integration of machine learning models with real-time monitoring data is a cornerstone of modern predictive maintenance frameworks, with studies demonstrating that algorithms like Random Forest can achieve high accuracy in predicting failure cycles based on evolving operational data [88]. This aligns with a broader systematic review of PdM in manufacturing, which highlights the role of artificial intelligence and real-time monitoring in transitioning from time-based to condition-based maintenance strategies, thereby significantly reducing unplanned downtime [89].

**Battery Safety Management.** For lithium-ion battery systems, where internal short-circuits are a primary failure mechanism leading to thermal runaway ([11], [90]), this governed measurement framework provides new diagnostic capabilities. It enables the direct observation of developing internal shorts through their characteristic exponential resistance decay signatures, offering an early warning indicator that conventional voltage and temperature monitoring cannot detect. This early detection capability facilitates preventive interventions before the onset of unstoppable exothermic reactions and opens the door to managed fault scenarios, where controlled discharge strategies for compromised cells can maintain limited functionality while managing fault energy within safe thermal limits. Research into next-generation power electronics materials emphasizes the critical importance of thermal management and reliability for high-energy-density systems, underscoring the value of any technology that can mitigate thermal runaway risks [91]. Furthermore, the general principles of prognostics and health management (PHM), which are central to predictive maintenance, are directly applicable to diagnosing incipient faults in complex electrochemical systems like batteries [89].

**Adaptive Grid Protection.** Modern power systems, characterized by high penetration of distributed energy resources (DERs) and power electronic interfaces, present complex protection challenges that conventional threshold-based systems are ill-equipped to handle. The governed measurement system enables the development of adaptive protection schemes that respond to the dynamic temporal characteristics of a fault -such as its $k$ -value and entropy profile -rather than relying on fixed current magnitude thresholds. This capability allows for highly accurate discrimination between genuine faults and legitimate transients, such as inrush currents from motor starts or switching artifacts from inverters, thereby reducing nuisance tripping while maintaining or even improving protection sensitivity. This is particularly critical for detecting high-impedance faults (HIFs) [92], which often draw currents below the pickup setting of conventional overcurrent relays yet pose significant safety hazards, including fire risks. The reliability of power electronic systems themselves is a key enabler for this adaptive grid architecture, with research dedicated to condition monitoring [93], [94], fault-tolerant strategies [95], and the robustness of wide-bandgap semiconductor devices in grid applications [96]. The analysis of enhanced switched impedance inverters, for instance, includes reliability assessment using Markov modeling, reflecting the grid's need for highly reliable and resilient power conversion components [97].

**Power Electronics Validation.** The power electronics industry faces significant challenges in validating the fault tolerance of wide-bandgap (WBG) semiconductor devices (SiC, GaN) and their associated protection systems. Conventional testing methodologies often rely on destructive single-pulse methods that provide limited insight into device behavior under sustained electrical and thermal stress. The governed short-circuit measurement system enables non-destructive characterization across the complete fault timeline, from initial transient through sustained operation. This provides invaluable data for device manufacturers seeking to optimize ruggedness characteristics and for system designers developing application-specific protection schemes that ensure reliability under extreme conditions. The intensive research into next-generation power electronics materials beyond SiC and GaN, such as B2O3 and BeO, seeks to overcome limitations like intrinsic defects and high fabrication costs, with a focus on properties critical for high-performance power devices [91]. Concurrently, the reliability of existing WBG devices is a major research focus, with special issues in leading journals dedicated to topics including failure mechanisms, robustness validation, and condition monitoring of power electronic components and systems [94].

**Table 12.** Implementation Pathways and Research Priorities

| Application Domain | Development Priority | Potential Impact | Implementation Timeline |
|---|---|---|---|
| **Battery Safety Systems** | High | Prevention of thermal runaway events | Near-term |
| **Adaptive Grid Protection** | High | Improved reliability and resilience | Mid-term |
| **Predictive Maintenance** | High | Reduced downtime and extended equipment life | Near-term |
| **Power Electronics Validation** | Medium | Enhanced device reliability | Near-term |
| **Quantum Sensor Integration** | Medium | Fundamental new measurement capabilities | Long-term |

The implementation pathways detailed in Table 12 provide a strategic roadmap for advancing this technology from laboratory validation to broad industrial deployment. The near-term focus on battery safety, predictive maintenance, and power electronics validation leverages the most immediate and high-impact opportunities for risk reduction and performance gains. The mid-term development of adaptive grid protection addresses the critical evolution of power systems, while the long-term vision of quantum sensor integration promises a future generation of measurement systems with fundamental new capabilities.

**5.13.4 Fundamental Advantages and Future Directions**
The diode-clamp circuit network establishes a superior approach to electrical fault analysis through its unique combination of measurement capabilities and operational characteristics. The system's capacity for continuous fault operation without destruction represents a paradigm shift from interruption-based to measurement-based fault management. The bounded operation, formalized through Proposition 1 and quantitatively described by the Modified Ohm's Law, ensures all electrical parameters remain within measurable ranges throughout fault duration.

The multi-resolution signal acquisition capability, validated through Nyquist compliance indices exceeding 300,000, ensures complete capture of fault dynamics across relevant timescales. This comprehensive temporal resolution enables detailed analysis of both rapid initiation transients and extended stabilization phases, providing insights previously inaccessible through conventional measurement approaches. The practical implementation advantages, utilizing commercially available components and straightforward topology, facilitate adoption across diverse educational, research, and industrial contexts.

Future research should explore integration with emerging quantum sensing technologies, particularly approaches leveraging nitrogen-vacancy centers in diamond for magnetic field measurement. The governed fault environment provides ideal conditions for evaluating quantum sensor performance under realistic electrical stress, potentially enabling new characterization approaches with quantum-limited sensitivity. This synergy between macroscopic electrical measurement and quantum sensing represents a promising frontier for both fields.

The governed short-circuit measurement system ultimately establishes a new paradigm for electrical engineering that transforms destructive events into measurable processes. This approach enables previously impossible investigations into fault physics while providing the empirical foundation for next-generation protection strategies, predictive maintenance systems, and safety management approaches. The technology's unique capabilities address critical limitations across electrical engineering practice while creating new opportunities for innovation in fault characterization, system resilience, and educational methodology.

## 6.0 Conclusion

This work establishes a fundamental paradigm shift in the science of electrical faults, transforming the short-circuit from a destructive, unmeasurable singularity into a governed, time-resolved physical process. The novel diode-clamp circuit architecture has enabled the first complete empirical characterization of a sustained short-circuit, definitively refuting the classical assumption of instantaneous resistance collapse and unbounded current. The experimental results provide unequivocal evidence that all short-circuit variables are bounded: the dynamic resistance decays exponentially to a finite, non-zero floor of approximately $(R_c^{min} = 5.61 m\Omega)$ at $(2.5V)$ and $(R_c^{min} = 5.30 m\Omega)$ at $(5.0V)$, the voltage stabilizes at a governed minimum near $(V_{min} \approx 0.07V)$, and the current is capped at a measurable maximum of approximately $(I_{max}^{clap} \approx 13.0A)$. These boundaries are not imposed by external interruption but are intrinsic to the physics of the governed fault path as constrained by the clamping network, a finding consistent across both nominal and extreme experimental configurations.

The introduction and experimental validation of the time-resolved Modified Ohm's Law provides the essential theoretical framework for this new paradigm. This formulation accurately describes the observed exponential current trajectories and bounded state, achieving a predictive fidelity with errors on the order of $(10^{-14}\%)$ when compared to experimental data. The Standard Ohm's Law, in contrast, is shown to diverge unrealistically, proving its inadequacy for modeling dynamic fault conditions. The exponential decay constant $(k \approx 1000 \text{ s}^{-1})$, extracted from the data, quantifies the millisecond-scale transition to a stable, sustained fault regime, explaining the remarkable temporal invariance of electrical parameters observed over durations of 3 -minutes, 10 -minutes, and 15 -minutes. This theoretical advancement bridges the gap between ideal circuit theory and the complex reality of non-equilibrium electrical processes.

A suite of novel, quantitatively validated metrics now provides a rigorous language for fault analysis. The Sustained Fault Efficiency (SFE), with values of $(SFE = 2.71)$ at $(2.5V)$ and $(SFE = 1.76)$ at $(5.0V)$, demonstrates that the governed fault state can dissipate more power than the nominal circuit operation, revealing a distinct high-energy regime that defies conventional dissipation models. The Transient Clamping Index (TCI) of approximately $(TCI \approx 137.7)$ confirms the rapid stabilization of the circuit. Most conclusively, the Sustained-to-Capacitive Energy Ratio (SCER) of $(\sim 1.53 \times 10^{12})$ provides definitive proof that the energy dissipated during a sustained fault, calculated as $(E_{\text{total, fault}} = 171 \text{J})$ over 180 -seconds, originates from the continuous power supply and the governed dynamics of the fault path. This ratio, exceeding one trillion, irrefutably negates the notion that the observed phenomena are a transient discharge of parasitic junction capacitance, whose total energy is a negligible $(E_{\text{max,capacitive}} \approx 1.12 \times 10^{-10} \text{J})$.

The multi-resolution signal analysis, conducted at $10ms$, $50ms$ and $100ms$ sampling rates with Nyquist Compliance Indices (NCI) exceeding (300,000), rigorously validates the system as the first true time-resolved short-circuit measurement basis. The low spectral entropy, dominant frequencies below ($0.16Hz$), and near-unity Time-Frequency Stability Metric (TFSM) confirm the capture of authentic physical processes without aliasing or significant transient loss. This fidelity is demonstrated across the operational range, with the circuit showing exceptional sensitivity to fault transients at the $10ms$ resolution for capturing fine-grained dynamics, while the $50ms$ and $100ms$ resolutions provide robust characterization of the sustained governed state. This generates the first AI-ready, empirically grounded dataset of electrical faults, providing the physically consistent, high-fidelity data required to train next-generation prognostic models for fault prediction, classification, and severity assessment.

The implications of this governed fault paradigm are profound and far-reaching. It exposes the critical vulnerabilities of conventional, threshold-based protection systems, which remain blind to the temporal evolution of sub-threshold faults that can lead to latent ignition and catastrophic thermal runaway. This work instead enables a future of intelligent, adaptive protection based on real-time analysis of fault dynamics, moving beyond simple binary trip decisions to a diagnostic management of electrical health. The circuit serves as a generative platform for designing fault-tolerant electronics, validating wide-bandgap semiconductor ruggedness, and developing solid-state breakers with integrated diagnostic capabilities. Overall, by re-contextualizing a short-circuit as a measurable process rather than a catastrophic event, this paper dissolves the traditional barrier between theoretical abstraction and physical reality. The established framework, encompassing a validated theoretical model, a suite of diagnostic metrics, and a generative measurement platform, provides the essential components for a fundamental restructuring of electrical safety engineering, power system design, and the development of resilient, adaptive, and intelligent energy systems.

**Acknowledgments.** This work was conducted without any funding.

**Conflicts of interest.** The authors declare no conflict of interest.

**Data availability statement.** The data used to support the findings of this study are included in the supplementary materials.

**Author contributions.** Conceptualization, organize: AMK and DMK; methodology: AMK; software: AMK; investigation: AMK and DMK; writing-the first draft preparation: AMK; writing-review and editing: AMK and DMK. All authors have read, revised and agreed to the published version of the manuscript.

**Appendix A. Overcoming Parasitic and Thermal Limitations**
**A.1. Parasitic Regimes in Conventional Clamping Circuits**
Traditional diode clamping circuits are architected for a singular purpose: the mitigation of high frequency, short-duration voltage transients, such as those from electrostatic discharge (ESD) [98] or inductive switching [99]. Their operational paradigm is fundamentally instantaneous and reactive. In a typical configuration, a capacitor couples the AC signal, and a diode clamps the voltage to a reference rail (e.g., $V_{CC}$ or GND), effectively adding a DC offset to the entire waveform [100]. The RC time constant is carefully selected to be significantly longer than the period of the input signal, ensuring the capacitor voltage does not discharge significantly between transient events. This conventional approach encounters two critical failures under sustained fault conditions:

1. *Parasitic Capacitance Discharge* [101]:- The core of a traditional AC-coupled clamp is its storage capacitor. During a sustained low-impedance fault, this capacitor would discharge its energy almost instantaneously-within milliseconds or less-into the fault path. This action provides a brief clamping action but is followed by a complete collapse of the clamping function. The circuit then behaves as a direct short, leading to the catastrophic current singularity predicted by standard Ohm's Law and the destruction of the diode and/or other components. The circuit is blind to the fault's evolution after this initial transient.

2. *Thermal Runaway and Sensitivity* [102]:- Conventional ESD clamping diodes are not designed for continuous power dissipation. When subjected to the sustained current of a fault, their junction temperature rises rapidly. Due to the negative temperature coefficient of the diode's forward voltage, increasing temperature lowers the voltage drop, which in turn allows for even more current to flow, creating a positive feedback loop known as thermal runaway [103], [104]. This process quickly destroys the diode, making any form of continuous monitoring impossible.

These limitations render traditional clamps entirely unsuitable for the time-resolved study of fault dynamics, as they are intrinsically designed to either suppress a transient momentarily or be sacrificed in the process.

**A.2. The Governed, Resistive Network Paradigm**

The clamped bidirectional short-circuit diode network presented in this paper operates on a fundamentally different principle: it establishes a governed, resistive voltage divider that is active throughout the entire fault duration. The topology of three 1N5408 diodes, as instantiated in "Circuit Block 1" ($CB_1$), does not function as a traditional AC-coupled clamp. Instead, it forms a non-linear resistive network that leverages the forward and reverse bias characteristics of the diodes to create a stable, bounded electrical environment.

This architecture deliberately avoids reliance on capacitive energy storage. The governing mechanism is the diodes' intrinsic forward voltage ($V_f$) and their dynamic on-resistance. During a fault, the network presents a small but finite and stable resistance-the Minimal Resistance Floor $R_c^{min} \approx 5.5\text{m}\Omega$ identified in our results. This resistance, in series with the adjusted reference resistance $R_0$, limits the current to a finite maximum $I_{max}^{clap} \approx 13.0\text{A}$ and maintains a non-zero voltage floor $V_{min} \approx 0.07\text{V}$. The circuit is therefore not "clamping" in the transient suppression sense but is actively "governing" the fault by enforcing a stable, high-current operating point defined by the Modified Ohm's Law.

**A.3. Overcoming Limitations of Conventional Circuits**

The experimental framework, enabled by this novel governing network, directly overcomes the limitations of conventional circuits:

- *Bypassing Parasitic Problems:-* By eliminating the need for a large coupling capacitor and operating the diode network in its sustained conduction regime, the architecture is immune to the parasitic discharge problem. The fault is powered continuously by the external supply, and the network's dynamic resistance defines the circuit's behavior, not a transient storage element. This is empirically validated by the Sustained-to-Capacitive Energy Ratio (SCER $\sim 10^{12}$), which proves the energy dissipated over minutes is trillions of times greater than what could be stored in any parasitic junction capacitance.
- *Suppressing Thermal Runaway:-* The use of 1N5408 power diodes, combined with the network's inherent current-limiting property, prevents thermal runaway. The power dissipation $P_{CSS}$ is stable and governed, allowing the diodes to reach a thermal equilibrium. This is evidenced by the flat temporal profiles of voltage and current recorded over 30-minutes durations, which would be impossible if the diodes were undergoing destructive thermal stress. The circuit design transforms the fault from a destructive event into a sustained, measurable state of operation.

**A.4. Distinction (Standard Clamped Vs. Bidirectional Governed Networks)**

The operational paradigm introduced in this paper represents a fundamental departure from established diode clamping methodologies. The critical distinction lies not in incremental improvement but in a complete re-imagination of the diode's role: from a transient suppressor to a governing element that enables sustained measurement. This shift is characterized by profound differences in operational mode, temporal domain, and the strategic exploitation of bias conditions, as systematically compared in Table A.4.1.

**Table A.4.1.** Comparative Analysis of Diode-Based Circuit Paradigms

| Feature | Standard Clamped Diode Circuit | This Work (Clamped Bidirectional Governed Network) |
|---|---|---|
| Primary Function | Transient voltage suppression; protection against ESD/spikes. | Creation of a governed, sustained fault environment for measurement. |
| Operational Time | Microseconds to milliseconds. | Minutes to tens of minutes (continuous). |
| Core Mechanism | Capacitive coupling and rapid diode switching to a reference rail. | A resistive network of diodes operating in forward and reverse bias to form a stable voltage divider. |
| Role of Diodes | To be non-conducting until a threshold is exceeded, then to briefly conduct. | To be continuously active, using their forward voltage and dynamic resistance to define the fault's electrical boundaries. |
| Exploitation of Bias | Uses unidirectional forward bias to shunt over-voltage transients. | Strategically uses reverse-biased diodes (e.g., $D_2$ in its idle state) as part of the network's impedance. These diodes can transition to conduction to absorb reverse current transients, enforcing bidirectional stability and enabling the observation of complex recovery phases. |
| Outcome | The fault is suppressed or the circuit is destroyed. Data is lost. | The fault is sustained, governed, and fully characterized. A rich, time-resolved dataset is generated. |

As Table A.4.1 elucidates, the experimental results presented in this paper overcome classical sensitivity and thermal runway challenges not by refining traditional clamping, but by superseding its underlying paradigm. The governed bidirectional network, analyzed through the lens of the Modified Ohm's Law, re-conceives a short-circuit not as a transient to be suppressed, but as a physical process to be measured. This foundational shift enables the first-ever continuous, time-resolved dissection of electrical fault dynamics over extended durations, establishing a new measurement basis for fault science.